\documentclass[12pt]{article}
\newlength{\abstractwidth}
\usepackage{epsf}

\renewcommand{\thefootnote}{\fnsymbol{footnote}}
\renewcommand{\thanks}[1]{\footnote{#1}}
\newcommand{\starttext}{
\setcounter{footnote}{0}
\renewcommand{\thefootnote}{\arabic{footnote}}}

\newcommand{\bea}{\begin{eqnarray}}
\newcommand{\eea}{\end{eqnarray}}
\newcommand{\ee}{\end{equation}}
\newcommand{\be}{\begin{equation}}

\def\cA{{\cal A}}
\def\cB{{\cal B}}
\def\cC{{\cal C}}

\def\cF{{\cal F}}
\def\cG{{\cal G}}
\def\cH{{\cal H}}
\def\cI{{\cal I}}
\def\cJ{{\cal J}}
\def\cK{{\cal K}}
\def\cL{{\cal L}}

\def\cN{{\cal N}}

\def\cT{{\cal T}}
\def\cU{{\cal U}}

\def\bC{{\bf C}}

\def\Re{{\rm Re}}
\def\Im{{\rm Im}}
\def\tr{{\rm tr}}

\def\half{ {1\over 2}}
\def\p{\partial}

\def\a{\alpha}
\def\b{\beta}
\def\tet{\vartheta}
\def\ep{\varepsilon}

\def\eps{\epsilon}
\def\g{\gamma}

\def\G{\Gamma}

\def\g{\gamma}

\def\ch{{\rm ch}}
\def\sh{{\rm sh}}

\def\tg{{\rm tg}}
\def\cotg{{\rm cotg}}

\def\no{\nonumber}

\begin{document}
\starttext
\setcounter{footnote}{0}

\begin{flushright}
UCLA/07/TEP/09 \\
 1 May 2007
\end{flushright}

\bigskip

\begin{center}

{\Large \bf Exact half-BPS  Type IIB interface  solutions  I: 

\medskip

Local solution and supersymmetric Janus}

\vskip .7in 

{\large  Eric D'Hoker,  John Estes and  Michael Gutperle}

\vskip .2in

 \sl Department of Physics and Astronomy \\
\sl University of California, Los Angeles, CA 90095, USA

\end{center}

\vskip .5in

\begin{abstract}

The complete Type IIB supergravity solutions with 16 supersymmetries are 
obtained on the manifold $AdS_4 \times S^2 \times S^2 \times \Sigma$ with 
$SO(2,3) \times SO(3) \times SO(3)$ symmetry in terms of two holomorphic 
functions on a Riemann surface $\Sigma$, which generally has a boundary.
This is achieved by reducing the BPS equations using the above symmetry requirements, 
proving that all solutions of the BPS equations solve the full Type IIB 
supergravity field equations, mapping the BPS equations onto a new 
integrable system akin to the Liouville and Sine-Gordon theories, and 
mapping this integrable system to a linear equation which can be solved 
exactly. Amongst the infinite class  of solutions, a non-singular Janus solution 
is identified which provides the  AdS/CFT dual of the maximally supersymmetric 
Yang-Mills interface theory discovered recently. The construction of general 
classes of globally non-singular solutions,
including fully back-reacted $AdS_5 \times S^5$ and supersymmetric
Janus doped with D5 and/or NS5 branes,
is deferred to a companion paper \cite{EDJEMG2}.

\end{abstract}

\newpage

\baselineskip=16pt
\setcounter{equation}{0}
\setcounter{footnote}{0}

\section{Introduction}
\setcounter{equation}{0}

A particularly interesting application of the AdS/CFT correspondence 
\cite{Maldacena:1997re, Gubser:1998bc, Witten:1998qj} (for reviews, see 
\cite{D'Hoker:2002aw, Aharony:1999ti}) is provided by conformal field theory (CFT)
in the presence of a planar interface or a planar defect.\footnote{We distinguish 
between the interface and defect theories as follows. Compared to the bulk theory, 
the  defect theory has extra degrees of freedom localized on the defect, while 
the interface does not.}
The addition of a planar interface to four-dimensional $\cN=4$ super Yang-Mills, 
(specified by interface couplings of local bulk operators which are supported only 
on the interface) already gives rise to a rich family of interface CFTs. 
In particular, it was shown in \cite{degsusy} that, while the conformal symmetry 
group $SO(2,4)$ of the $\cN=4$ theory is always reduced to the conformal group 
$SO(2,3)$ of the planar interface, the 32 conformal supersymmetries of the bulk 
theory may be reduced to either 0,  4, 8, or 16 conformal supersymmetries,
and maximal internal symmetry groups of $SO(6)$, $SU(3)$, $SU(2) \times U(1)$
and $SO(3) \times SO(3)$ respectively. 

\smallskip

The AdS/CFT duals of conformal interface and defect theories reflect the residual 
conformal group $SO(2,3)$ of the planar interface, and correspond to Type IIB 
superstring theory (or its Type IIB supergravity limit) on a warped space containing $AdS_4$, 
since the isometry group of $AdS_4$ is precisely $SO(2,3)$. 
For example, the intersection of D3 and probe D5 branes produces AdS/CFT duals to 
planar defect theories,  where the extra degrees of freedom are produced by the  
dynamics of open strings spanned between the various intersecting branes 
\cite{Karch:2000gx, DeWolfe:2001pq,Erdmenger:2002ex,Bachas:2001vj,Yamaguchi:2003ay,Skenderis:2002vf}.

\smallskip

The original Janus solution of \cite{Bak:2003jk} is AdS/CFT dual to the interface 
Yang-Mills theory with 0 supersymmetries listed at the end of the first paragraph.
(see  \cite{Bak:2004yf,Sonner:2005sj,Bak:2006nh,Hirano:2006as,Bak:2007jm}  
for other developments  on the Janus solution).
The Janus solution is a 1-parameter family of dilatonic deformations of 
$AdS_5 \times S^5$ in which the entire  internal symmetry $SO(6)$ is preserved,
but supersymmetry is completely broken. Nonetheless, Janus is stable against 
all small and a certain class of  large perturbations \cite{Freedman:2003ax,Cheng:2005wk}.   
Its geometry is 
$AdS_4 \times {\bf R} \times S^5$, where ${\bf R}$ parametrizes the varying dilaton, 
and is of co-homogeneity 1.
The AdS/CFT dual interface theory is pure $\cN=4$ super-Yang-Mills 
on either side of the interface, across which the gauge coupling varies discontinuously.  
Several dynamical problems in the interface Yang-Mills theory, such as the persistence
of the interface conformal symmetry at the quantum level, may be addressed by 
directly exploiting the dynamics of the bulk theory~\cite{Clark:2004sb,Papadimitriou:2004rz}.

\smallskip

A 2-parameter family of supersymmetric Janus solutions to Type IIB supergravity 
was obtained in \cite{D'Hoker:2006uu} (see also \cite{Clark:2005te}). 
With its 4 supersymmetries,  and $SU(3)$ internal 
symmetry, it emerged as a natural AdS/CFT dual to the interface theory with 4 
supersymmetries listed at the end of the first paragraph. Its geometry is now
$AdS_4 \times {\bf R} \times CP_2 \times _1 S^1$, and is of co-homogeneity 1.
Here, $CP_2 \times _1 S^1$ is topologically $S^5$, but isometric only 
under the $SU(3) \times U(1)$ subgroup of the isometry group $SU(4)$ of $S^5$.
This space was encountered earlier in the context of supergravity solutions in 
\cite{Romans:1984an,Pope:1984jj,Pilch:2000fu}. 

\smallskip

The initial motivation for the present work was to obtain a Janus solution of Type IIB 
supergravity which is dual to the interface Yang-Mills theory with 16 supersymmetries, 
listed at the end of the first paragraph. 
The geometry of the solution is in part determined by the conformal $SO(2,3)$, and
the internal $SO(3) \times SO(3)$ symmetries of the Yang-Mills interface theory,
which require a manifold $AdS_4 \times M_6$ where $M_6$ has $SO(3) \times SO(3)$
isometry and the product is warped over $M_6$. There are many possible such 
$M_6$ spaces. The particular reduction of $SO(6)$ internal symmetry on the six 
scalars of the Yang-Mills theory, obtained in \cite{degsusy}, lead one to conclude 
that $M_6$ is a warping of
$S^2 \times S^2$, which manifestly exhibits the desired $SO(3) \times SO(3)$ isometry.

\smallskip

The initial motivation described above, namely a search for a Janus 
solution with 16 supersymmetries, thus leads one to consider Type IIB 
supergravity on the following spaces,
\bea
AdS_4 \times S^2 \times S^2 \times \Sigma 
\eea
with $SO(2,3) \times SO(3) \times SO(3)$ isometry. 
In general,  the product spaces are warped over the two-dimensional  
parameter space $\Sigma $, which is a Riemann surface with boundary, 
and these spaces are of co-homogeneity~2.
A further motivation for considering Type IIB solutions on these spaces
derives from the similarity of this problem to the one of ``bubbling AdS space 
and 1/2 BPS geometries" of \cite{Lin:2004nb} (see also \cite{Liu:2006pd}).
The Killing spinors and  the reduced BPS equations for this  case were calculated 
by Gomis and R\"omelsberger \cite{Gomis:2006cu}, but the only explicit solution 
obtained there was $AdS_5 \times S^5$.

\smallskip

{\sl In the present paper, we shall derive all Type IIB supergravity solutions
with 16 supersymmetries and $AdS_4 \times S^2 \times S^2 \times \Sigma $
space-time geometry with $SO(2,3) \times SO(3) \times SO(3)$ symmetry, 
in terms of  two harmonic functions $h_1$ and $h_2$ 
on $\Sigma $. In general, these
solutions have varying dilaton $\phi$ and non-vanishing 3-form field strengths.
For example, the dilaton field for the general solution takes the following form,}
\bea
\label{dilsol1}
e^{4 \phi} =
{ 2 h_1 h_2 |\p_w h_2 |^2 - h_2^2 (\p_w h_1 \p_{\bar w} h_2 + \p_w h_2 \p_{\bar w} h_1 )
\over 
2 h_1 h_2 |\p_w h_1 |^2 - h_1^2 (\p_w h_1 \p_{\bar w} h_2 + \p_w h_2 \p_{\bar w} h_1 )}
\eea
{\sl for any local complex coordinate $w$ on $\Sigma $. Other fields are given
by analogous explicit expressions in terms of $h_1$ and $h_2$, which will
be derived and presented in section 9.}

\smallskip

Some of these solutions are everywhere non-singular, while others have singularities.
The analysis in this paper is mostly restricted to the local structure of the solutions
and we defer to a companion paper \cite{EDJEMG2} the study of global properties and 
singularities such as those of the D5 and NS5 brane type.  Amongst the regular solutions, 
we readily identify in section 10 of this paper one family which is of the Janus type.
By construction, this solution has 16 supersymmetries and $SO(3) \times SO(3)$
internal symmetry, as was hoped for.

\smallskip

The complete and exact solution to the reduced BPS equations is constructed by
mapping the BPS equations onto a seemingly new integrable system, which is akin to
the Liouville and Sine-Gordon theories. Its field equation is given by,
\bea
\label{univ0}
\p_{\bar w} \p_w \tet + {1 \over \cos \mu} 
\left ( e^{-i \tet} \, \p_{\bar w}  \tet \, \p_w \lambda 
    + e^{i \tet } \, \p_w \tet \, \p_{\bar w} \bar \lambda \right )
    - 2 {\sin \mu \over \cos ^2 \mu}  \p_w \lambda \, \p_{\bar w} \bar \lambda \, \cos \tet
    =0
\eea
Here $\tet$ is the field of the integrable system,  $\lambda $ is any holomorphic 
function of the complex coordinate $w$, and $\mu$ is a real harmonic function 
defined by  $i \mu = \lambda - \bar \lambda$. The field $\tet$ is simply related 
to the dilaton by $e^{2 i \tet} = \sh (2 \phi + 2 \lambda)/\sh (2 \phi + 2 \bar \lambda) $.
The equation (\ref{univ0}) is invariant under conformal reparametrizations,
just as Liouville theory is. Choosing the conformal coordinate to coincide with 
$\lambda$ gives a non-translation-invariant equation, akin to Liouville
theory in a non-translation invariant ground state, as was examined in \cite{Liouville1,
Liouville2}.

\smallskip

Remarkably, the system (\ref{univ0}) is completely integrable. Actually, even better,
it may be mapped onto a linear equation which can be solved exactly, and 
whose general solution may be exhibited in explicit form, just as in Liouville theory
\cite{Liouville1}.

\smallskip

The remainder of this paper is organized as follows. In section 2, the interface Yang-Mills 
theory with the maximal number of 16 supersymmetries, and in section 3, Type IIB 
supergravity are briefly reviewed, mostly to fix notations. In section 4, the 
$AdS_4 \times S^2 \times S^2 \times \Sigma $ Ansatz is implemented
on all the Type IIB supergravity fields, and in section 5, the BPS equations are reduced
on this Ansatz. This reduction was already carried out by Gomis and R\"omelsberger 
\cite{Gomis:2006cu}; the derivation given here is included in order to clarify a number 
of important issues and to give the proper $S$-duality interpretation of the reality 
conditions which are key to obtaining a full solution to the BPS equations.

\smallskip

In section 6, it is shown that every Type IIB solution with 16 supersymmetries may
be mapped, using the $SL(2,{\bf R})$ S-duality of Type IIB supergravity, onto
a solution in which the axion vanishes, and the 3-form field strengths, as well as
the supersymmetry generating spinors obey certain reality conditions. 
In section 7, it is shown that the fully reduced BPS equations consist of two 
first order differential equations for the dilaton $\phi$ and the Weyl factor
$\rho $ of the metric on $\Sigma $, as well as two arbitrary holomorphic 
functions on $\Sigma $. It is further shown that this system of 
differential equations is automatically integrable.  In section 8, the Bianchi identities
and field equations are reduced to the $AdS_4 \times S^2 \times S^2 \times \Sigma $ 
Ansatz, and are shown to hold whenever $\phi$ and $\rho$ are solutions
to the BPS system of first order equations. 

\smallskip

In section 9, a first change of variables is used to map the system onto the 
integrable system (\ref{univ0}), for which the reduced BPS system constitutes 
a  B\"acklund pair.  A second change of variables is used
to map this integrable system onto a set of linear equations, which is then solved
in terms of two holomorphic functions, or equivalently, two harmonic functions $h_1$ 
and $h_2$, on $\Sigma $. The exact solution for the 
dilaton $\phi$, the metric $\rho$, as well as all the other geometrical data entering the 
solution are obtained explicitly. In section 10, the Janus solution with 16 
supersymmetries 
is identified and shown to be everywhere regular.

\smallskip

The general solutions obtained in this paper will be the starting point in a 
companion paper \cite{EDJEMG2} for the construction of infinite classes of 
non-singular solutions corresponding to back-reacted solutions of $AdS_5 \times S^5$ 
and Janus doped with D5 and NS5 branes.
These solutions generalize the supersymmetric Janus 
solution found in this paper. Instead of two there can be $2g+2$ asymptotic 
$AdS_5 \times S^5$ regions where the dilaton approaches (in general) different values. 
In addition, there are non-trivial NSNS and RR 3-form fluxes present in these solutions. 
In certain limits the geometry has singularities which correspond to probe D5 and NS5 branes. 
The AdS/CFT duals correspond to generalized interface Yang-Mills theories. 

\smallskip

There is a closely related supergravity solution which has  
$SO(2,1)\times SO(3)\times SO(5)$ symmetry and is described by an 
$AdS_2 \times S^2 \times S^4 \times \Sigma $ Ansatz.  
The gravitational solution describes the fully back reacted  geometry 
dual to  half-BPS Wilson loops \cite{Lunin:2006xr,Gomis:2006sb,Yamaguchi:2006te}. 
A detailed analysis of this solution  applying  the methods of this paper 
can be found in a further companion paper \cite{EDJEMG3}.

\newpage 

\section{Interface Yang-Mills with maximal supersymmetry}
\label{two}
\setcounter{equation}{0}

The Yang-Mills theory with planar interface and maximal supersymmetry has 8 
Poincar\'e supersymmetries, an additional 8 supersymmetries in the conformal limit, 
and $SO(3) \times SO(3) \sim SU(2) \times SU(2)$  R-symmetry.  This reduced R-symmetry 
canonically splits the scalar multiplet  into two triplets,
which we shall denote by $\phi^i$ and $\tilde \phi^i$, with $i \in \{1,3,5\}$
for $\phi^i$ and $i \in \{2,4,6\}$ for $\tilde \phi^i$. Under $SO(3)\times SO(3)$
the triplet $\phi$ transforms as $({\bf 3}, {\bf 1})$, while $\tilde \phi$ transforms
as $({\bf 1}, {\bf 3})$. The bulk Lagrangian is given by
\bea
\cL_{bulk} &=& 
 - {1 \over 4 g^2} \tr \left ( F^{\mu \nu } F_{\mu \nu } \right ) 
- {1 \over 2 g^2} \tr (D^\mu \phi^i D_\mu \phi^i )
- {g^2 \over 2} \tr (D^\mu \tilde \phi^i D_\mu \tilde \phi^i )
\no \\ &&
+ { 1 \over 4 g^2} \tr ( [\phi^i, \phi^j ]  [\phi ^i, \phi ^j ] )
+ { g^2 \over 2 } \tr ( [\tilde \phi^i, \phi^j ]  [\tilde \phi ^i, \phi ^j ] )
+ { g^6 \over 4} \tr ( [\tilde \phi^i, \tilde \phi^j ]  [\tilde \phi ^i, \tilde \phi ^j ] )
\no \\ &&
- {i \over 2 g^2} \tr \left ( \bar \psi \gamma ^\mu D_\mu \psi \right )
+ {i \over 2 g^2} \tr \left ( D_\mu \bar \psi \gamma ^\mu \psi \right )
+ {1 \over 2 g^2} \tr \left ( \psi ^t \cC  \rho^i [\phi ^i, \psi] 
+ \psi^\dagger \cC (\rho^i)^*  [\phi ^i, \psi ^* ] \right )
\no \\ &&
+ {1 \over 2} \tr \left ( \psi ^t \cC  \rho^i [\tilde \phi ^i, \psi] 
+ \psi^\dagger \cC (\rho^i)^*  [\tilde \phi ^i, \psi ^* ] \right )
\eea
and the interface Lagrangian is given by
\bea
\cL_{interface} = 
{(\partial_\pi g) \over g^3} \tr \bigg (  {i \over 2} \psi^t  \cC \psi 
+ {i \over 2} \psi^\dagger \cC \psi^*
- {2 \over 3} i  g^6 \epsilon ^{ijk} \tilde \phi^i [\tilde \phi^j ,\tilde \phi^k]  \bigg)\label{interfl}
\eea
Here, the Yang-Mills coupling $g(x^\pi)$ is a function of the coordinate
$x^\pi$ transverse to the interface. The interface theory which is AdS/CFT dual to the 
supersymmetric Janus solution has conformal symmetry, achieved by choosing 
$g(x^\pi)$ to be a step function. For this choice, the interface term (\ref{interfl}) 
is localized at $x^\pi=0$ and the superconformal symmetry respecting the location 
of the interface is restored (for notation and details, see \cite{degsusy}).

\smallskip

Notice that, in both the bulk and interface Lagrangians, the scalar triplets
$\phi$ and $\tilde \phi$ enter with different scalings of the gauge coupling~$g$.\footnote{
The bulk Lagrangian may be put in a more standard from by scaling
the scalar fields as $\tilde \phi^i \rightarrow g^{-2} \tilde \phi^i$ at the cost of introducing interface operators of the form $(\p_\pi g)^2 \tilde \phi^i \tilde \phi^j$.}
The space of interface theories is parametrized by the gauge coupling
and the interface couplings $U \, \in \, SO(6)/\left (SO(3) \times SO(3) \right )$, which rotate the 
embedding of $SO(3) \times SO(3)$ in $SO(6)$. Theories for different $U$ 
are physically equivalent, although described by a different set of couplings. 

The interface theories are different in character from the defect CFT discussed 
in the AdS/CFT context in 
\cite{Karch:2000gx,DeWolfe:2001pq,Erdmenger:2002ex,Bachas:2001vj,Yamaguchi:2003ay}. 
In an interface theory there are no new degrees of freedom (e.g. hypermultiplets 
coming from open strings localized at brane intersections) living on the interface 
other than the ones already present in the bulk.

\newpage

\section{Type IIB supergravity}
\setcounter{equation}{0}
\label{three}

For completeness, we briefly review the Type IIB supergravity  Bianchi identities 
and field equations, as well as the supersymmetry variations, all for vanishing 
fermion fields. Our  conventions are those of \cite{D'Hoker:2006uu,Schwarz:1983qr} 
(see also  \cite{Howe:1983sr}). The bosonic fields are: the metric $g_{MN}$; 
the complex axion-dilaton scalar $B$;  the complex 2-form $B_{(2)}$ 
and the real 4-form $C_{(4)}$. We introduce composite fields
in terms of which the field equations are expressed simply, as follows,
\bea
\label{sugra1}
P  = f^2 d B \hskip 0.45in & \hskip 1in &  f^2=(1-|B|^2)^{-1}
\no \\
Q  =  f^2 \Im( B d  \bar B) &&
\eea
and the field strengths $F_{(3)} = d B_{(2)}$, and
\bea
\label{GF5}
G & = &  f(F_{(3)} - B \bar F_{(3)} )
\no \\
F_{(5)} & = & dC_{(4)} + { i \over 16} \left ( B_{(2)} \wedge \bar F_{(3)} 
- \bar B_{(2)} \wedge  F_{(3)} \right )
\eea
The scalar field $B$ is related to the complex string coupling $\tau$,
the axion $\chi$, and dilaton $\Phi$ (for notational convenience we use 
$\phi \equiv \Phi/2$ for the dilaton field) by
\bea
\label{Btau}
B = {1 +i \tau \over 1 - i \tau } \hskip 1in \tau =  \chi + i e^{- 2\phi}
\eea
In terms of the composite fields $P,Q$, and $G$, there are Bianchi identities
given as follows,
\bea
0 &=& dP-2i Q\wedge P 
\label{bianchi1} \\
0 &=& d Q + i P\wedge \bar P
\label{bianchi2} \\
0 &=& d G - i Q\wedge G +  P\wedge \bar G
\label{bianchi3} \\
0 &=& d F_{(5)} -  {i\over 8} G \wedge \bar G
\label{bianchi4}
\eea
The field strength $F_{(5)}$ is required to be self-dual,
\bea
\label{SDeq}
 F_{(5) } = * F_{(5) }
\eea
The field equations are given by,
\bea
0 & = & 
\nabla ^M P_M - 2i Q^M P_M 
+ {1\over 24} G_{MNP }G^{MNP}
\label{Peq}
\\
0 & = & 
\nabla ^P  G_{MNP } -i Q^P G_{MNP}
- P^P \bar G_{MNP }
+ {2\over 3} i F_{(5)MNPQR }G^{PQR}
\label{Geq}
\\
0 & = & R_{MN } 
- P_M  \bar P_N  - \bar P_M  P_N 
- {1\over 6} (F_{(5)}^2)_{MN }
\no \\ && \hskip .5in
- {1\over 8} (G_M {} ^{PQ } \bar G_{N PQ }
+ {\bar G_M} {} ^{ PQ } G_{N PQ }) 
+{1\over 48 } g_{MN } G^{PQR } \bar G_{PQR }
\label{Eeq}
\eea
The fermionic fields are the dilatino $\lambda$ and the gravitino $\psi_M$,
both of which are complex Weyl spinors with opposite 10-dimensional
chiralities, given by $\Gamma_{11} \lambda =\lambda$, and $\Gamma_{11}
\psi_M  =-\psi_M$. The supersymmetry variations of the fermions  are\footnote{Throughout,
we shall use the notation $\G \cdot T \equiv \G^{M_1 \cdots M_p} T_{M_1 \cdots M_p}$
for the contraction of any antisymmetric tensor field $T$ of rank $p$ and the
$\G$-matrix of the same rank.} 
\bea
\label{BPS}
\delta\lambda
&=& i (\G \cdot P) \cB^{-1} \ep^* 
-{i\over 24} (\G \cdot G) \ep
\label{susy1} \\
\delta \psi_M 
&=& D _M  \ep 
+ {i\over 480}(\G \cdot F_{(5)})  \Gamma_M  \ep
-{1\over 96}\left ( \Gamma_M (\G \cdot G)
+ 2 (\G \cdot G) \G^M \right ) \cB^{-1} \ep^* 
\no
\eea
where ${\cal B}$ is the charge conjugation matrix of the Clifford algebra.\footnote{It is 
defined by $\cB \cB^*=I$ and $\cB \Gamma ^M \cB^{-1} = (\Gamma ^M)^*$; 
see Appendix A  for our $\Gamma$-matrix conventions.
Throughout, complex conjugation will be denoted by {\sl bar} for functions,
and by {\sl star} for spinors.} The BPS equations are obtained by setting 
$\delta \lambda = \delta \psi _M=0$.

\subsection{$SU(1,1)$ symmetry}

Type IIB supergravity is invariant under $SU(1,1) \sim SL(2,{\bf R})$ symmetry,
which leaves $g_{\mu \nu}$ and $C_{(4)}$ invariant, acts by M\"obius
transformation on the field $B$, and linearly on $B_{(2)}$, 
\bea
\label{stransf}
B           & \to & B^s = {u B  + v \over \bar v B + \bar u}
\no \\  
B_{(2)} & \to & B_{(2)}^s = u B_{(2)} + v \bar B_{(2)}
\eea
with $u,v \in {\bf C}$ and $\bar u u - \bar v v=1$. In this non-linear realization of 
$SU(1,1)$, the field $B$ takes values in the coset  $SU(1,1) /U(1)_q$,
and the fermions $\lambda$ and $\psi_\mu$ transform linearly under
the isotropy gauge group $U(1)_q$ with composite gauge field $Q$. The transformation 
rules for the composite fields are \cite{D'Hoker:2006uu}, 
\bea
\label{su11a}
P & \to & P^s = e^{2 i \theta} P
\no \\
Q & \to & Q^s = Q + d \theta
\no \\
G & \to & G^s = e^{i \theta} G
\eea
where the phase $\theta$ is defined by
\bea
\label{su11b}
e^{i \theta} = \left ( {v \bar B + u \over \bar v B + \bar u} \right )^\half
\eea
In this form, the transformation rules clearly exhibit the $U(1)_q$ gauge 
transformation that accompanies the  global $SU(1,1)$ transformations.

\newpage

\section{The two-parameter Ansatz}
\setcounter{equation}{0}
\label{four}

We seek a general Ansatz in Type IIB supergravity with the following symmetry,
\bea
SO(2,3) \times SO(3) \times SO(3)
\eea
which may be viewed as the bosonic subgroup of $OSp(2,2|4)$.
The factor $SO(2,3)$ requires the geometry to contain $AdS_4$,
while the factor $SO(3) \times SO(3)$ could be accommodated by either 
$S^2 \times S^2$ or $S^3$. 

\smallskip

Given that our initial motivation was
the construction of a Janus solution with 16 supersymmetries, and  dual
to the interface Yang-Mills theory with maximal supersymmetry of section \ref{two},
the case of $S^3$ is excluded. This is because the 6 scalar fields $\phi^i$
and $\tilde \phi^j$, with $i=1, 3, 5$ and $j=2, 4, 6$ are grouped in two
independent sets which immediately suggests $S^2 \times S^2$. 
Two dimensions remain undetermined by the symmetries alone, so
that the most general space of interest to us will be of the form,
\bea
AdS_4 \times S^2_1 \times S^2 _2 \times \Sigma
\eea
where $\Sigma$ stands for the two-dimensional space,
over which the above products are warped.
In order for the above space to be a Type IIB supergravity geometry, 
$\Sigma$ must carry an orientation
as well as a Riemannian metric, and is therefore a 
Riemann surface, generally with boundary. The subscripts 1 and 2 label the two-spheres.

\subsection{Ansatz for the Type IIB fields}

The Ansatz for the metric is
\bea
ds^2 = f_4^2 ds^2 _{AdS_4} + f_1 ^2 ds^2 _{S_1^2} + f_2 ^2 ds^2 _{S^2_2}
+ ds^2 _\Sigma
\eea
where $f_1,f_2,f_4$ and $ds^2 _\Sigma$ are  functions on $\Sigma$. We
introduce an orthonormal frame,
\bea
\label{frame1}
e^m & = & f_4 \, \hat e^m \hskip 1in m=0,1,2,3
\no \\
e^{i_1} & = & f_1 \, \hat e^{i_1} \hskip 1in i_1 =4,5
\no \\
e^{i_2} & = & f_2 \, \hat e^{i_2} \hskip 1in i_2 =6,7
\no \\
e^a \, &  & \hskip 1.4in a=8,9
\eea
where $\hat e^m$, $\hat e^{i_1}$, $\hat e^{i_2}$ and $e^a$ refer to orthonormal 
frames for the spaces $AdS_4$, $S_1^2$, $S_2^2$ and $\Sigma $ respectively. 
In particular, we have\footnote{The convention of summation over repeated 
indices will be used throughout whenever no confusion is expected to arise,
with the ranges of the various indices following the pattern of the frame in (\ref{frame1}).}
\bea
ds^2 _{AdS_4} & = & \eta _{mn} \, \hat e^m \otimes \hat e^n
\no \\
ds^2 _{S_1^2} & = & \delta _{i_1 j_1} \, \hat e^{i_1} \otimes \hat e^{j_1}
\no \\
ds^2 _{S_2^2} & = & \delta _{i_2 j_2} \, \hat e^{i_2} \otimes \hat e^{j_2}
\no \\
ds^2 _\Sigma & = & \delta _{ab} \, e^a \otimes e^b 
\eea
where $\eta = {\rm diag} [-+++]$. The complex dilaton/axion field $P$, and the 
connection $Q$ are  1-forms, and their structure is simply given as follows,
\bea
\label{PQdef}
P& = & p_a e^a
\no \\
Q & = & q_a e^a
\eea
Throughout, we shall view the $P$ and $Q$-forms as given in terms  of 
the dilaton/axion field $B$, as in (\ref{sugra1}), so that $P$ and $Q$
are not independent fields. Thus, they will always automatically satisfy 
their Bianchi identities,
\bea
\label{dilbianchi}
d P - 2 i Q \wedge P & = & 0
\no \\
dQ + i P \wedge \bar P & = & 0
\eea
This approach will allow us  to dispense with 
the $Q$ field and show that every half-BPS solution in fact arises as 
a $SU(1,1)$ transformation of a solution with vanishing axion.

\smallskip

Finally, the anti-symmetric tensor forms $F_{(5)}$ and $G$ are  given by
\bea
F_{(5)} & = & f_a (- e^{0123a} + \ep^a {}_b \,  e^{4567b} )
\hskip 1in \ep^{89}=1
\no \\
G & = & g_a e^{45a} + i h_a e^{67a}
\eea
Here, $f_a, q_a$ are real, while $g_a, h_a, p_a$ are complex.
It will be useful to introduce 1-forms for these reduced fields as well,
\bea
\cF & \equiv & f_a e^a \hskip 1in * _2 \cF = \ep ^a {}_b \, f_a \, e^b 
\no \\
\cG & \equiv & g_a e^a
\no \\
\cH & \equiv & h_a e^a
\eea
so that we have equivalently,
\bea
F_{(5)} & = & - e^{0123} \wedge \cF + e^{4567} \wedge *_2 \cF
\no \\
G & = & e^{45} \wedge \cG + i e^{67} \wedge \cH
\eea
Here, $*_2$ denotes the Poincar\'e dual on $\Sigma$ with respect to
the metric $ds_\Sigma ^2$.

\newpage 

\section{Reduced BPS equations with 16 supersymmetries}
\setcounter{equation}{0}
\label{five}

Solutions of the form given by the Ansatz of subsection 4.1 which preserve 16 
supersymmetries correspond to supergravity fields for which the BPS 
equations  $\delta \lambda = \delta \psi _M=0$ in  (\ref{BPS}) have 16 
independent solutions $\ep$.
Whenever the dilaton is subject to a non-trivial space-time variation,
$\p_M \phi \not=0$, the dilatino BPS equation will allow for at most 16
independent supersymmetries $\ep$. Therefore, the gravitino BPS equation
should not impose any further restrictions on the number of supersymmetries,
but should instead simply give the space-time evolution of $\ep$.
As a result, at any given point in the space $\Sigma$, $\ep$
must be a Killing spinor on each of the spheres $S_1^2 \times S_2^2$,
as well as on $AdS_4$.  

\smallskip

The analysis in this section is similar to the one employed in  
\cite{Gomis:2006cu}, and we use a closely related notation.
The method of bilinears in the Killing spinors, pioneered in \cite{gauntlett},
is not needed here, and the corresponding results will be derived 
systematically from the reduced BPS equations instead.  
To illustrate our method, and  for the sake of additional clarity and 
completeness, the derivation will be presented
here in detail.

\subsection{Using Killing spinors}

Killing spinors on $AdS_4 \times S^2_1 \times S_2^2$ are non-vanishing solutions to
the equations,
\bea
\label{KS}
\left ( \hat \nabla _m  - \half \eta_1 \gamma _m \otimes I_2 \otimes I_2 \right ) 
\chi ^{\eta _1, \eta _2, \eta _3} _{\eta _0}
    & = & 0 \hskip 1in m=0,1,2,3
\no \\
\left ( \hat \nabla _{i_1} - {i \over 2} \eta_2 I_4 \otimes \gamma _{i_1} \otimes I_2 \right ) 
\chi ^{\eta _1, \eta _2, \eta _3} _{\eta _0}
    & = & 0 \hskip 1in i=4,5 
\no \\
\left ( \hat \nabla _{i_2} - {i \over 2} \eta_3 I_4 \otimes I_2 \otimes \gamma _{i_2} \right ) 
\chi ^{\eta _1, \eta _2, \eta _3} _{\eta _0}
    & = & 0 \hskip 1in i=6,7 
\eea
Our conventions for the Clifford algebra are given in Appendix A.
The spinors $\chi ^{\eta _1, \eta _2, \eta _3} _{\eta _0}$ are 16-dimensonal. 
The covariant derivatives $\hat \nabla _m$, $\hat \nabla_{i_1}$,  and $\hat \nabla_{i_2}$ 
act in the Dirac spinor representations for  $AdS_4$, $S^2_1$, and $S^2_2$, 
with respect to the canonical spin connections associated 
with the frames  $\hat e^m$, $\hat e^{i_1}$ and $\hat e^{i_2}$. 
Once the integrability conditions, $\eta _1^2 = \eta _2 ^2 = \eta _3 ^2=1$, 
for (\ref{KS}) are satisfied, the solution spaces are of maximal dimension, namely 16.

\smallskip

Since the chirality matrix for each Killing spinor equation (respectively $\g_{(1)}$,
$\g_{(2)}$, and $\g_{(3)}$) commutes with the corresponding covariant derivative,
but not with the entire equation in (\ref{KS}), the chirality matrices will map between
two linearly independent solutions. This is explained in detail in Appendix B,
where the geometry of Killing spinors on $S^2$ and $AdS_4$ is reviewed.
Thus, we may use  $\eta _1,\eta_2, \eta_3$ to label the linearly
independent solutions to the Killing spinor equations. The Killing equation 
for $AdS_4$, however, has 4 linearly independent solutions, and the label 
$\eta _1$ alone does not suffice to label these solutions uniquely. The 4
solutions consist of 2 degenerate solutions for each chirality, and this 
degeneracy is uniquely specified by the extra label $\eta _0 =\pm 1$.
In total, the solutions of the $AdS_4$ Killing equation are uniquely labeled by the pair 
$(\eta _0, \eta _1)$. To economize notation,  the index $\eta _0$
will be not be exhibited, with the understanding that the solution space for 
$\chi ^{\eta _1, \eta _2 , \eta _3}$ remains 16-dimensional.

\smallskip

For any one of the chirality matrices $\gamma _{(s)}$, for $s=1,2,3$, the product
$\gamma _{(s)} \chi$ satisfies (\ref{KS}) with the opposite 
value of $\eta _s$. We may therefore identify the corresponding spinors,
\bea
\label{chiralities}
\gamma _{(1)} \chi ^{\eta _1, \eta _2, \eta _3} 
    & = & \chi ^{-\eta _1, \eta _2, \eta _3}
\no \\
\gamma _{(2)} \chi ^{\eta _1, \eta _2, \eta _3}
    & = & \chi ^{\eta _1,  -\eta _2, \eta _3}
\no \\
\gamma _{(3)} \chi ^{\eta _1, \eta _2, \eta _3}
& = & \chi ^{\eta _1, \eta _2,  -\eta _3}
\eea
To examine the Killing spinor properties, we begin by decomposing the 32
component (complex) spinor $\ep$ onto the $\Sigma$-independent basis of spinors 
$\chi ^{\eta _1,  \eta _2, \eta _3} _{\eta _0}$, with coefficients which are 
$\Sigma$-dependent  2-component spinors $\zeta _{\eta _1, \eta _2, \eta _3}$,
\bea
\ep = \sum _{\eta _1, \eta _2, \eta _3} \chi ^{\eta _1, \eta _2, \eta _3} \otimes 
\zeta _{\eta _1, \eta _2, \eta _3}
\eea
The 10-dimensional chirality condition $\G^{11} \ep = - \ep$ reduces to
\bea
\label{chiral1}
\g_{(4)} \zeta _{- \eta _1, - \eta _2, - \eta _3} = - \zeta _{\eta_1, \eta _2, \eta _3}
\eea
where $\g_{(4)}$ is the chirality matrix associated with $\Sigma$; see Appendix D
for its detailed expression. The Killing spinor equations are invariant under charge 
conjugation $\chi \to \chi ^c$,
with 
\bea
\left ( \chi ^c \right ) ^{\eta _1, \eta _2 , \eta _3}  = B_{(1)} \otimes B_{(2)} \otimes B_{(3)} 
\left ( \chi ^{\eta _1 , \eta _2, \eta _3} \right )^*
\eea
where $B_{(1)}, \, B_{(2)}, \,  B_{(3)}$ are the charge conjugation matrices on
the Dirac algebras for $AdS_4$, $S_1^2$ and $S_2^2$ respectively.
Since $(B_{(1)} \otimes B_{(2)} \otimes B_{(3)})^2 = I_{16}$, we 
may impose, without loss of generality, the reality condition  $\chi ^c = \pm \chi$ 
on the basis. The sign assignments  are related by (\ref{chiralities}), and found to be
\bea
\label{real1}
B_{(1)} \otimes B_{(2)} \otimes B_{(3)} 
\left ( \chi ^{\eta _1 , \eta _2, \eta _3} \right )^* = \eta _1 \eta _2 \eta _3
 \chi ^{\eta _1 , \eta _2, \eta _3}
\eea
Upon imposing the reality condition (\ref{real1}) on the basis of spinors $\chi$,
and the chirality condition (\ref{chiral1}) on $\zeta$, and recalling that 
$\chi^{\eta_1 \eta _2 \eta _3}$ has double degeneracy due to the suppressed 
quantum number $\eta _0$, we indeed recover 16 complex components for
the spinor $\ep$.

\smallskip

Following \cite{Gomis:2006cu}, we introduce a matrix notation in the
8-dimensional space of $\eta$ by,
\bea
\tau ^{(ijk)}  \equiv  \tau^i \otimes \tau ^j \otimes \tau ^k \hskip 1in i,j,k =0,1,2,3
\eea
where $\tau^0=I_2$, and $\tau^i$ with $i=1,2,3$ are the  Pauli matrices in the standard
basis. Multiplication by $\tau ^{(ijk)}$ is defined as follows,
\bea
(\tau ^{(ijk)} \zeta  )_{\eta _1 , \eta _2 , \eta _3}
& \equiv & \sum _{\eta _1 ', \eta _2 '. \eta _3'} 
(\tau ^i)_{\eta _1 \eta _1'} (\tau ^j)_{\eta _2 \eta _2'} (\tau ^k)_{\eta _3 \eta _3'}
\zeta _{\eta _1', \eta _2 ', \eta _3'}
\eea
Henceforth, we shall use matrix notation for $\tau$ and suppress the indices $\eta$.

\subsection{The reduced BPS equations}

With the help of the Ansatz for the Type IIB fields produced in subsection 4.1,
the BPS equations (\ref{BPS}) may be reduced and presented using the notations
introduced in the preceding subsection.  The explicit reduction is
carried out in Appendix C.  The dilatino BPS equation is given by,
\bea
\label{dilatino1}
(d) & \quad& 0 = p_a  \sigma^a \sigma^2 \zeta ^*
+ {1\over 4} \left ( g_a \tau^{(322)}  
- i h_a \tau^{(333)} \right ) \sigma^{a} \zeta 
\hskip 2.2in
\eea
while the gravitino equation decomposes into a system of 4 equations, 
\bea
\label{gravitino1}
(m) &\quad& 
0 = - {i \over 2 f_4} \tau^{(211)}  \zeta 
+  {D_a f_4 \over 2 f_4}  \sigma^{a} \zeta
+ {1 \over 2}  f_a \tau^{(100)}  \sigma^{a} \zeta 
+ {1 \over 16}  \left ( g_a \tau^{(322)}  
- i h_a \tau^{(333)}  \right ) \sigma^{a} \sigma^2 \zeta ^*
\no\\
(i_1) &\quad& 
0 = {1 \over 2 f_1} \tau^{(021)}  \zeta 
+  {D_a f_1 \over 2 f_1}  \sigma^{a} \zeta
- {1 \over 2}  f_a \tau^{(100)}  \sigma^{a} \zeta
+ {1 \over 16} \left ( - 3 g_a \tau^{(322)}   
- i  h_a \tau^{(333)}  \right ) \sigma^{a}  \sigma^2 \zeta ^*
\no\\
(i_2) &\quad& 
0 = {1 \over 2 f_2} \tau^{(002)}  \zeta 
+  {D_a f_2 \over 2 f_2}  \sigma^{a} \zeta
- {1 \over 2}  f_a \tau^{(100)}  \sigma^{a} \zeta
+ {1 \over 16} \left ( g_a \tau^{(322)}   
+ 3 i  h_a \tau^{(333)} \right ) \sigma^{a} \sigma^2 \zeta ^*
\no\\
(a) &\quad& 
0 = D_a \zeta  + {i \over 2} \hat \omega_a  \sigma^{3} \zeta
- {i \over 2} q_a  \zeta + {1 \over 2} f_b 
\tau^{(100)}  \sigma^{b} \sigma_a  \zeta
+ {1 \over 16} \left ( - 3 g_a \tau^{(322)} 
+ g_b \tau^{(322)}  \sigma^{ab} \right )  \sigma^2 \zeta ^*
\no\\&& \hskip 2in
+ {1 \over 16} \left ( 3i h_a \tau^{(333)}   
- i  h_b \tau^{(333)}  \sigma^{ab} \right ) \sigma^2 \zeta ^*
\eea
The derivatives $D_a$ are defined with respect to the frame $e^a$, so that 
$e^a D_a = d$, the total differential on $\Sigma$. 
Also, we denote the Dirac matrices $\gamma^a$ on $\Sigma$
simply by  $\sigma ^a$, in a slight abuse of notation where $\sigma ^8 = \sigma ^1$
and $\sigma ^9 = \sigma ^2$, in accord with the conventions of Appendix A.

\subsection{Symmetries of the reduced BPS equations}

The reduced BPS equations exhibit continuous as well as discrete symmetries,
which will be exploited to further reduce them. The continuous
symmetries are as follows.

\smallskip

Local frame rotations of the frame $e^a$ on $\Sigma$ generate a gauge symmetry
$U(1)_c$, whose action on all fields is standard.

\smallskip

The axion/dilaton field $B$ transforms non-linearly under the continuous
$S$-duality group  $SU(1,1)$ of Type IIB supergravity.
As was discussed at the end of section 3, $B$ takes values in the coset 
$SU(1,1)/U(1)_q$,  and $SU(1,1)$ transformations on the fields are 
accompanied by local $U(1)_q$ gauge transformations, given in (\ref{su11a})
and (\ref{su11b}), 
\bea
U(1)_q & \hskip .6in &
\zeta \to e^{i \theta/2} \zeta 
\no \\
    & \hskip .6in & 
    q_a \to q_a +  D_a \theta   
\no \\
    & \hskip .6in & 
    p_a \to e^{2 i \theta} p_a
\no \\ &&
g_a \to e^{ i \theta} g_a
\no \\ &&
 h_a \to e^{ i \theta} h_a
\eea
The real function $\theta$ depends on the  $SU(1,1)$ transformation,
as well as on the field $B$.

\subsubsection{Discrete symmetries}

The reduced BPS equations are also invariant under three commuting involutions.
The first two do not mix $\zeta$ and $\zeta ^*$ and leave the fields 
$f_a, p_a, q_a, g_a, h_a$ unchanged. They are defined by,
\bea
\cI \zeta & = & - \tau ^{(111)} \sigma ^3 \zeta
\no \\
\cJ \zeta & = & \tau ^{(032)} \zeta
\eea
Both $\cI$ and $\cJ$ commute with the symmetries $U(1)_q$ and $U(1)_c$.

\subsubsection{Complex conjugation}

The third involution $\cK$ amounts to complex conjugation. This operation acts 
non-trivially on all complex fields, and its action on $\zeta$ depends on the basis 
of $\tau$-matrices. In a basis in which both $\sigma ^2$  and $\tau^2$ are purely 
imaginary, the involution $\cK$ has the following form. Taking the complex
conjugates of $p_a, g_a, h_a$, letting $q_a \to - q_a$ and mapping
$\zeta \to -i \, \tau ^{(020)} \sigma ^2 \zeta ^*$ will leave the BPS equations invariant. 

\smallskip

Complex conjugation, defined this way, however, does not commute with the 
$SU(1,1)$ transformations, since $\zeta$ transforms under $SU(1,1)$ by a local $U(1)_q$ gauge transformation. Therefore, we 
relax the previous definition of complex conjugation, and allow for complex 
conjugation modulo a $U(1)_q$ gauge transformation with phase $\theta$,
\bea
\cK \zeta & = & e ^{ i \theta} \tau ^{(020)} \sigma ^2 \zeta ^*
\no \\
\cK q_a & = &  - q_a + 2 D_a \theta
\no \\
\cK p_a  & = & + e^{4 i \theta} \bar p_a 
\no \\
\cK g_a & = & - e^{2 i \theta } \bar g_a 
\no \\
\cK h_a & = & - e^{2 i \theta } \bar h_a 
\eea
which continues to be a symmetry of the BPS equations.\footnote{The factor of $i$ in the transformation rule
for $\zeta$ has been absorbed into the compensating $U(1)_q$ transformation.}
The need for such a compensating gauge transformation should be clear from the 
fact that $\zeta $ and $\zeta ^*$ transform with opposite phases under $U(1)_q$.
On the other hand, $\cK$ commutes with the group $U(1)_c$ of frame rotations.

\subsubsection{Restricting chirality in Type IIB}

In Type IIB theory only a single chirality is retained, so we have the condition
\bea
\label{chiralityconstraint}
\cI \zeta = - \tau ^{(111)} \sigma ^3 \zeta =  \zeta
\eea
This subspace is invariant under the remaining  involutions, since  $\cJ$ and $\cK$ 
commute with $\cI$. 

\newpage 

\section{Reality properties of the supersymmetric solution}
\setcounter{equation}{0}
\label{six}

In this section, we shall establish that the BPS equations restrict $\zeta$ to belong 
to a single one of the eigenspaces of $\cJ$, but not both, and to 
a single one of the eigenspaces of $\cK$, but not both. These results lead to
a further reduction of the BPS equations.\footnote{In \cite{Gomis:2006cu},
this further reduction was achieved upon the additional use of the closure of
the supersymmetry algebra. Here, it is shown that this is in fact unnecessary and 
that the entire further reduction of the BPS equations follows directly from
the BPS equations themselves.} In particular, we shall show that
every solution with 16 supersymmetries may be mapped by an $SU(1,1)$
transformation onto a solution with vanishing axion field and real $g_a, h_a$.
The  Janus solution with 4 supersymmetries,
obtained in \cite{D'Hoker:2006uu}, exhibits an analogous reality property.

\smallskip

The restrictions of $\zeta$ to definite eigenspaces of $\cJ$ and $\cK$ may 
be established directly from the BPS equations, by showing that they imply a certain
number of {\sl bilinear constraints} on $\zeta$ (and $\zeta ^*$) which are 
independent of reduced fields $f_1,f_2,f_4, f_a, g_a, h_a, p_a, q_a$.

\subsection{Restriction to a single eigenspace of $\cJ$}

The restriction for $\cJ$ is obtained as follows; the detailed arguments are 
presented  in Appendix~D. Contracting the dilatino BPS equation $(d)$ of 
(\ref{dilatino1}) on the left by  $\zeta ^\dagger T \sigma ^{0,3}$ for certain $\tau$
matrices $T$, and using the assumption $\p_a B \not=0$, leads to a first set of constraints,
\bea
\label{Tcon}
\zeta ^\dagger T \sigma ^a \zeta =0 
\hskip .8in 
T \in \cT = \left  \{ \tau^{(310)}, ~  \tau^{(301)}, ~  \tau^{(201)}, ~ \tau^{(210)} \right \} 
\hskip .5in
\eea
and $a=1,2$. Contracting the gravitino BPS equations $(m)$, $(i_1)$ 
and $(i_2)$ of (\ref{gravitino1}) on the left by  $\zeta ^\dagger T \sigma ^{0,3}$, 
with $T \in \cT$, and using the vanishing of terms involving $g_a,h_a$ due to 
(\ref{Tcon}), leads to a second set of constraints,
\bea
\label{Ucon}
 \zeta ^\dagger U \sigma ^p   \zeta =0 
& \hskip .5in & 
U \in \cU =  \Big \{
\tau^{(001)}, \,  \tau^{(010)} , \, 
\tau^{(231)} , \,  \tau^{(220)} , \, 
\tau^{(212)} , \,  \tau^{(203)}, \, 
\no \\ && \hskip .8in 
\tau^{(110)}, \,  \tau^{(101)} , \, 
\tau^{(320)} , \,  \tau^{(331)} , \, 
\tau^{(303)} , \,  \tau^{(312)} \, 
\Big  \}
\eea
for $p=0,3$. In Appendix \ref{appE}, a detailed derivation of the  
solution to both sets of bilinear constraints is given. The general solution may
be expressed as the projection condition onto a single eigenspace of $\cJ$, 
\bea
\label{cJsol}
\cJ \zeta = \tau ^{(032)} \zeta = \nu \zeta
\eea
where $\nu$ is either $+1$ or $-1$. The constraints (\ref{Tcon}) and (\ref{Ucon})
are automatically satisfied once (\ref{cJsol}) is, since $\tau ^{(032)}T$
and $\tau^{(032)} U$ anticommute with the chirality constraint (\ref{chiralityconstraint}).

\subsection{Restriction to a single eigenspace of $\cK$}

The use of the $T$-matrices in (\ref{Tcon}), and the $U$-matrices in (\ref{Ucon}), 
has allowed us to obtain relations between 
bilinears in $\zeta$ in which the dependence on both $g_a$ and $h_a$ 
was eliminated. Further useful information may be obtained from relations
in which the dependence on either $g_a$ or $h_a$, but not both, is eliminated.
This is achieved by contracting the equations $(m)$, $(i_1)$ and $(i_2)$ 
respectively by $\zeta ^\dagger M_g \sigma ^p$ and $\zeta ^\dagger M_h \sigma ^p$,
where $p=0,3$, and  the $\tau$-matrices $M_g$ and $M_h$ are Hermitian and satisfy,
\bea
\left ( M_g \tau ^{(333)} \right ) ^t = + M_g \tau ^{(333)}
& \hskip 1in &
\left ( M_h \tau ^{(333)} \right ) ^t = - M_h \tau ^{(333)}
\no \\
\left ( M_g \tau ^{(322)} \right ) ^t = - M_g \tau ^{(322)}
& \hskip 1in &
\left ( M_h \tau ^{(322)} \right ) ^t = + M_h \tau ^{(322)}
\eea
Non-trivial relations are obtained only if the corresponding matrices $M_g$
and $M_h$ commute with $\tau ^{(032)}$, and if the products 
$M_g \tau ^{(211)}$ and $M_h \tau ^{(211)}$ commute with $\tau^{(111)}$.
(It then follows that $M_g \tau ^{(021)}$, $M_g \tau ^{(002)}$,
$M_h \tau ^{(021)}$, and $M_h \tau ^{(002)}$ commute with $\tau^{(111)}$.)
Finally, using the restrictions on $\zeta$ under the involutions $\cI$ and 
$\cJ$, we may consider $M_g$ and $M_h$ modulo equivalence under multiplication by
$\tau ^{(032)}$ and $\tau ^{(111)} \sigma ^3$. This leaves unique solutions,
\bea
M_ g & = & \tau ^{(002)}
\no \\
M_h & = & \tau ^{(021)}
\eea
We shall analyze the case $M_h$ in detail, and simply quote the results from
$M_g$. We start by multiplying the $(m)$, $(i_1)$, and $(i_2)$ equations in
$(\ref{gravitino1})$ by $\zeta^\dagger \tau^{(021)} \sigma ^p$, for $p=0,3$, to get
\bea
\label{bilin8}
0 & = & - {1 \over 2 f_4} \zeta^\dagger \tau^{(230)} \sigma ^p \zeta 
+  {\p_a f_4 \over 2 f_4}  \zeta^\dagger \tau^{(021)} \sigma ^p \sigma^{a} \zeta
+ {1 \over 2}  f_a \zeta^\dagger \tau^{(121)} \sigma ^p \sigma^{a} \zeta 
+ {i \over 16}  g_a \zeta^\dagger \tau^{(303)} \sigma ^p \sigma^{a} \sigma^{2} \zeta ^* 
\no\\ && \no \\
0 & = & {1 \over 2 f_1} \zeta^\dagger \sigma ^p \zeta 
+  {\p_a f_1 \over 2 f_1}  \zeta^\dagger \tau^{(021)} \sigma ^p \sigma^{a} \zeta
- {1 \over 2}  f_a \zeta^\dagger \tau^{(121)} \sigma ^p \sigma^{a} \zeta
- {3i \over 16}  g_a \zeta^\dagger \tau^{(303)} \sigma ^p \sigma^{a} \sigma^{2} \zeta ^* 
 \\ && \no \\
0 & = & {i \over 2 f_2} \zeta^\dagger \tau^{(023)} \sigma ^p \zeta 
+  {\p_a f_2 \over 2 f_2} \zeta^\dagger \tau^{(021)} \sigma ^p \sigma^{a} \zeta
- {1 \over 2}  f_a \zeta^\dagger \tau^{(121)} \sigma ^p \sigma^{a} \zeta
+ {i \over 16}  g_a \zeta^\dagger \tau^{(303)} \sigma ^p \sigma^{a} \sigma^{2} \zeta ^* 
\quad \no
\eea
For $p=0$, the imaginary part of the first and second equations of (\ref{bilin8}), 
and for $p=3$ the real part of the third equation of (\ref{bilin8}) obey the first 
two equations below,
\bea
\label{b1}
\Im \bigg( i g_a \zeta^\dagger \tau^{(303)} \sigma^{a} \sigma^{2} \zeta ^* \bigg) 
& = & 0
\no \\
\Re \bigg(i g_a \zeta^\dagger \tau^{(303)} \sigma^3 \sigma^{a} \sigma^{2} \zeta ^* \bigg) 
& = & 0
\no \\
\Im \bigg( h_a \zeta^\dagger \tau^{(331)} \sigma^{a} \sigma^{2} \zeta ^* \bigg) &=& 0
\no \\
\Re \bigg(h_a \zeta^\dagger \tau^{(331)} \sigma^3 \sigma^{a} \sigma^{2} \zeta ^* \bigg) &=& 0
\eea
where the last two equations result from the analysis of the case $M_g$.
Using the first line of (\ref{b1}) in the imaginary part of the third equation of (\ref{bilin8}) 
for $p=0$, and the second line of (\ref{b1}) in the real parts of the first two equations 
of (\ref{bilin8}) for $p=3$  gives the first three equations below, 
\bea
\label{b2}
\zeta^\dagger \tau^{(023)} \zeta & = & 0
\no \\
\zeta^\dagger  \sigma^3 \zeta &=& 0
\no \\
\zeta^\dagger \tau^{(230)} \sigma^3 \zeta &=& 0
\no \\
\zeta ^\dagger \tau ^{(213)} \sigma ^3 \zeta & = & 0
\eea
The analysis for $M_g$  yields the first,  second and fourth equations in (\ref{b2}).

\smallskip

It will be convenient to use the following rotated basis for the $\tau$-matrices,
\bea
\label{taubasis}
\tau ^1 = \left ( \matrix{1 & 0 \cr 0 & -1 \cr} \right )
\hskip .6in 
\tau ^2 = \left ( \matrix{0 & -i \cr i & 0 \cr} \right )
\hskip .6in 
\tau ^3 = \left ( \matrix{0 & -1 \cr -1 & 0 \cr} \right )
\eea
Notice that the transposition and complex conjugation properties of these matrices are
identical to those in the standard basis of Pauli matrices, so the equations (\ref{dilatino1}) 
and (\ref{gravitino1})  continue to hold unchanged in this basis.

\medskip

The bilinear constraints of (\ref{b2}) are solved by the following complex conjugation relation,
\bea
\label{conj}
\sigma ^2 \zeta^* = e^{-  i \theta} \tau ^{(020)}  \zeta 
\eea
where $\theta$ is an arbitrary phase function on $\Sigma$, which is not fixed by 
(\ref{b2}). 
This result  is easily verified by using (\ref{conj})
in the form $\zeta ^\dagger = e^{-  i \theta} \zeta ^t \sigma ^2 \tau ^{(020)} $ to
eliminate $\zeta ^\dagger$ in (\ref{b2}) and then verifying that each of the four
equations is of the form $\zeta ^t M \zeta =0$ with $M$ anti-symmetric. In fact,
one may check that (\ref{conj}) is the most general solution to (\ref{b2}),
by decomposing $\zeta$ in components, and using the restrictions 
$\cI \zeta = \zeta$ and $ \cJ \zeta = \nu \zeta$.
Equation (\ref{conj}) is just the condition that we restrict to $\cK \zeta = \zeta$,
i.e. to a single eigenspace of the involution $\cK$.

\subsection{Reality constraints}

Having solved completely the bilinear constraints (\ref{b2}), it remains to solve  (\ref{b1}). 
To do so, we use (\ref{conj}) to recast $\sigma ^2 \zeta ^*$ in terms of $\zeta$, so that 
these constraints become, 
\bea
\Im \left ( i e^{- i \theta} \, g_a \, \zeta ^\dagger \tau ^{(323)} \sigma ^a \zeta \right )  
& = & 0
\no \\
\Im \left ( i e^{-  i \theta} \, g_a \, \ep ^{ab} \zeta ^\dagger \tau ^{(323)} \sigma _b \zeta \right )  
& = & 0
\no \\
\Im \left ( i e^{- i \theta} \, h_a \, \zeta ^\dagger \tau ^{(311)} \sigma ^a \zeta \right ) 
& = & 0
\no \\
\Im \left ( i e^{- i \theta} \, h_a \, \ep ^{ab} \zeta ^\dagger \tau ^{(311)} \sigma _b \zeta \right ) 
& = & 0
\eea
The bilinears $\zeta ^\dagger \tau ^{(323)} \sigma ^a \zeta
= \nu \zeta ^\dagger \tau ^{(311)} \sigma ^a \zeta$ are real and non-vanishing, 
(the latter will be verified once the solution is obtained)  so that
\bea
\label{ghreal}
\Im \left ( i e^{- i \theta} g_a \right ) & = & 0
\no \\
\Im \left ( i e^{- i \theta} h_a \right ) & = & 0
\hskip 1in a=1,2
\eea
Finally, we perform the same elimination of $\zeta ^\dagger$, using (\ref{conj}), also
in the dilatino BPS equation (\ref{dilatino1}) which, after multiplying through by 
$4 \tau^{(020)}$, becomes, 
\bea
4 e^{- i \theta } p_a \sigma ^a \zeta + g_a \tau ^{(302)} \zeta + h_a \tau ^{(313)} \zeta =0
\eea
Contracting to the left in turn by $\zeta ^\dagger $ and $\zeta ^\dagger \sigma ^3$,
and using (\ref{ghreal}), we obtain, 
\bea
\label{preal}
\Im (p_a e^{-2 i \theta} ) =0
\eea
Here, we have assumed that $\zeta ^\dagger \sigma ^a \zeta $ does not
identically vanish, as will be verified from the solution later on. 

\subsection{$SU(1,1)$ map to solutions with vanishing axion}

Equation (\ref{preal}) implies that the dilaton/axion 1-form $P$
satisfies $P = e^{2 i \theta} \tilde P$, where $\tilde P$ is a real form.
Using the second Bianchi identity in (\ref{dilbianchi}), it follows that $dQ=0$,
so that $Q$ is pure gauge. Additionally, from the $SU(1,1)$ transformation
laws (\ref{su11a}) and (\ref{su11b}), it follows that the phase $\theta$
is to be interpreted as the accompanying $U(1)_q$ gauge transformation
of an $SU(1,1)$ transformation that maps the solution to the BPS equations
onto a solution for which $ P$ is real, and $Q=0$.

\smallskip

The fact that the BPS equations result in a reality condition on the 
supergravity fields and allow any solution to the BPS equations to be 
mapped onto a solution with vanishing axion (i.e. real $P$ and $Q=0$)
is familiar from the study of the Janus problem with 4 supersymmetries
in \cite{D'Hoker:2006uu}), where an analogous result holds.

\smallskip

Performing now this $SU(1,1)$ transformation on all fields, we have 
$e^{- i \theta } =i$
(up to an immaterial choice of sign) so that the reality conditions become, 
\bea
\bar p_a = p_a & \hskip 1in & \bar g_a  = g_a \hskip 1in a=8,9
\no \\
q_a=0 & \hskip 1in & \bar h_a  = h_a
\eea
Complex conjugation is now  a symmetry with 
$\sigma ^2 \zeta ^* =  i  \tau ^{(020)} \zeta$.

\subsection{The BPS equations reduced by $\cI$, $\cJ$, and $\cK$ projections}

In view of the involution constraints 
\bea
\cI \zeta & = & - \tau ^{(111)} \sigma ^3 \zeta = \zeta
\no \\
\cJ \zeta & = & \tau ^{(032)} \zeta = \nu \zeta
\no \\
\cK \zeta & = & - i \tau ^{(020)} \sigma ^2 \zeta ^* = \zeta
\eea
 the non-vanishing components of 
$\zeta$ may be parametrized as follows,
\bea
\label{xi8}
\zeta _{+++-} = +i \nu \, \zeta _{+---} & = & \alpha ~ e^{i \theta/2} 
\no \\
\zeta _{---+} = -i \nu \, \zeta _{-+++} & = & \bar \beta \, e^{i \theta/2} 
\no \\
\zeta _{+-++} = +i \nu \, \zeta _{++-+} & = & \bar \alpha \, e^{i \theta/2} 
\no \\
\zeta _{-+--} = -i \nu \, \zeta _{--+-} & = & \beta ~ e^{i \theta/2} 
\eea
Here, the first 3 indices on $\zeta$ refer to its $\eta$-assignments, and the last refers
to its eigenvalue under $\sigma ^3$, while $\alpha, \beta \in \bC$. 
The overall constant phase is the one that resulted from the reality condition, 
and is given by $e^{-i \theta }=i$.

\smallskip

We shall analyze the BPS equations for a single chirality $\g_{(4)}=-1$; the 
opposite chirality equation is just the complex conjugate thereof. 
To do so, we use the complex frame\footnote{Frame indices are defined 
by $e^z = (e^8 + i e^9)/2$ and $e^{\bar z} = (e^8 - i e^9)/2$, and are  to 
be contracted with the flat Euclidean metric with non-zero components 
$\delta _{z\bar z} = \delta _{\bar z \, z} = 2$.} $e^z$ and $e^{\bar z}$ 
on $\Sigma$.
We begin by eliminating $\sigma ^2 \zeta ^* = i \tau ^{(020)}  \zeta$ in favor
of $\zeta$ in all equations of (\ref{dilatino1}) and (\ref{gravitino1}). 
To make the $\tau$-matrices act simply, we make use of the relation 
$\tau ^{(032)} \zeta = \nu \zeta$ in the first terms of equations $(m)$ and $(i_2)$
in (\ref{gravitino1}), and recast them in the following form, $\tau ^{(211)} \zeta =
\nu \tau ^{(223)} \zeta$ and $\tau ^{(002)}\zeta = \nu \tau ^{(030)} \zeta$.

\smallskip

The components of $\zeta$ may be regrouped in terms of their dependence
on $\alpha$ and $\beta$ in terms of spinors $\xi$ and $\xi^*$ whose 
$\g_{(4)}$-eigenvalues are $-1$ and $+1$ respectively,
\bea
\label{albetzeta}
\xi =  \left ( \matrix{ \alpha \, \cr  \beta  \, \cr } \right )  
\hskip 1in 
\xi ^* =   \left ( \matrix{ \bar  \alpha \cr \bar  \beta  \cr } \right ) 
\eea

The $\tau$-matrices in the basis (\ref{taubasis})  may
be  represented in terms of $\sigma$-matrices acting 
on the spinors $\xi$ and $\xi^*$ in the standard basis,
according to the following rule, 
\bea
\tau ^{(011)} = \pm \sigma ^3 & \hskip 1in & \tau ^{(223)} = i \sigma ^2
\no \\
\tau ^{(302)} = - \sigma ^2 & \hskip 1in & \tau ^{(021)} = -i \sigma ^3
\no \\
\tau ^{(100)} = \sigma ^3 \ \ & \hskip 1in & \tau ^{(030)} = - \sigma ^0
\eea
where $\pm$ refers to the action of the matrix on $\zeta _\mp$.
The BPS equations, reduced by the restrictions from the 
involutions $\cI$, $\cJ$, and $\cK$, then become,
\bea
\label{summary4}
(d) & \quad& 
0 =  4 i p_z   \xi -  \left (  g_z \sigma ^2 - h_z \sigma^1 \right )  \xi
\no \\ && \no \\
(m) && 
0 = { \nu \over 2 f_4} \sigma ^2  \xi^*
+  {D_z f_4 \over 2 f_4} \xi
+ {1 \over 2}  f_z \sigma ^3  \xi
- {i \over 16}  
     \left ( g_z \sigma ^2 +  h_z \sigma ^1 \right )  \xi
\no\\ && \no \\
(i_1) && 
0 = - {i \over 2 f_1} \sigma ^3  \xi^*
+  {D_z f_1 \over 2 f_1} \xi
- {1 \over 2}  f_z \sigma ^3 \xi
- {i \over 16}  
     \left ( - 3 g_z \sigma ^2 + h_z \sigma ^1 \right )  \xi
\no\\ && \no \\
(i_2) && 
0 =  - {\nu \over 2 f_2} \xi^*
+  {D_z f_2 \over 2 f_2}  \xi
- {1 \over 2}  f_z \sigma ^3 \xi
- {i \over 16}  
     \left ( g_z \sigma ^2 - 3 h_z \sigma ^1 \right ) \xi
\no\\ && \no \\
(a+) && 
0 =  D_z \xi^* + {i \over 2} \hat \omega_z   \xi^*  +  f_z \sigma ^3 \xi ^*
+ {i \over 4}   \left ( g_z \sigma ^2 - h_z \sigma ^1 \right )  \xi^*
\no\\ && \no \\
(a-) && 
0 = D_z \xi   - {i \over 2} \hat \omega_z   \xi
+ {i \over 8}   \left ( g_z \sigma ^2 + h_z \sigma ^1 \right ) \xi
\eea

\subsection{Normalizations of $f_1, f_2, f_4$}

We shall now extract all information contained in equations $(d)$, $(m)$, $(i_1)$, 
and $(i_2)$.
To do this, for each equation, we form two functionally independent linear 
combinations; since each equation is a 2-component matrix, these two
linear combinations will fully capture the contents of the corresponding equation.

\medskip

The first linear combinations are obtained by contracting the equations $(m)$, 
$(i_1)$, and $(i_2)$ on the left respectively
by $ 2 \xi ^\dagger$, $2 \xi ^\dagger \sigma ^1$, and $\xi ^\dagger \sigma ^2$.
The first term in each resulting equation vanishes by antisymmetry of $\sigma ^2$,
and we obtain,
\bea
\label{mii}
(m) && 
0 =
  {D_z f_4 \over  f_4} \xi ^\dagger \xi
+   f_z \xi ^\dagger \sigma ^3  \xi
- {i \over 8}  
    \xi ^\dagger  \left ( g_z \sigma ^2 +  h_z \sigma ^1 \right )  \xi
\no\\ 
(i_1) && 
0 =
  {D_z f_1 \over  f_1} \xi ^\dagger \sigma ^1 \xi
+   f_z  \xi ^\dagger  \sigma ^3 \sigma ^1 \xi
- {i \over 8}  \xi ^\dagger 
     \left ( - 3 i g_z \sigma ^3 + h_z  \right )  \xi
\no\\ 
(i_2) && 
0 =    {D_z f_2 \over f_2}  \xi ^\dagger \sigma ^2 \xi
+   f_z \xi ^\dagger \sigma ^3  \sigma ^2 \xi
- {i \over 8}  \xi ^\dagger
     \left ( g_z  + 3 i h_z \sigma ^3 \right ) \xi
\eea
The following combinations for $p=0,1,2$ are 
calculated using the equations $(a\pm)$,  
\bea
D_z \left ( \xi ^\dagger \sigma ^p \xi \right )
=
- f_z \xi ^\dagger \sigma ^3 \sigma ^p \xi 
- {i \over 8} \xi ^\dagger \left ( \sigma ^p (g_z \sigma ^2 + h_z \sigma ^1) 
- 2 (g_z \sigma ^2 + h_z \sigma ^1)
\sigma ^p \right ) \xi
\eea
They combine with (\ref{mii}) to give
\bea
D_z \left ( f_4 ^{-1}  \xi ^\dagger \xi \right )=
D_z \left ( f_1 ^{-1}  \xi ^\dagger \sigma ^1 \xi \right )=
D_z \left ( f_2 ^{-1}  \xi ^\dagger \sigma ^2 \xi \right )=0
\eea
so that these ratios are constants. The BPS equations are linear in $\xi, \xi^*$
and invariant under scaling by a real constant, which allows us  to normalize the 
relation involving $f_4$, as follows,
\bea
\label{metricbilin}
\xi ^\dagger \xi = f_4
\hskip 1in
\xi ^\dagger \sigma ^i \xi = \lambda _i f_i
\hskip 1in i=1,2
\eea
where $\lambda_1, \lambda _2$ are real constants.

\medskip

The second set of linear combinations is obtained by contracting the equations 
$(d)$, $(m)$, $(i_1)$, and $(i_2)$ on the left respectively by $  \xi ^t \sigma ^2$, 
so as to  eliminate the derivative terms in $f_i$. We also make use of (\ref{metricbilin})
to obtain,
\bea
\label{norm1}
(d) && 0= g_z \, \xi ^t \xi + i h_z \, \xi^t \sigma ^3 \xi \hskip 1.6in
\no\\
(m) && 
0 = i \nu  
- f_z \xi ^t \sigma ^1 \xi 
+ { 1 \over 8} \xi ^t (g_z - i h_z \sigma ^3 ) \xi
\no\\ 
(i_1) && 
0 =
  -i \lambda _1
-  f_z \xi ^t \sigma ^1 \xi 
- { 1 \over 8} \xi ^t (-3 g_z - i h_z \sigma ^3 ) \xi
\no\\ 
(i_2) && 
0 =   
 -i \nu  \lambda _2
-  f_z \xi ^t \sigma ^1 \xi 
- { 1 \over 8} \xi ^t (g_z + 3 i h_z \sigma ^3 ) \xi
\eea
In the sum $2(m) - (i_1) - (i_2)$, all dependence on $f_z$, $g_z$, $h_z$ cancels, 
and  using (\ref{metricbilin}), we obtain $  \nu \lambda _1 +  \lambda _2=-2$.
From the sum $(i_1)-(i_2)$, and using $(d)$, we find  
$\nu \lambda _1 = \lambda _2$, so that $\lambda _1 = - \nu$, and $ \lambda _2 = - 1$.
Putting all together, the final expressions are,
\bea
\label{metricbilin2}
f_4 & = & \xi ^\dagger \xi \hskip 0.5in       = \alpha \bar \alpha + \beta \bar \beta
\no \\
f_1 & = & - \nu \xi ^\dagger \sigma ^1 \xi 
\hskip 0.12in 
= - \nu ( \alpha \bar \beta + \beta \bar \alpha)
\no \\
f_2 & = & - \xi ^\dagger \sigma ^2 \xi  
\hskip 0.2in 
= i (  \beta \bar \alpha - \alpha \bar \beta )
\eea
These relations completely solve the $(i_1)$ and $(i_2)$ equations, so that
only the $(d)$, $(m)$ and $(a\pm)$ equations remain to be solved.

\subsection{Consistency}

To solve for the reality conditions in subsection 6.3, we had made the 
assumption that $\zeta ^\dagger \tau ^{(323)} \sigma ^a \zeta =
\nu \zeta ^\dagger \tau ^{(311)} \sigma ^a \zeta$ is non-vanishing.
In terms of the parametrization of $\zeta$ by $\a$ and $\b$, 
this quantity takes on the following form, 
$\zeta ^\dagger \tau ^{(323)} \sigma ^z \zeta = - 2i \a \b$. 
Generically, this quantity must be non-vanishing on regular solutions,
since its vanishing would imply that $f_1=f_2=0$ identically.
Thus, our earlier non-vanishing assumption is consistent.

\subsection{The reduced BPS equations in conformal coordinates}

We choose conformal complex coordinates $w$ on $\Sigma$, such that the 
metric takes the form, $ds^2_\Sigma =  4 \rho ^2 d\bar w \, d w$. The frames, derivatives,  
and connection are then given by
\bea
e^z = \rho \, dw
& \qquad \qquad D_z = \rho ^{-1} \p_w 
\qquad \qquad & 
\hat \omega _z = + i \rho ^{-2} \p_w \rho 
\no \\
e^{\bar z} = \rho \, d\bar w
& \qquad \qquad 
D_{\bar z} = \rho ^{-1} \p_{\bar w} 
\qquad \qquad & 
 \hat \omega _{\bar z} = - i \rho ^{-2} \p_{\bar w}   \rho
\eea
Notice that $z$ and $\bar z$ are {\sl frame indices}, whence the extra factor 
of $1/\rho$ in $D_z , \, D_{\bar z}$. 

\smallskip

With the help of the relation (\ref{albetzeta}), we may express $\xi$ in terms 
of $\alpha, \beta , \bar \alpha $ and $\bar \beta $ in the remaining equations. 
These equations are: the dilatino equation $(d)$, the differental
equations $(a+)$ and $(a-)$  of (\ref{summary4}), and the single 
equation $(m)$ of (\ref{norm1}). Expressed in conformal complex coordinates 
$w$, these equations become, 
\bea
\label{fullBPS}
(d) & \qquad & 
4 p_z \alpha + (g_z -i h_z) \beta =0
\no \\ && 
4 p_z \beta - (g_z+ih_z) \alpha =0
\no \\ 
(m) & \qquad &
i \, \nu - 2 \alpha \beta f_z 
+ {1 \over 8} (g_z-i h_z) \alpha ^2 + {1 \over 8} (g_z+ih_z) \beta ^2 =0
\no \\ 
(a+) & \qquad &
{1 \over \rho}  \p_w \bar \alpha   - {1 \over 2 \rho^2} (\p_w \rho)  \bar \alpha 
+ f_z \bar \alpha  + {1 \over 4} (g_z-ih_z) \bar \beta  =0
\no \\ &&
{1 \over \rho}  \p_w \bar \beta  - {1 \over 2 \rho^2 } (\p_w \rho) \bar \beta 
- f_z \bar \beta  - {1 \over 4} (g_z + ih_z) \bar \alpha  =0
\no \\ && \no \\
(a-) & \qquad &
{1 \over \rho} \p_w \alpha  + {1 \over 2 \rho^2 } (\p_w \rho ) \alpha 
+  {1 \over 8} (g_z + ih_z) \beta =0
\no \\ &&
{1 \over \rho}  \p_w \beta + {1 \over 2 \rho^2 } (\p_w \rho ) \beta 
  - {1 \over 8} (g_z - ih_z) \alpha =0
\eea
This system will be the starting point of our construction of the complete
and exact solution.

\subsection{Constant dilaton implies $AdS_{5}\times S^{5}$}

In this subsection we show that the only solution of the BPS equations 
(\ref{fullBPS}) with constant dilaton is $AdS_{5}\times S^{5}$.  
The argument is presented independently from the general solution which 
will be derived in the next section.

\smallskip

A constant dilaton implies $p_{z}=0$. It follows from the fact that the metric factors 
(\ref{metricbilin2}) cannot be identically zero, and equations $(d)$ and $(a-)$  
of (\ref{fullBPS}) that the 3-form fluxes have to vanish, i.e. $g_{z}=h_{z}=0$.
Solving equation  $(a-)$ implies
\bea
\rho \alpha^{2} & = & \bar A(\bar w)
\no \\
\rho \beta^{2} & = & i \bar B(\bar w)
\eea
where $A(w)$ and $B(w)$ are purely holomorphic functions of $w$.
The phase is chosen for later convenience. From the  difference of the $(a+)$ 
equations it follows that
\be
\partial_{w}\left( {A B \over \rho^{4}}\right)=0
\ee
and hence $\rho^{4}=|AB|^{2}$. It is therefore possible to make a holomorphic 
change of coordinates and set $\rho=1$. This implies $A=1/B$. The remaining 
equation of $(a-)$ leads to an equation for $A(w)$
\be
{\partial_{w}A}+\nu A =0 
\ee
which is solved by $A=e^{-\nu w}$ and hence
\bea
\alpha & = & e^{-\nu w/2}
\no \\
\beta  & = &  i e^{\nu w/2}
\eea
The metric factors are given by
\bea
f_{1} & = & + 2 \sin y 
\no \\
f_{2} & = & - 2 \cos  y
\no \\
f_{4} & = & + 2 \, \ch x 
\eea
where  the coordinates $x,y$ are related to $w$ by 
$w=x+i y$ and take values in the strip $x\in[-\infty,\infty], y\in[0,\pi]$. 
Hence the  ten-dimensional metric is given by
\be
ds^{2}= 4 \Big( dx^{2}+ \ch^{2}x \;ds_{AdS_{4}}^{2} +dy^{2} 
+  \cos^{2} y\; ds_{S^2}^{2}+\sin^{2} y \;ds_{S^2}^{2}\Big)
\ee
which is indeed $AdS_{5 }\times S^{5}$.  Since a constant dilaton implies $AdS_5 \times S^5$, we shall henceforth assume that the dilaton has a non-trivial variation over $\Sigma$. 

\newpage

\section{The BPS equations form an integrable system}
\setcounter{equation}{0}
\label{seven}

In this section, we shall  show that the BPS equations form an integrable system. 
The solutions to this system will automatically solve the Bianchi identities and 
field equations of Type IIB supergravity as discussed in section 8.  As a first step,
we solve the $(d)$ and $(a-)$ part of the system,
and then use its solution to reduce further the remaining $(m)$ and $(a+)$
equations to a system of first order equations on two real scalar fields, 
the dilaton $\phi$ and the $\Sigma$-metric factor $\rho$. We shall then 
show that this system is automatically integrable.

\smallskip

First, we shall view the dilatino equations $(d)$ as determining $g_z$ and $h_z$
in terms of $\alpha$, $\beta$ and $p_z$ (with corresponding equations for their 
complex conjugates),
\bea
\label{ghzee}
g_z - i h _z & = & - {4 \alpha \over  \beta} ~ p_z
\no \\
g_z + i h _z & = & +  {4 \beta \over  \alpha } ~ p_z
\eea
and the $(m)$ equation as determining $f_z$,
\bea
\label{fzee}
f_z = {i \nu \over 2 \alpha \beta} 
- {\alpha ^4 - \beta ^4 \over 4  \alpha ^2 \beta ^2} p_z
\eea
These equations may be used to eliminate 
$f_z$, $g_z$ and $h_z$ from the remaining equations, $(a+)$ and $(a-)$. 
We recast the resulting equations in the following form, ready for later use,
\bea
\label{derplus}
(a+) & \qquad &
\p_w \ln \left ( {\bar \alpha \over \bar \beta} \right ) 
- {1 \over \rho} \left ( {\alpha \bar \beta \over \bar \alpha \beta} 
- {\bar \alpha \beta \over \alpha \bar \beta} \right ) p_z  
+  {i \nu \rho \over  \alpha \beta} 
- {\alpha ^4 - \beta ^4 \over 2  \alpha ^2 \beta ^2} \rho p_z
= 0
\no \\ &&
\p_w \ln ( \bar \alpha \bar \beta ) - \p_w \ln \rho  
- {1 \over \rho} \left ( {\alpha \bar \beta \over \bar \alpha \beta} 
+ {\bar \alpha \beta \over \alpha \bar \beta} \right ) p_z  = 0
\\ && \no \\
\label{derminus}
(a-) & \qquad &
\p_w \alpha + {1 \over 2 \rho} (\p_w \rho ) \alpha 
+  {1 \over 2}  { \beta ^2 \over \alpha} \rho \, p_z =0
\no \\ &&
\p_w \beta  + {1 \over 2 \rho} (\p_w \rho ) \beta 
+ {1 \over 2}   {\alpha ^2 \over \beta} \rho \, p_z =0
\eea

\subsection{Solution to the $(a-)$ system}

Multiplying the first equation of (\ref{derminus})  by $2 \rho \alpha$ and the second 
by $2 \rho \beta$, we obtain equivalently, 
\bea
\p_w \left ( \rho \alpha^2 \right ) 
+  \rho \beta ^2 \, \rho p_z & = & 0
\no \\
\p_w \left ( \rho \beta ^2 \right ) 
+  \rho  \alpha ^2 \rho p_z & = & 0
\eea
Adding and subtracting both, we get 
\bea
\label{aplus1}
\p_w  \ln \left ( \rho (\alpha ^2 + \beta ^2 ) \right ) + \rho p_z & = & 0
\no \\
\p_w  \ln \left ( \rho (\alpha ^2 - \beta ^2 ) \right ) - \rho p_z & = & 0
\eea
It follows immediately from these equations that $\rho p_a$ is the gradient
of a scalar function. Inspection of (\ref{sugra1}) and (\ref{Btau}) reveals
that this scalar function is none other than  $\phi$ (related to the 
dilaton $\Phi$ in standard normalization by $\phi = \Phi /2$), so that we
deduce from the BPS equations the relation
\bea
\label{pphi}
\rho p_z = \p_w \phi
\eea
The system (\ref{aplus1}) may now be solved completely in terms of two
(locally) holomorphic functions  $\kappa ( w)$  and $\lambda (w)$,
\bea
\label{albetsol0}
\rho (\alpha ^2 + \beta ^2) & = & \bar \kappa \,
 e^ { - \bar \lambda - \phi  }
\no \\
\rho (\alpha ^2 - \beta ^2) & = & \bar \kappa\,
e^{ + \bar \lambda  +\phi }
\eea 
From the $(m)$ equation, it is manifest that $\alpha$ and $\beta$ are spinors
with respect to the $SO(2)$ frame group of $\Sigma$ with weight $(-1/2,0)$ 
in a convention in which $e^z$ has weight $(-1,0)$. Since $\rho$ has weight
$(1/2, 1/2)$, we conclude that $\kappa$ is actually a form of weight 
$(1/2, -1/2) \sim (1,0)$.
These relations may be solved for $\alpha$ and $\beta$, as follows,
\bea 
\label{albetsol1}
\alpha & = & \left ( {\bar \kappa \over \rho} \right )  ^\half  \ch (\phi + \bar \lambda )^\half
\no \\
\beta & = & i  \left ( {\bar \kappa \over \rho} \right )  ^\half  \sh (\phi + \bar \lambda )^\half
\eea
Here, we have adopted a definite sign choice for each square root.
The parametrization in terms of $\kappa$ and $\lambda$ is convenient since 
one natural combination will involve only $\phi$ and $\lambda$, while
another will involve only $\rho$ and $\kappa$. They are given by\footnote{The 
corresponding equation (5.16) of \cite{Gomis:2006cu} is incorrect.}
\bea
\label{albetsol2}
{\alpha ^2 \over \beta^2} & = & 
{ 1+ e^{2 \phi + 2 \bar \lambda} \over 1- e^{2 \phi + 2 \bar \lambda} }
\no \\
 \bar \kappa ^2 & = &
\rho ^2 \left (\alpha ^4 - \beta ^4 \right )  
\eea
Another equation, which may be directly deduced from (\ref{albetsol1}), will be useful as well,
\bea
\label{albetsol3}
4 \rho ^2 \alpha ^2 \beta ^2 =  \bar \kappa^2 \left ( e^{-2 \phi - 2 \bar \lambda}
- e^{2 \phi + 2 \bar \lambda} \right )
\eea

\subsection{Solution to the $(a+)$ system}

The $(a+)$ system will be solved as follows. Equations (\ref{albetsol1}) will be
viewed as giving $\alpha$ and $\beta$ in terms of $\phi$, $\rho$ and the 
holomorphic functions $\kappa$ and $\lambda$, and will be 
used to eliminate $\bar \alpha$ and $\bar \beta$ from the $(a+)$ system.
The combinations of $\p_w$-derivatives of logarithms occurring in 
 (\ref{derplus}) may be computed from the solution of the $(a-)$
equations, by taking the derivatives of the complex conjugates to the first relation in 
(\ref{albetsol2}) and the equation of (\ref{albetsol3}),
\bea
\label{logder2}
\p_w \ln \left ( { \bar \alpha  \over \bar \beta  } \right )
& = & 
\half \left ( {\bar \alpha ^2 \over \bar \beta ^2 } - {\bar \beta ^2 \over \bar \alpha ^2 } \right )
\left ( \p_w \phi + \p_w \lambda  \right )
\no \\
\p_w \ln \left ( \rho \bar \alpha \bar \beta  \right )
& = &
\p_w \ln \kappa 
- \half \left ( {\bar \alpha ^2 \over \bar \beta ^2 } + {\bar \beta ^2 \over \bar \alpha ^2 } \right )
\left ( \p_w \phi + \p_w \lambda  \right )
\eea
Eliminating now from both $(a+)$ equations in (\ref{derplus}), $p_z$ using (\ref{pphi}),
the $w$-derivatives of  $\ln \bar \alpha $ and $\ln \bar \beta $ in (\ref{derplus}) using 
(\ref{logder2}), and the remaining algebraic dependence of $\alpha$, $\beta$ and their 
complex conjugates, using (\ref{albetsol0}) -- (\ref{albetsol3}), we obtain
a system of first order differential equations for $\phi$ and $\rho$ only, with 
$\kappa$ and $\lambda$ viewed as given  holomorphic functions, 
\bea
\label{inteq1}
\left [ {1 \over \sh (2 \phi + 2 \lambda )} - {1 \over \sh (2 \phi + 2 \bar \lambda )} -
 {2 \, \sh(\lambda - \bar \lambda ) \over | \sh (2 \phi + 2 \lambda ) |}  \right ] \p_w \phi
= {\sqrt{2}  \nu \, \bar \kappa^{-1}  \, \rho ^2  \over  \sh (2 \phi + 2 \bar \lambda)^\half } 
- {\p_w \lambda  \over \sh (2 \phi + 2  \lambda )} \qquad
\eea
\bea
\label{inteq2}
{2 \over \rho} \p_w \rho = \p_w \ln \kappa
+ {\ch (2 \phi + 2 \lambda ) \over \sh (2 \phi + 2 \lambda )}
( \p_w \phi + \p_w \lambda) 
+ 2 { \sh (2 \phi + \lambda + \bar \lambda ) \over |\sh (2 \phi + 2 \lambda)|}  \p_w \phi 
\eea
One can now use the first equation to eliminate $\p_w \phi$ from the second 
equation, so that we get two first order equations separately for $\p_w\phi$ and 
$\p_w \ln \rho$. Each of these equations has a complex conjugate giving the 
$\p_{\bar w}$-derivatives, and there will be two integrability conditions. 

\subsection{Relation to a new integrable system}
\label{intsyst}

To begin, we recast (\ref{inteq2}) in
terms of its complex conjugate equation, rearranged as follows,
\bea
\label{inteq2a}
\p_{\bar w} \ln \left (  
{\sqrt{2}  \nu \rho ^2  \over \bar \kappa \, \sh (2 \phi + 2 \bar \lambda)^\half }  \right )
=
 2 { \sh (2 \phi + \lambda + \bar \lambda ) \over |\sh (2 \phi + 2 \lambda)|}   \p_{\bar w} \phi 
\eea
From this form of (\ref{inteq2}), we see that $\rho$ and $\kappa$ may be 
eliminated  between (\ref{inteq1}) and (\ref{inteq2}), by eliminating 
the combination $\sqrt{2}  \nu \rho ^2 \bar \kappa ^{-1} 
\sh (2 \phi + 2 \bar \lambda)^{-\half}$
between (\ref{inteq1}) and (\ref{inteq2a}). The resulting equation involves only
$\phi$ and $\lambda$, though this ``simplification" has been achieved at
the cost of obtaining a second order partial differential equation, given by
\bea
\label{univ1}
&&
\p_w  \p_{\bar w} \phi 
- 2 {\sh (4 \phi + 2 \lambda + 2 \bar \lambda) \over | \sh (2 \phi + 2 \lambda)|^2 }
    \p_w \phi \p_{\bar w} \phi 
\\ && \hskip .6in 
=
{\p_{\bar w} \phi \, \p_w  \lambda \over \ch ( \lambda -  \bar \lambda)} \, 
    { \sh (2 \phi + 2 \bar \lambda) \over  \sh (2 \phi + 2 \lambda)}
\left ( - {1 \over \sh (\lambda - \bar \lambda) }  -  
    { \sh (2 \phi + 2  \lambda)^\half  \over  \sh (2 \phi + 2 \bar \lambda)^\half } \right )
    + {\rm c.c.}
\no 
\eea
Note that (\ref{univ1}) is real.
Although the equation looks daunting, we shall show that it is integrable and
better even, that its general solution may be obtained in analytic form.

\subsection{Integrability}

In this subsection, we shall show that the system of first order differential
equations (\ref{inteq1}), (\ref{inteq2}) and their complex conjugates,
form an integrable system for any choice of holomorphic functions 
$\kappa$ and $\lambda$. Equation (\ref{univ1}), which was shown to
be a consequence of the first order system (\ref{inteq1}) and (\ref{inteq2}),
will be used in the process.

\smallskip

Integrability in $\rho$ amounts to the reality of $\p_{\bar w} \p_w \ln \rho^2$.
This quantity may be obtained directly by taking the $\p_{\bar w}$-derivative 
of (\ref{inteq2}). By construction, the resulting integrability equation 
does not involve $\rho$ or $\kappa$, and actually coincides with the 
second order equation (\ref{univ1}), which we have already show to be
a consequence of the system (\ref{inteq1}) and (\ref{inteq2}). Thus, 
integrability in $\rho$ holds automatically for any holomorphic $\kappa$ 
and $\lambda$.

\smallskip

Integrability in $\phi$ may be verified by taking the $\p_{\bar w}$ derivative
of (\ref{inteq1}). As was already established in subsection \ref{intsyst}, the resulting
equation, after elimination of $\rho$ and $\kappa$ is precisely the second
order equation (\ref{univ1}). The fact that (\ref{univ1}) emerges  as a real
equation directly guarantees that $\p_{\bar w} \p_w \phi$ is real, and thus
that the system (\ref{inteq1}) and (\ref{inteq2}) is integrable in $\phi$.

\newpage

\section{Reduced Bianchi Identities and Field Equations}
\setcounter{equation}{0}
\label{eight}

In this section, we shall derive the reduced Bianchi identities, and field equations, 
and show that they are satisfied for any field configuration that satisfies the 
BPS equations 
for 16 supersymmetries.  As shown in subsection 6.4, every solution of the BPS equations may be 
transformed to a solution with vanishing axion field under an $SU(1,1)$ 
transformation. 
Thus, we may restrict to the case of vanishing axion, without loss of generality. 
The solutions with non-vanishing axion may be obtained by making 
$SU(1,1)$ transformations. 

\subsection{The reduced Bianchi identities}

Using differential form notation, it is straightforward to reduce the Bianchi
identities on the Ansatz defined in section 3. We find, 
\bea
\label{redBianchi}
0 & = &  d P - 2 i Q \wedge P 
\no \\
0 & = &  dQ + i P \wedge \bar P 
\no \\
0 & = &  d\cG + 2 (d \ln f_1) \wedge \cG  -i Q \wedge \cG + P \wedge \bar \cG 
\no \\
0  & = & d\cH + 2 (d \ln f_2) \wedge \cH  -i Q \wedge \cH - P \wedge \bar \cH 
\no \\
0 & = &  d  \cF + 4 (d \ln f_4) \wedge \cF 
\no \\
0 & = &  d (*_2 \cF) + 2 \left ( d \ln (f_1 f_2) \right ) \wedge *_2\cF 
- {1 \over 8} \left ( \cG \wedge \bar \cH  + \bar \cG \wedge \cH \right ) 
\eea
The reduced Bianchi identities now simplify as one may set $Q=0$
in (\ref{redBianchi}).

\subsection{Derivation of the reduced field equations}

In this subsection we reduce the Type IIB field equations of
(\ref{Peq}), (\ref{Geq}), and (\ref{Eeq}) to the two-parameter Ansatz 
of section \ref{four} for $Q=0$.

\subsubsection{The dilaton field equation}

It is straightforward to derive the dilaton equation, using the convention 
$f^2 d B = d\phi$, 
\bea
D^a D_a \phi + 2 (D^a \phi) D_a \ln (f_1f_2 f_4)
+ {1 \over 4} (g_a g^a - h_a h^a) =0
\eea
In local conformal coordinates, this becomes,
\bea
\label{dil5}
\p_{\bar w} \p_w \phi + (\p_{\bar w} \phi ) \p_w \ln (f_1 f_2 f_4^2)
+ (\p_w \phi )\p_{\bar w} \ln (f_1 f_2 f_4^2) 
+{1 \over 4} \rho ^2 (g_z g_{\bar z} - h_z h_{\bar z}) =0
\eea

\subsubsection{The $G$-field equation}

To reduce the field equations of the antisymmetric tensor field $B_{(2)}$,
it is convenient to first recast (\ref{Geq}) in terms of differential 
forms,\footnote{Our conventions for the Poincar\'e dual are given via the 
following pairing relation between two arbitrary rank $p$ differential
forms $S_{(p)}$ and $T_{(p)}$, by  $S_{(p)} \wedge *T_{(p)} 
= {1 \over p!} S_{(p)}^{a_1 \cdots a_p} T_{{(p)}a_1 \cdots a_p} e^{0123456789}$. 
In particular, we have $** S_{(p)}= (-1)^{p+1} S_{(p)}$, and the duals
$* e^{01234} = - e^{56789}$, and $*e^{6789} = e^{012345}$, 
$* e^{45a} = + \ep ^{ab} \, e^{012367b}$, $* e^{01234589} = - e^{67}$,
$ * e^{67a} = + \ep ^{ab} \, e^{012345b}$, and $* e^{01236789} = - e^{45}$,
which are useful in deriving the $G$-equations.}
\bea
 * d(* G) + (i_P \bar G) - 4 i \, (i_G F_{(5)}) =0
\eea
Here $i_V G$ stands for the contraction of $G$ with $V$.
Some useful intermediary results for this calculation are as follows,
\bea
i_P \bar G & = & p^a g_a e^{45} - i p^a h_a e^{67}
\no \\
i_G F_{(5)} & = & f_a \ep ^{ab} g_b e^{67} + i f_a \ep ^{ab} h_b e^{45}
\eea
The $G$-field equation then reduces to the following two real equations,
\bea
D^a g_a + 2 g_a D^a \ln (f_2 f_4^2) 
- p^a g_a - 4 f_a \ep ^{ab} h_b & = &0
\no \\
D^a h_a + 2 h_a D^a \ln (f_1 f_4^2) 
+ p^a h_a + 4 f_a \ep ^{ab} g_b & = &0
\eea
In conformal gauge, and after multiplication by $\rho^2$, this simplifies to
\bea
\p_w (\rho g_{\bar z}) + 2 \rho g_{\bar z} \p_w \ln (f_2 f_4^2) 
- (\p_w \phi ) \rho  g_{\bar z}  + 4i \rho f_z \, \rho h_{\bar z} + c.c.  & = &0
\no \\
\p_w (\rho h_{\bar z}) + 2 \rho h_{\bar z} \p_w \ln (f_1 f_4^2)
+ (\p_w \phi) \rho  h_{\bar z}  - 4i \rho f_z \, \rho g_{\bar z} + c.c. & = &0 \qquad 
\eea

\subsubsection{Einstein's equations}

The Einstein equations, respectively for the components $mn$,
$i_1j_1$, $i_2j_2$, and $ab$, are as follows, (all other components must vanish by 
$SO(2,3) \times SO(3) \times SO(3)$ symmetry),
\bea
\label{Einstein}
0 & = & 
- {3 \over f_4^2} 
- 3 { |D_a f_4 |^2 \over f_4 ^2} 
- 2 {D^a f_4 D_a (f_1 f_2) \over f_1 f_2 f_4}
- {D^a D_a f_4  \over f_4} 
+ 4 f_a f^a + {1 \over 8} g_a g^a + {1 \over 8} h_a h^a
\no \\ 
0 & = & 
 {1 \over f_1^2} 
-  { |D_a f_1 |^2 \over f_1 ^2} 
- 4 {D^a f_4 D_a f_1 \over f_1  f_4}
- 2 {D^a f_1 D_a f_2 \over f_1  f_2}
- {D^a D_a f_1 \over f_1} 
- 4 f_a f^a - {3 \over 8} g_a g^a + {1 \over 8} h_a h^a
\no \\ 
0 & = &  
{1 \over f_2^2} 
-  { |D_a f_2 |^2 \over f_2 ^2} 
- 4 {D^a f_4 D_a f_2 \over f_2  f_4}
- 2 {D^a f_1 D_a f_2 \over f_1  f_2}
- {D^a D_a f_2 \over f_2} 
- 4 f_a f^a + {1 \over 8} g_a g^a - {3 \over 8} h_a h^a
\no \\ 
0 & = & 
- 2 {D_b D_a f_4 \over f_4}  
- {D_b D_a f_1 \over f_1}  
- {D_b D_a f_2 \over f_2}  + \half R^{(2)} \delta _{ab}
-  D_a \phi D_b \phi
\no \\ && \hskip .3in 
- 2 \delta _{ab} f_c f^c + 4 f_a f_b
+ {1 \over 16} \delta _{ab} \left ( g_c g^c + h_c h^c \right ) 
- {1 \over 4} g_a g_b - {1 \over 4} h_a h_b
\eea
With respect to the frame rotation group $SO(2)$ of $\Sigma$, the first three equations 
are weight $(0,0)$, while the last contains both weights $(0,0)$, and $(2,0)$. It will be 
useful to separate these two parts in the last equation. The weight $(2,0)$ part is given by
\bea
\label{spin2}
2 {D_z ^2 f_4 \over f_4}  
+ {D_z^2 f_1 \over f_1}  
+ {D_z^2 f_2 \over f_2} +  (D_z \phi )^2
- 4 f_z^2
+ {1 \over 4} g_z^2 + {1 \over 4} h_z^2 =0
\eea
and its complex conjugate, 
It may be viewed as the {\sl constraint} of vanishing spin 2 parts of the stress tensor
on $\Sigma$. The spin 0 part is given by
\bea
\label{spin0}
- 2 {D^a D_a f_4 \over f_4}  
- {D^a D_a f_1 \over f_1}  
- {D^a D_a f_2 \over f_2}  + R^{(2)} 
-  D_a \phi D^a \phi
- {1 \over 8}  \left ( g_c g^c + h_c h^c \right ) =0
\eea

\subsection{Derivation of Bianchi Identities and Field Equations from BPS Equations}

The Bianchi identity for the dilaton of (\ref{redBianchi}), namely $dP=0$, 
was already derived from the BPS equations in (\ref{pphi}). The derivation 
of the remaining Bianchi identities and field equations from the BPS equations is
considerably more involved. Below, we shall present the analytic derivations
of the dilaton field equation (\ref{dil5}) and the spin 2 constraint (\ref{spin2})
of the Einstein equations directly from the BPS equations. The remaining
Bianchi identities and field equations were verified to follow from the BPS 
equations  using Mathematica. General arguments that the BPS equations 
imply the field equations are given in \cite{papadopoulos}.

\subsubsection{The Dilaton equation}

To work out the dilaton equation (\ref{dil5}), we first derive the following quantities,
\bea
\label{intres2}
\rho ^2 (g_z g_{\bar z} - h_z h_{\bar z} )
& = & 16 {\sh (2 \phi + \lambda + \bar \lambda) \over |\sh (2 \phi + 2 \lambda) |} 
\, \p_w \phi \, \p _{\bar w} \phi
\no \\
\p_w \ln (f_1 f_2 f_4^2)  
& = & -2 \p_w \ln \rho ^2 + 2 \p_w \kappa + \p_w \ln \sh (\lambda - \bar \lambda)
\no \\ && \hskip .2in 
+ \p_w \ln \left \{ \ch (2 \phi + \lambda + \bar \lambda) + |\sh (2 \phi + 2 \lambda) | \right \}
\eea
The first equation is obtained from (\ref{ghzee}) by eliminating $\alpha$ and $\beta$
using (\ref{albetsol0}) -- (\ref{albetsol3}). The second equation is obtained starting 
from (\ref{metricbilin2}) to express $f_1,f_2,f_4$ in terms of $\alpha, \beta, \bar \alpha,
\bar \beta$, and then using (\ref{derplus}) and (\ref{derminus}) to compute the 
derivatives of $\alpha, \beta, \bar \alpha, \bar \beta$, after which all algebraic
dependence on these variables is eliminated in favor of $\phi, \rho, \kappa, \lambda$
using (\ref{albetsol0}) -- (\ref{albetsol3}).
Using now (\ref{inteq2}) to eliminate $\rho$ and $\kappa$ from the above
expression, we obtain,
\bea
\p_w \ln (f_1 f_2 f_4^2)  
& = & 
-\p_w \ln \sh (2 \phi + 2 \lambda)
- 4 { \sh (2 \phi + \lambda + \bar \lambda) \over |\sh (2 \phi + 2 \lambda) |} \p_w \phi +
 \p_w \ln \sh (\lambda - \bar \lambda)
\no \\ && \hskip .2in 
+ \p_w \ln \left \{ \ch (2 \phi + \lambda + \bar \lambda) + |\sh (2 \phi + 2 \lambda) | \right \}
\eea
Substituting these expressions back into the dilaton equation (\ref{dil5}), it is
clear that we obtain a second order partial differential equation that involves 
only $\phi$ and $\lambda$. Not surprisingly, this equation  coincides with 
(\ref{univ1}), which in turn was already shown to follow from the system of 
first order equations (\ref{inteq1}) and (\ref{inteq2}). Thus, the dilaton
equation will be satisfied as soon as the system of equations (\ref{inteq1}) and 
(\ref{inteq2})  is satisfied.

\subsubsection{The constraint equation}

The weight $(2,0)$ constraint equation (\ref{spin2}) is also a consequence of the 
BPS equations.
To see this, we need to compute second order derivatives, using the formula,
\bea
\rho ^2 D_z ^2 f_i = \p_w ^2 f_i - (\p_w \ln \rho ^2) \p_w f_i \hskip 1in i=1,2,4
\eea
We first derive formulas for the first derivatives of the $f_i$ functions, 
using the reduced BPS equations  of (\ref{ghzee}), (\ref{fzee}), and (\ref{derplus}), and find, 
\bea
\label{fder}
\p_w f_1 & = & \left ( + \p_w \phi - A_+ \right ) f_1 + { \rho \over 2 \alpha \beta} f_2
\no \\
\p_w f_2 & = &  \left ( - \p_w \phi - A_+ \right ) f_2 - { \rho \over 2 \alpha \beta} f_1
\no \\
\p_w f_4 & = &  +  A_+  f_4 -  {i  \nu \rho \over 2 \alpha \beta} \sqrt{f_4^2 - f_1^2 - f_2^2}
\eea
where we use the following objects,
\bea
A_+ & \equiv &
 - {1 \over 4} \p_w \ln \bigg (  \sh (2\phi + 2 \bar \lambda) \bigg )
 ~ = \half \p_w \ln (\rho \alpha \beta)
\no \\
A_- & \equiv & 
 - {1 \over 4} \p_w \ln \left ( { \sh (\phi + \bar \lambda) \over \ch (\phi + \bar \lambda) } \right )
 ~ ~ \, = - \rho f_z + { i \nu \rho \over 2 \alpha \beta}
\eea
Furthermore, the second derivatives may be computed with the help of 
the reduced BPS equations (\ref{ghzee}), (\ref{fzee}), and (\ref{fder}) and  (\ref{derplus}). 
After considerable  simplifications, we find,
\bea
 4 \rho ^2 {D_z ^2 f_4 \over f_4}  
+ 2 \rho ^2 {D_z^2 f_1 \over f_1}  
+ 2\rho ^2 {D_z^2 f_2 \over f_2} 
= 4 (\p_w \phi )^2 + 8 A_+^2 - 2 { \rho ^2 \over \alpha ^2 \beta ^2}
- 8 i { \nu \rho \over \alpha \beta} A_-
\no
\eea
We shall use also the following equations, derived from (\ref{ghzee}) and (\ref{fzee}), 
\bea
- \half \rho ^2 \left ( g_z ^2 + h_z^2 \right ) 
& = & 8 (\p_w \phi)^2
\no \\
- 8 \rho ^2 f_z^2 
& = & - 8 A_-^2 + 8 {i \nu \rho \over \alpha \beta} A_- + 2 {\rho^2 \over \alpha ^2 \beta ^2}   
\eea
Putting all together, and using $8A_+^2 - 8A_-^2 = 2 (\p_w\phi)^2$, the constraint 
(\ref{spin2}) is found to be satisfied, and thus (\ref{spin2})  follows from the BPS equations.

\newpage

\section{Complete Analytic Solution}
\setcounter{equation}{0}

The system of first order equations of (\ref{inteq1}) and (\ref{inteq2})
for the unknown real scalar functions $\phi$ and $\rho$ appears  
formidable. Nonetheless, we shall succeed in constructing a sequence of 
local changes of variables by which this system is exactly mapped into a 
system of linear equations. These linear equations will be 
solved exactly in terms of the 2 holomorphic functions $\kappa$
and $\lambda$, appearing already in (\ref{inteq1}) and (\ref{inteq2}).

\smallskip

The task of finding simultaneous changes of variables for $\phi$ and $\rho$
which simplify the first order equations is made easier by first searching for
a helpful change of variables for $\phi$ only. This is possible, because 
we have already shown that the first order system (\ref{inteq1}) and (\ref{inteq2})
implies a single second order partial differential equation, (\ref{univ1}), which involves 
only $\phi$, but not $\rho$. It is in this equation that we shall identify a first
change of variables, just for $\phi$.

\subsection{A new field for the dilaton $\phi$}

The key complication in (\ref{univ1}) is the appearance of a square root
of a ratio of hyperbolic functions on the right hand side of (\ref{univ1}). 
To uniformize this square root, we define the new real field $\tet$ by
\bea
\label{theta}
e^{2 i \tet } \equiv {\sh (2 \phi + 2 \lambda) \over \sh (2 \phi + 2 \bar \lambda)}
\eea
In terms of $\tet$, equation (\ref{univ1}) simplifies considerably, and becomes,
\bea
\label{univ2}
\p_{\bar w} \p_w \tet + {1 \over \cos \mu} 
\left ( e^{-i \tet} \, \p_{\bar w}  \tet \, \p_w \lambda 
    + e^{i \tet } \, \p_w \tet \, \p_{\bar w} \bar \lambda \right )
    - 2 {\sin \mu \over \cos ^2 \mu}  \p_w \lambda \, \p_{\bar w} \bar \lambda \, \cos \tet
    =0
\eea
This equation is of the Liouville or sine-Gordon type \cite{Liouville1,Liouville2}. 
Alternatively, it may be 
recast in the form of a current conservation equation, 
\bea
\p_{\bar w} \left ( \p_w \tet - 2  {\p_w \mu \over \cos \mu} e^{- i \tet} \right )
+
\p_w \left ( \p_{\bar w} \tet - 2  {\p_{\bar w} \mu \over \cos \mu} e^{+ i \tet} \right )=0
\eea
where we use the notation $\lambda - \bar \lambda = i \mu$, with $\mu$ real and harmonic.
Intermediate steps in this calculation are considerably simplified with the help 
of the following equation,
\bea
\label{inter3}
|\sh (2 \phi + 2 \lambda)|^2 
    & = & { (\sin 2 \mu )^2 \over 4 \sin (\tet + \mu ) \sin (\tet - \mu)}
\eea
and the derivatives
\bea
\label{inter4}
\p_w \phi = - {\sin 2 \mu \, \p_w \tet \over 4 \sin (\tet +\mu) \sin (\tet - \mu)} 
-{i \over 2} \p_w \mu  + {\sin 2 \tet \, \p_w \mu \over 4 \sin (\tet +\mu) \sin (\tet - \mu)}
\eea

\subsection{A new field for the metric $\rho$}

Having identified a change of variables for the dilaton $\phi$ that significantly
simplifies (\ref{univ1}), we shall carry out the same change
of variables for the dilaton in the first order system (\ref{inteq1}), (\ref{inteq2}) 
as well, leaving the metric function $\rho$ unchanged for the time being.
Equation (\ref{inteq1}) becomes, 
\bea
\label{fo1}
\p_w \tet -  \left ( e^{-i \tet} + i \sin \mu \right ) {\p_w \mu \over \cos \mu}
= {i \nu \sqrt{2} \rho ^2  \over  \bar \kappa} 
\left ( { \sin \mu - \sin \tet \over \cos \mu \sin \mu} \right )
\sh (2\phi + 2 \lambda)^\half 
\eea
where we have left the factor $\sh(2 \phi + 2 \lambda)^\half $ on the right hand side 
unconverted; it will combine with other factors later.
Equation (\ref{inteq2}) becomes,
\bea
\label{fo2}
\p_w \ln  \rho ^2 & = &
\left ( {i \over 2} - { \sin 2 \tet + 4 \sin \mu \, \cos \tet \over 4 \sin (\tet + \mu)
\sin (\tet - \mu) } \right ) \p_w \tet
\no \\ &&
+ \left ( 
 {\cos 2 \tet - \cos 2 \mu \over 2 \sin 2 \mu}  - i { \sin 2 \tet \over  \sin 2 \mu}
-i {\cos \tet \over \cos \mu} \right ) \p_w \mu
\no \\ &&  
+ {\sin 2 \tet \over \sin (\tet + \mu) \sin (\tet - \mu) } 
\left \{  {\cos \tet \over 2 \cos \mu} +  {\sin 2 \tet \over 4 \sin 2 \mu} \right \}\p_w \mu 
\eea
Equation (\ref{fo2}), though apparently complicated, provides a clue as to 
how $\rho^2$ should be redefined. The strategy will be to multiply 
$\rho^2$ by a factor which absorbs {\sl all} the terms proportional to $\p_w \tet$ 
on the right hand side of (\ref{fo2}), except for the term ${i \over 2} \p_w \tet$.
The change of variables that effects this is given by
\bea
\label{changerho}
\rho^8 &=& {\hat \rho^8 \over 16} \, \kappa ^4 \bar \kappa^4 \, (\sin 2 \mu )^2\, 
{\sin \tet + \sin \mu \over (\sin \tet - \sin \mu )^3} 
\eea
The factor $(\sin 2 \mu)^{- \half}$ has been included for later convenience.

\subsection{The first order system in terms of the new fields}

In terms of $\tet $ and $\hat \rho$, the system of first order equations 
(\ref{inteq1}) and (\ref{inteq2}) or equivalently, equations (\ref{fo1}) and (\ref{fo2}) 
now simplifies considerably and becomes,
\bea
\p_w \tet - \left ( e^{-i \tet} + i \sin \mu \right ) {\p_w \mu \over \cos \mu}
& = &
- i \nu \hat \rho ^2 \, \kappa \, e^{ + i  \tet /2}
\no \\
\p_w \tet - 2 e^{-i \tet }  {\p_w \mu \over \cos \mu} 
& = & 
- 2 i \, \p_w \ln \hat \rho ^2
\eea
An alternative way of writing the second equation is obtained by eliminating 
$\p_w \tet$ between both equations,
\bea
\label{fob}
\p_w \tet & = & 
    \left ( e^{-i \tet} + i \sin \mu \right ) {\p_w \mu \over \cos \mu}
    - i \nu \hat \rho ^2 \, \kappa \, e^{ + i  \tet /2}
\no \\
\p_w \ln \hat \rho ^2 & = &
- { i \over 2} {e^{-i \tet} \over \cos \mu} \p_w \mu 
+ \half \p_w \ln \cos \mu + {\nu \over 2} \hat \rho ^2 \, \kappa \, e^{ + i \tet /2}
\eea
It is readily checked that these two equations, together with their complex 
conjugate equations, forms a system of equations that is integrable in both
the real functions $\tet$ and $\ln \hat \rho^2$.

\subsection{The first order system in terms of  a single complex field}

By taking the sum of the second equation in (\ref{fob}) with $i/2$ times the first
equation, we eliminate the $e^{-i \tet} \p_w \mu /\cos \mu$ term from both equations
and we are left with
\bea
\label{foc}
\p_w \ln \left ( \hat \rho ^2 e^{i\tet /2} \right )
= \p_w \ln \cos \mu + \nu \hat \rho ^2 \, \kappa \, e^{ + i \tet /2}
\eea
Thus, the natural variable is the complex combination $\hat \rho ^2 e^{i\tet /2}$,
for which (\ref{foc}) gives a first integral. Actually, a slightly 
more convenient combination is the following,
\bea
\label{psidef}
\psi \equiv {\cos \mu \over \hat \rho ^2} e^{-i \tet /2} 
\eea
In terms of this new variable, the system (\ref{fob}) becomes,
\bea
\label{inteq8}
\p_w \psi & = & 
    - \nu \, \kappa \, \cos \mu
\no \\
\p_w \bar \psi & = & 
    i \psi \, {\p_w \mu \over \cos \mu} - \bar \psi \, {\sin \mu \over \cos \mu} \p_w \mu 
\eea
Thus, the change of variables from $(\phi, \tilde \rho^2)$ to $(\psi, \bar \psi)$
maps the original first order system into a system of linear equations.

\subsection{Integration of the first order system}

The system is actually even better than linear, since its first equation in (\ref{inteq8}) 
may be integrated by quadrature alone. To see this, write all components in terms
of the (locally) holomorphic functions $ \kappa, \lambda$,
\bea
\label{psi}
\p_w \psi = - {\nu  \over 2} \, \kappa \, \left ( e^{ + \lambda - \bar \lambda}
+ e^{ - \lambda + \bar \lambda} \right )
\eea 
We introduce the following (locally) holomorphic scalar functions $\cA (w)$ and $\cB (w)$,
as primitives of the (locally) holomorphic exponentials,\footnote{Notice that the sign
$\nu$ merely changes the sign of both $\cA$ and $\cB$.}
\bea
\label{cAcB}
\p_w \cA (w) & = & - {\nu \over 2} \, \kappa (w) \, e^{  + \lambda (w)} 
\no \\
\p_w \cB (w) & = &  - {\nu \over 2} \, \kappa (w) \, e^{  - \lambda (w)} 
\eea
In terms of these functions, the general solution of (\ref{psi}) is readily 
written down explicitly,
\bea
\label{psipart1}
\psi (w, \bar w) & = & 
    e^{- \bar \lambda} (\bar w) \, \cA (w) + e^{+\bar \lambda} (\bar w) \, \cB (w)
    + \overline{ \varphi (w) }
\no \\
\bar \psi (w, \bar w) & = & 
    e^{- \lambda} (w) \, \overline{\cA (w)} + e^{+ \lambda} ( w) \, \overline{\cB (w)}
    + \varphi (w)
\eea
where $\varphi (w)$ is a holomorphic function which remains yet to be determined
by the second equation in (\ref{inteq8}). Substituting (\ref{psipart1}) into the second 
equation in (\ref{inteq8}), we obtain the following equation for $\varphi$,
\bea
&&
\left ( e^{\lambda - \bar \lambda} + e^{- \lambda + \bar \lambda} \right ) \p_w \varphi
-\varphi   \left ( e^{\lambda - \bar \lambda} - e^{- \lambda + \bar \lambda} \right ) \p_w \lambda 
- 2 \bar \varphi \, \p_w \lambda 
\no \\ && \hskip 1in =
2 e^{-\bar \lambda} (\cA + \bar \cA) \p_w \lambda 
+ 2 e^{+\bar \lambda} (\cB - \bar \cB) \p_w \lambda
\eea
The inhomogeneous solution is readily identified as $- e^{-\lambda } \cA + e^\lambda \cB$, 
so we redefine $\varphi$ 
in terms of a new holomorphic function $\varphi _0$, by
\bea
\varphi & \equiv & \varphi _0 - e^{-\lambda } \cA + e^\lambda \cB
\no \\
\bar \varphi & \equiv & \bar \varphi _0 - e^{- \bar \lambda } \bar \cA + e^{\bar \lambda} \bar \cB
\eea
where $\varphi_0$ now satisfies the homogeneous equation, 
\bea
\left ( e^{\lambda - \bar \lambda} + e^{- \lambda + \bar \lambda} \right ) \p_w \varphi_0
-\varphi _0  \left ( e^{\lambda - \bar \lambda} - e^{- \lambda + \bar \lambda} \right ) \p_w \lambda 
- 2 \bar \varphi _0 \, \p_w \lambda =0
\eea
To solve this equation, we set
$\varphi _0 = e^\lambda \varphi _1$,
where $\varphi_1$ is again  holomorphic, and satisfies the equation,
\bea
\label{varphi1}
2 ( \varphi _1 - \bar \varphi _1) \p_w \lambda 
+ \left ( 1 + e^{2 \lambda - 2 \bar \lambda} \right ) \p_w \varphi _1 =0
\eea
Taking the $\p_{\bar w}$ derivative of the entire equation, and rearranging factors, we get
\bea
{e^{2 \lambda } \p_w \varphi _1 \over \p_w \lambda} = - 
{e^{2 \bar \lambda } \p_{\bar w} \bar \varphi _1 \over \p_{\bar w} \bar \lambda}
\eea
The left hand side is holomorphic, while the right hand side is anti-holomorphic. 
The above equality then requires that
both ratios be constant and purely imaginary, a number we shall denote by $2i r_1$
with $r_1$ real. The remaining equation is then
\bea
\p_w \varphi _1 = 2 i r_1 e^{-2 \lambda } \p_w \lambda
\eea
whose general solution is given by $\varphi _1 = -i r_1 e^{-2 \lambda} + r_2$, with 
$\p_w r_2=0$. Since $\varphi_1$ is holomorphic, $r_2$ must be a complex 
constant. Substituting this result into the full equation (\ref{varphi1})
for $\varphi_1$, we find the additional requirement that $r_2$ must be real.
The most general solution for $\varphi_0$ is thus,
$\varphi _0 = -i r_1 e^{- \lambda} + r_2 e^\lambda$ with $r_1, r_2 $ real.
The constants $r_1$ and $r_2$ can be absorbed into the functions
$\cA$ and $\cB$, so that the most general solution for $\psi$ is given by
\bea
\psi = e^{-\bar \lambda} (\cA - \bar \cA) + e^{+\bar \lambda} (\cB + \bar \cB)
\eea
It will often be convenient to express the results directly in terms of 
two real harmonic functions $h_1$ and $h_2$, instead of the holomorphic 
functions $\cA$ and $\cB$. The relation between the two sets of functions 
is as follows,
\bea
\label{ABh1h2}
\cA - \bar \cA = i h_1 & \hskip 1in & \psi = + i h_1 e^{-\bar \lambda} + h_2 e^{\bar \lambda}
\no \\
\cB + \bar \cB = h_2 & \hskip 1in & \bar \psi = - i h_1 e^{- \lambda} + h_2 e^{ \lambda}
\eea
In general, these harmonic functions are independent of one another just as
 the holomorphic functions $\kappa$ and $\lambda$ were independent.

\subsection{Explicit solution for the dilaton}

The dilaton is given in terms of the variable $\tet$, 
which in turn is given by
\bea
e^{i \tet} = {\bar \psi \over \psi} =
{- i e^{-\lambda} h_1 + e^\lambda h_2
\over 
+i e^{- \bar \lambda} h_1 + e^{\bar \lambda} h_2}
\eea
The dilaton field $\phi$ is related to $\tet$  via (\ref{theta}), or 
equivalently, via
\bea
e^{4 \phi + 2 \lambda + 2 \bar \lambda} 
= {\sh (i \tet + \lambda - \bar \lambda) \over \sh (i \tet - \lambda + \bar \lambda)}
\eea
From  $\cA$ and $\cB$ in (\ref{cAcB}) , we derive  formulas
for  $\kappa$ and $e^{2 \lambda}$ in terms of $h_1$ and $h_2$,
\bea
\label{h1h2lambda}
\kappa ^2 = 4 i \p_w h_1 \p_w h_2 \hskip 1in 
e^{2 \lambda} = i \, {\p_w h_1 \over \p_w h_2}
\eea
The dilaton solution may be expressed solely in terms of the 
harmonic functions $h_1$ and $h_2$,
\bea
\label{dilsol}
e^{4 \phi} =
{ 2 h_1 h_2 |\p_w h_2 |^2 - h_2^2 W \over  2 h_1 h_2 |\p_w h_1 |^2 - h_1^2 W}
\eea
where the following combination $W$ will occur ubiquitously,
\bea
W  \equiv  \p_w h_1 \p_{\bar w} h_2 + \p_w h_2 \p_{\bar w} h_1 
\eea
This then gives the result for the dilaton, announced in (\ref{dilsol1}) of the 
Introduction.  Note that it is possible for the right hand side of (\ref{dilsol})
to be negative.  Therefore, in order for it to be a valid solution of the 
BPS equations (\ref{inteq1}), (\ref{inteq2}),
as well as Type IIB supergravity, there is an additional restriction on the harmonic functions
$h_1$ and $h_2$.  They must be chosen so that (\ref{dilsol}) is positive on the right hand side.
Construction of such harmonic functions is non-trivial.  The Janus solution, will be given
in section 10, while an infinite class of such harmonic functions is constructed in a companion paper 
\cite{EDJEMG2}.

\subsection{Explicit solution for the metric factors}

The metric factor $\rho^2$ is readily calculated by taking the norm of $\psi$ in (\ref{psidef}), 
and using the conversion formula (\ref{changerho}), and is found to be,
\bea
\label{rhosol1}
\rho^8 &=& { W^2
\over h_1^3 h_2^3 }
\bigg( 2 h_1  |\p_w h_2|^2 - h_2 W \bigg) \bigg( 2  h_2 |\p_w h_1|^2 - h_1 W \bigg)
\eea

The metric factors $f_1$, $f_2$, and $f_4$ are given in terms of the 
spinor variables $\alpha$ and $\beta$ by (\ref{metricbilin2}),
which in turn are given via (\ref{albetsol1}) in terms of $\phi$ and 
$\kappa$ and $\lambda$.
The latter are obtained in terms of the harmonic functions $h_1$ and $h_2$ 
using (\ref{cAcB}). It will sometimes be useful to keep the dilaton $\phi$ 
and $\Sigma$-metric $\rho^2$, 
as their presence will often allow for simplification in the metric factors.
As a result, we have the following simple combinations for $\alpha$ and $\beta$,
\bea
\nu  \rho \, \bar \alpha ^2 & = & -i \, e^\phi \p_w h_1  -  e^{-\phi} \p_w h_2 
\no \\
\nu  \rho \, \bar \beta ^2 & = & +i \, e^\phi \p_w h_1  -  e^{-\phi} \p_w h_2 
\eea
and the metric factors are given by,
\bea
\rho \, f_1 &=& 
    - 2 \nu\,  \Re \left ( e^{-2 \phi} |\p_w h_2|^2 - e^{2 \phi} |\p_w h_1|^2 -  i W \right )^{\half}
\no\\
\rho \, f_2 &=& 
- 2 \, \Im \, \left ( e^{-2 \phi} |\p_w h_2|^2 - e^{2 \phi} |\p_w h_1|^2 -  i W \right )^{\half}
\no \\
\rho \, f_4 &=& 
\left  | e^{-\phi} \p_w h_2  - i \, e^\phi  \p_w h_1 \right |
+ \left  | e^{-\phi} \p_w h_2  + i \, e^\phi  \p_w h_1  \right | 
\eea
Note that the following bilinears in $f_1$ and $f_2$ are especially simple,
\bea
\rho ^2 f_1 f_2 & = & - 2 \nu W
\no \\
\rho ^2 (f_1 ^2 - f_2 ^2) & = & 
4 e^{-2 \phi} |\p_w h_2|^2 - 4 e^{2 \phi} |\p_w h_1|^2
\eea
For completeness, formulas for the metric factors $f_1$, $f_2$, and $f_4$ 
expressed directly in terms of the harmonic functions $h_1$ and $h_2$ 
are presented in Appendix E.

\subsection{The 3-form field strengths}

To compute the fluxes of the 3-form field strength $F_{(3)} = dB_{(2)}$, it will
be useful to also have an explicit expression for the 2-form potential $B_{(2)}$.
The form $F_{(3)}$ decomposes into the real  NSNS form $H_{(3)}$
and the real RR form $C_{(3)}$, 
\bea
F_{(3)} =   H_{(3)} + i C_{(3)}
\eea
Identifying with the form of the Ansatz,
in conformal coordinates $w$ on $\Sigma $, we have,
\bea 
H_{(3)} & = &  e^{+\phi} \, g_a \, e^{45a}  =
\hat e^{45} \wedge e^{+\phi} \, f_1 ^2 \, \left (  \rho  g_z  \, dw + \rho g_{\bar z} \, d\bar w \right )
\no \\
C_{(3)} & = &  e^{-\phi} \, h_a \, e^{67a} =
\hat e^{67} \wedge e^{-\phi} \, f_2 ^2 \, \left (  \rho  h_z  \, dw + \rho h_{\bar z} \, d\bar w \right )
\eea
The forms $\hat e^{45}$ and $\hat e^{67}$ are the volume forms on the 
two unit spheres, as such they are automatically closed forms. Closure of $H_{(3)}$ and 
$C_{(3)}$ thus requires that, 
\bea
\label{etader}
e^{+\phi} \, f_1 ^2 \, \left (  \rho  g_z  \, dw + \rho g_{\bar z} \, d\bar w \right ) 
& = & d b _1
\no \\
e^{-\phi} \, f_2 ^2 \, \left (  \rho  h_z  \, dw + \rho h_{\bar z} \, d\bar w \right )
& = & d b _2
\eea
for two (locally defined) real functions $b _1$ and $b _2$. In order to
evaluate line integrals of these currents and compute the associated charges, 
we calculate $b_1$ and $b_2$.

\subsubsection{Calculation of $b _{1,2}$}

The calculations of $b_1$ and $b_2$ proceed in parallel. 
The starting points are the formulas 
of (\ref{ghzee}) which express $g_z$ and $h_z$ in terms of $\a$ and $\b$.
Using the explicit expression for $f_{1,2}$ in terms of $\a$ and $\b$, we find,
\bea
\rho g_z e^{+\phi} f_1^2 & = & - 2 (\a^2 - \b ^2) \bar \a \bar \b \, e^{+\phi} \left (
{\a \bar \b \over \bar \a \b} + {\bar \a \b \over \a \bar \b} + 2 \right ) \p _w \phi
\no \\
\rho h_z e^{-\phi} f_2^2 & = & - 2i (\a^2 + \b ^2) \bar \a \bar \b \, e^{-\phi} \left (
-{\a \bar \b \over \bar \a \b} - {\bar \a \b \over \a \bar \b} + 2 \right ) \p _w \phi
\eea
Using the second equation in (\ref{derplus}) to eliminate the ratios in the 
parentheses, and using (\ref{albetsol0}) to eliminate the factors $\a^2 \pm \b^2$,
and recognizing that all terms in the parenthesis arise from a derivation, we have
\bea
\rho g_z e^{+\phi} f_1^2 & = &  \p_w \left ( 
- 2 \rho ^{-1}  \bar \a \bar \b \,  \bar \kappa \, e^{+2 \phi  + \bar \lambda} \right )
\no \\
\rho h_z e^{-\phi} f_2^2 & = & \p_w \left (
2 i  \rho^{-1}   \bar \a \bar \b \, \bar \kappa \, e^{-2 \phi  - \bar \lambda}  \right ) 
\eea
This result is not yet of the form (\ref{etader}) since the arguments of the 
derivatives are complex. For (\ref{etader}) to work, it must be that the imaginary
parts of the arguments are actually harmonic. We separate the real and imaginary
parts as follows,
\bea
- {2 \over \rho}   \bar \a \bar \b \,  \bar \kappa \, e^{+2 \phi  + \bar \lambda} 
& = & b _{1+} + i b _{1-}
\no \\
i {2  \over \rho}   \bar \a \bar \b \,  \bar \kappa \, e^{-2 \phi  - \bar \lambda}
& = & b _{2+} + i b _{2-}
\eea
where $b _{1\pm}$ and $b _{2\pm}$ are real. To calculate $b_1$ and $b_2$, 
we make use of formulas (\ref{albetsol1}) to express $\a$ and $\b$ in terms of $\phi$
and $\lambda$, and formula (\ref{theta}) to express the phase in terms of the function $\tet$.
Using formula (\ref{changerho}) to further express $ \rho^2$ in terms of $\hat \rho ^2$, 
we find the following expressions, from which $b_1$ and $b_2$ can be readily evaluated,
\bea
{ 1 \over \rho } \, \bar \a \bar \b \, \bar \kappa \, e^{+2 \phi + \bar \lambda} 
& = &
- { i \over \hat \rho ^2} { \cos \half (\tet + \mu) \over \cos \half (\tet - \mu)}
\, e^{- \lambda + i \tet/2}
\no \\
{ 1 \over \rho } \, \bar \a \bar \b \, \bar \kappa \, e^{-2 \phi - \bar \lambda} 
& = &
- { i \over \hat \rho ^2} { \sin \half (\tet - \mu) \over \sin \half (\tet + \mu)}
\, e^{+ \lambda + i \tet/2}
\eea
Putting all together,  and using formula (\ref{psidef}) to express $\hat \rho$ and the 
phase $\tet$ in terms of the single complex variable $\psi$, we get
\bea
b _{1-} & = &  {1 \over \cos \mu} \left ( \psi e^{-\lambda} + \bar \psi e^{-\bar \lambda} \right )
\no \\
b _{2-} & = &  {i \over \cos \mu} \left ( \psi e^\lambda + \bar \psi e^{\bar \lambda} \right )
\eea
Finally, using the expression (\ref{ABh1h2}) 
for $\psi$ in terms of the harmonic functions $h_1$ and $h_2$,
we arrive at the final formulas for the imaginary parts,
\bea
b _{1-} & = & +2 h_2 = 2 (\cB + \bar \cB)
\no \\
b _{2-} & = & - 2 h_1 = 2i (\cA - \bar \cA)
\eea
Since we have 
\bea
\p_w h_1 = - i \p_w \tilde h_1 & \hskip 1in & \tilde h_1 =  \cA + \bar \cA
\no \\
\p_w h_2 = - i \p_w \tilde h_2 & \hskip 1in & \tilde h_2 =  i (\cB - \bar \cB)
\eea
the imaginary parts $i b _{1,2-}$ may be recast in terms of differentials
of real functions only, 
\bea
\label{etader1}
b _1   & = & b _{1+} + 2 \tilde h_2
\no \\
b _2  & = & b _{2+} - 2 \tilde h_1
\eea
The same steps used to compute the imaginary parts may also be used
to simplify the real parts and express them solely in terms of the harmonic
functions $h_1$ and $h_2$. We omit the details of the calculation, and 
only quote the result,
\bea
b _{1+} & = &  {2 i  h_1 h_2 (\p_w h_1 \p_{\bar w} h_2 - \p_{\bar w} h_1 \p_w h_2)
 \over 2 h_2 |\p_w h_1|^2 - h_1 W}
\no \\
b _{2+} & = &  {2 i  h_1 h_2 (\p_w h_1 \p_{\bar w} h_2 - \p_{\bar w} h_1 \p_w h_2)
 \over 2 h_1 |\p_w h_2|^2 - h_2 W}
\eea
The contributions $b_{1+}$ and $b_{2+}$ are well-defined single-valued
functions on $\Sigma$ (since by construction $h_1$, $h_2$ are single-valued, as well as their derivatives), 
but the harmonic duals $\tilde h_1$ and $\tilde h_2$ are not, generally,
single-valued. As a result, the calculation of a flux through a 3-cycle 
can be greatly simplified. For example, consider a 3-cycle $M_3$ with the 
following  homological decomposition, 
\bea
M_3 = C_1 \times (S_1^2)^{n_1} \cup C_2 \times (S_2^2)^{n_2}
\eea
where $C_1$ and $C_2$ are closed curves in $\Sigma$ and $n_1$
and $n_2$ are integers representing the number of times $M_3$ contains 
the spheres $S_1^2$ and $S_2^2$.  The flux through this cycle is given by
\bea
\oint _{M_3} F_{(3)} = 8 \pi n_1 \oint _{C_1} d\tilde h_2 - 8 \pi n_2 i \, \oint _{C_2} d \tilde h_1
\eea
The contour integrals over $C_1$ and $C_2$ will reduce to residue calculations.

\subsection{The 5-form field strength}

The conserved flux of the 5-form field strength is given by the cohomology of $dC_{(4)}$.
Therefore, we shall compute $C_{(4)}$ for our Ansatz and solutions. The starting 
point is its expression in terms of $F_{(5)}$ and the 3-form fields,
\bea
dC_{(4)} = F_{(5)} - { i \over 16} \left ( B_{(2)} \wedge \bar F_{(3)} -
\bar B_{(2)} \wedge F_{(3)} \right )
\eea
In view of the results obtained in the preceding section, we have 
\bea
B_{(2)} & = & b_1 \, \hat e^{45} + i b_2 \, \hat e^{67}
\no \\
F_{(3)} & = & e^\phi g_a e^a f_1 ^2 \hat e^{45} + i e^{-\phi} h_a e^a f_2^2 \hat e^{67}
\no \\
F_{(5)} & = & - f_a e^a f_4^4 \hat e^{0123} + f_a \ep ^{ab} e^b f_1 ^2 f_2^2 \hat e^{4567}
\eea
From the structure of this Ansatz, it is immediate that $C_{(4)}$ takes on 
the following form,
\bea
C_{(4)} & = & - j_1 \hat e^{0123} + j_2 \hat e^{4567}
\eea
where the real functions $j_1$ and $j_2$ are determined by the differential equations,
\bea
d j_1 & = &  f_a e^a f_4^4
\no \\
dj_2 & = & f_a \ep ^{ab} e^b f_1 ^2 f_2 ^2 
+ {1 \over 8} e^{+ \phi } g _a e^a f_1^2 b_2
- {1 \over 8} e^{- \phi} h_a e^a f_2^2 b_1
\eea
Closure of these 1-forms on the right hand side of the equations is precisely the 
contents of the two Bianchi identities for the 5-form, and was established using
Mathematica. The form of these functions will not be needed in the sequel,
so we shall limit the calculation to $j_1$.

\subsubsection{Calculation of $j_1$}

The starting point is the following expression for $\rho f_z$, 
\bea
2 \rho f_z =  
\left ( {\a \bar \b \over \bar \a \b } -  {\bar \a \b \over  \a \bar \b }\right ) \p _w \phi
+ \p _w \ln {\bar \beta \over \bar \alpha}
\eea
It is obtained starting from the two $(a+)$ equations in (\ref{fullBPS}), 
in which $g_z \pm i h_z$ are eliminated using (\ref{ghzee}) and $p_z$
is eliminated from the result using (\ref{pphi}).
The strategy is to first convert this quantity to $\phi, \rho, \lambda$, then to $\theta , \mu$
and finally to $\psi, \bar \psi$ and $h_1, h_2$. First, we use (\ref{albetsol1})
to express the right hand side in terms of $\phi$ and $\lambda$,
\bea
2 \rho f_z =  {\p_w \phi + \p_w \lambda \over \sh (2 \phi + 2 \lambda)}
 - {2 \sh (\lambda - \bar \lambda ) \p_w \phi \over  | \sh (2 \phi + 2 \lambda ) |} 
\eea
On the other hand, squaring $f_4$ and expressing the result 
in terms of $\phi, \lambda, \rho$, we have 
\bea
f_4 ^2  = \left ( |\a|^2 + |\b |^2 \right )^2 & = &
{ e^{\kappa + \bar \kappa} \over \rho ^2} \left ( \ch (2 \phi + \lambda + \bar \lambda)
+ | \sh (2 \phi + 2 \lambda ) | \right )
\eea
Next, we covert these formulas to $\tet, \mu$, which were defined by (\ref{theta}) and
$\lambda - \bar \lambda = i \mu$, using (\ref{inter3}).
Converting all but the derivatives of $\phi$ and $\lambda$, we find, 
\bea
\label{f4s}
2 \rho f_z f_4^4 = { 2 \sin (\tet - \mu ) \sin (\theta + \mu)
\over \hat \rho ^4 (\sin \mu)^2 }
\left [ e^{-i \tet} (\p_w \phi + \p _w \lambda) - 2 i (\sin \mu) \p_w \phi \right ]
\eea
Converting now also the derivatives, using (\ref{inter4}), we get
\bea
2 \rho f_z f_4^4 & = & { 1 \over 2 \hat \rho ^4 (\sin \mu)^2 }\bigg [
e^{-i \tet} \left ( - (\sin 2 \mu) \p_w \tet - i e^{2 i \tet} \p_w \mu + i (\cos 2 \mu) \p_w \mu 
\right )
\no \\ && \qquad 
- 2i (\sin \mu) \left ( - (\sin 2 \mu) \p_w \tet + i e^{-2i \tet} \p_w \mu - i (\cos 2 \mu) \p_w \mu
\right ) \bigg ]
\eea
Converting this expression to $\psi, \bar \psi$, using (\ref{psidef}), we find,
\bea
2 \rho f_z f_4^4 & = & { 2 \over  (\sin 2 \mu)^2 }\bigg [
-i (\sin 2 \mu) (\psi \p_w \psi - \psi ^2 \p_w \bar \psi /\bar \psi)
- i \bar \psi ^2 \p_w \mu 
\no \\  && \hskip 0.7in  
+ i \psi ^2 (\cos 2 \mu) \p_w \mu
- 2 (\sin \mu) (\sin 2 \mu) (\bar \psi \p_w \psi - \psi \p_w \bar \psi)
\no \\ && \hskip 0.7in
+ 2 (\sin \mu) \psi ^3 \p_w \mu /\bar \psi 
- 2 (\sin \mu) (\cos 2 \mu) \psi \bar \psi \p_w \mu \bigg ]
\eea
Next, we use the field equation for $\p_w \bar \psi$ of (\ref{inteq8}), to eliminate 
all terms which are not of second order in $ \psi$ and/or $\bar \psi$. 
All terms but one may be expressed as total derivatives of a real function,
\bea
2 \rho f_z f_4^4 = \p_w \left ( { i \bar \psi ^2 - i \psi ^2 \over \sin 2 \mu} 
- 2 {\psi \bar \psi \over \cos \mu} \right )
+ 3i {\psi ^2 \p_w \mu \over (\cos \mu)^2}
\eea
The last term may be recast as follows,
\bea
{\psi ^2 \p_w \mu \over (\cos \mu)^2} 
= \p_w \left ( \psi ^2 \tg \mu  -i h_1^2 e^{-2 \bar \lambda} -i  h_2^2 e^{2 \bar \lambda}\right ) 
+ 2 h_1 \p_w h_2 - 2 h_2 \p_w h_1 
\eea
where we have used the field equation for $\p_w \psi$ of (\ref{inteq8}), as well
as (\ref{cAcB}) and (\ref{ABh1h2}) to express $\kappa$ in terms of the harmonic 
functions $h_1$ and $h_2$.
Defining a new holomorphic function, $\cC$ by
\bea
\label{cC}
\p_w \cC \equiv  \cA \p_w \cB - \cB \p_w \cA
\eea
allows us to recast the expression in its final  form,
\bea
2 \rho f_z f_4^4 = \p_w \left ( 6 \cC + 2 h_1 h_2 \cotg \mu \right )
\eea
Here, we have simplified the argument by using the following identity, 
\bea
 { i \bar \psi ^2 - i \psi ^2 \over \sin 2 \mu} 
- 2 {\psi \bar \psi \over \cos \mu} + 3 i \psi ^2 \tg \mu 
+ 3 h_1^2 e^{- 2 \bar \lambda} + 3 h_2^2 e^{2 \bar \lambda}
= 2 h_1 h_2 \cotg \mu
\eea
We thus obtain an explicit formula for $j_1$,   
\bea
j_1 = 3\cC +3 \bar \cC + h_1 h_2 \cotg \mu 
= 3\cC + 3\bar \cC + i h_1 h_2
 {\p_w h_1 \p _{\bar w} h_2 - \p_{\bar w} h_1 \p_w h_2
\over 
\p_w h_1 \p _{\bar w} h_2 + \p_{\bar w} h_1 \p_w h_2}
\eea
In evaluating closed contour integrals of $d j_1$, only the $\cC$ 
terms contribute as the other term is single-valued,
\bea
\oint _C d j_1 = 3 \oint _C d(\cC + \bar \cC)
\eea

\subsection{Transformation rules}

It will be helpful to derive the effect of  simple operations on $h_1$ and $h_2$. 
The first transformation is a constant shift in the dilaton,
leaving all other supergravity fields unchanged, 
\bea
\label{dilshift}
h_1 \rightarrow e^{- \phi_0}  h_1
\qquad
h_2 \rightarrow e^{\phi_0}  h_2 \hskip 1in \phi \to \phi + \phi _0
\eea
The second transformation is a common scaling by a real positive constant $\Lambda^2$,
\bea
\label{radiusshift}
h_1 \rightarrow \Lambda^2 h_1
\qquad
h_2 \rightarrow \Lambda^2 h_2
\eea
which transforms the fields as
\bea
\rho \rightarrow \Lambda \rho
\qquad
f_4 \rightarrow \Lambda f_4
\qquad
f_{1,2} \rightarrow \Lambda f_{1,2}
\no\\
g_z \rightarrow  g_z/\Lambda
\qquad
h_z \rightarrow  h_z/\Lambda
\qquad
f_z \rightarrow  f_z/\Lambda
\eea
with all other fields, including the dilaton, left invariant.  
The third transformation  is a strong-weak duality,
\bea
h_1 \leftrightarrow h_2 \hskip 1in 
\phi  \rightarrow - \phi
\hskip 1in
g_z \leftrightarrow h_z
\eea
Finally, the effect of sign reversal of $\nu$ is given by
\bea
h_{1 (2)} \rightarrow - h_{1 (2)}
\qquad
h_{2 (1)} \rightarrow h_{2 (1)}
\qquad
\nu \rightarrow - \nu
\eea
under which the fields transform as
\bea
f_z \rightarrow - f_z
\qquad
g (h)_z \rightarrow - g (h)_z
\eea

\newpage

\section{The Half-BPS Janus solution}
\setcounter{equation}{0}

In this section, we shall first recover the $AdS_5 \times S^5$ solution (with constant 
dilaton and vanishing $G$-field) from the general $AdS_4 \times S^2 \times S^2 \times \Sigma$
solution derived in the preceding sections. A simple deformation of the harmonic functions $h_1$ and $h_2$
of the $AdS_5 \times S^5$ solution will produce a family of regular solutions with varying 
dilaton and non-zero $G$-field. This solution is naturally identified with the 
generalization of the Janus solution that possesses 16 supersymmetries, 
predicted to exist in \cite{degsusy} on the basis of its dual interface CFT.

\subsection{The $AdS_5 \times S^5$ solution}

The $AdS_5 \times S^5$ solution has constant dilaton $\phi_0$, 
and is obtained via a linear combination of exponentials with opposite 
arguments,
\bea
\label{AdSsol}
\cA =  e^w - e^{-w}   & \hskip .8in & 
    h_1 =  -i e^w + i e^{-w} + i e^{\bar w} -i e^{-\bar w} 
\no \\
\cB =  e^w + e^{-w}   &  &
    h_2 =  e^w + e^{-w} + e^{\bar w} + e^{-\bar w} 
\eea
Here, we have used the transformation properties of (\ref{dilshift}) to shift 
the dilaton to 0 value, and the dilaton equation (\ref{dilsol}) is indeed satisfied with $\phi = 0$;
the $\Sigma$-metric $\rho$ is constant in these coordinates; and the metric functions are,
\bea
\label{AdSsol1}
\rho ^2 f_1 f_2 & = & -8\,  i \nu   \, \sh (w - \bar w)
\no \\
\rho ^2 (f_1 ^2 - f_2^2) & = &
- 4   \, \ch (w - \bar w) 
\no \\
\rho f_4 & = & 4\,   \ch (w + \bar w)
\eea
The domain of variation of $w$ may be figured out from the fact that the 
sphere $S^5$ arises from $S^2 \times S^2$ varying on an interval
with one $S^2$ vanishing at one end, and the other $S^2$ vanishing
at the other end of the interval. Therefore, the domain must be 
\bea
\label{domain}
\Sigma = \left \{ w \in {\bf C} ; ~ 0 \leq \Im (w) \leq {\pi \over 2} \right \}
\eea
with $\Re(w)$ running over the entire ${\bf R}$.

\subsection{More general solutions with exponentials}

Next, we shall seek regular solutions of the Janus type. This means that 
the solution will have two asymptotic regions, with the dilaton tending towards 
distinct constant values $\phi _\pm$ in each asymptotic region.  More properly, 
Janus consists of a family of solutions parametrized by the difference 
$\phi _+ - \phi _-$  between the dilaton values in these two different regions. 
This family contains $AdS_5 \times S^5$ for the special value $\phi _+ - \phi _- =0$.

\smallskip

The behavior of Janus, described above,  suggests that Janus should 
correspond to harmonic functions $h_1$ and $h_2$ given by a family of 
deformations of the harmonic functions of (\ref{AdSsol}) of 
the $AdS_5 \times S^5$ solution. 
Thus, we are led to  seek solutions of the following type,
\bea
\cA & = & a_+ e^w + a_- e^{-w}
\no \\
\cB & = & b_+ e^w + b_- e^{-w}
\eea
The corresponding harmonic functions are given by
\bea
h_1 & = & -i a_+ e^w -i a_- e^{-w} + i \bar a_+ e^{\bar w} + i \bar a_- e^{- \bar w}
\no \\
h_2 & = & b_+ e^w + b_- e^{-w} + \bar b_+ e^{\bar w} + \bar b_- e^{- \bar w}
\eea
These are harmonic functions on the domain $\Sigma$ without poles. 
They define a family of solutions to our equations with $AdS_5\times S^5$
as a special point in this family, corresponding to $a_\pm =\mp e^{-2 \phi_0} $ 
and $b_\pm=1$. We shall now show that the dilaton tends to 
a constant as $\Re (w) \to \pm \infty$, under certain conditions on $a_\pm$ 
and $b_\pm$.

\subsubsection{Asymptotics}

As $\Re(w) \to \pm \infty$, the leading behavior of the harmonic 
functions is given by,
\bea
h_1 & \to & -i a_\pm e^{\pm w} + i \bar a_\pm  e^{\pm \bar w}
\no \\
h_2 & \to & b_\pm e^{\pm w} + \bar b_\pm e^{\pm \bar w}
\eea
and the leading behavior of the dilaton is readily evaluated using formula (\ref{dilsol}), 
\bea
 e^{4 \phi } & \to &
  { b_\pm ^2 e^{\pm 2 (w- \bar w)} - \bar b_\pm ^2 
\over a_\pm ^2 e^{\pm 2 (w -  \bar w) } - \bar a_\pm ^2 }
\eea
For generic $a_\pm$ and $b_\pm$, this behavior involves a non-trivial 
dependence on $w-\bar w$, even in the limit as $\Re(w) \to \pm \infty$,
which is generically singular whenever $e^{w-\bar w}$ coincides with the 
phase of $a_\pm $ or $b_\pm$.
If, however, we choose $b_+^2/a_+^2$ and $b_-^2/a_-^2$ to be real constants, 
then the residual dependence cancels, and the limits are regular and constant.
Positivity of the exponential on the left hand side requires $b_+^2/a_+^2$
and $b_-^2/a_-^2$ be positive.

\subsubsection{Restricted family of solutions}

We shall solve the asymptoticity conditions, arrived at in the preceding paragraph,
as follows,
\bea
a_+ & = &  r_+ b_+ \hskip 1in r_+ ^2 = e^{- 4 \phi _+}
\no \\
a_- & = &  r_- b_-
\hskip 1in r_- ^2 = e^{- 4 \phi _-}
\eea
By shifting $w$ by the constant $\half \ln b_+/b_-$, and defining 
$b \equiv (b_+ b_-)^\half$,
we reduce the harmonic expressions to a simpler form, 
\bea
h_1 & = & -i r_+ b \,  e^w - i r_- b \, e^{-w} + i r_+ \bar b \, e^{\bar w} + i r_- \bar b \, e^{- \bar w}
\no \\
h_2 & = & b\,  e^w + b\,  e^{-w} + \bar b\,  e^{\bar w} + \bar b \,  e^{- \bar w}
\eea
We now make use of the transformations (\ref{dilshift}) and (\ref{radiusshift}) to further
reduce the harmonic functions.  Picking $\Lambda^2 = (\sqrt{r_+} |b|)^{-1}$ and
$e^{2 \phi_0} = r_+$, we are left with the maximally reduced form of the harmonic functions,
\bea
\label{h1h2genb}
h_1 & = & -i  b \,  e^w + i r b \, e^{-w} + i  \bar b \, e^{\bar w} - i r \bar b \, e^{- \bar w}
\no \\
h_2 & = & b\,  e^w + b\,  e^{-w} + \bar b\,  e^{\bar w} + \bar b \,  e^{- \bar w}
\eea
where we have 
\bea
r = - r_- / r_+ = \pm e^{2 \phi _+ - 2 \phi _-} 
\eea
and we simply take $b$ be its phase, $b = e^{i \varphi}$ for $\varphi$ real.
Dependence on the norm of $b$ may be restored using transformation (\ref{radiusshift}).
The minus sign has been introduced so that the $AdS_5 \times S^5$ 
solution with constant dilaton throughout corresponds to $r=+1$, and $b=1$.
The dilaton is given by
\bea
\label{dilsol9}
e^{4 \phi - 4 \phi_+} =
{h_2 \over h_1} \times 
{ \p_w h_2 (h_1 \p_{\bar w} h_2 - h_2 \p_{\bar w} h_1) + c.c.
\over 
\p_w h_1 (h_2 \p_{\bar w} h_1 - h_1 \p_{\bar w} h_2) + c.c.}
\eea
where we have explicitly included the dilaton shift parameter $\phi_+$.

\subsection{The Half-BPS Janus solution}

Only for the choice $b= 1$ (or, equivalently, $b=-1$) is the above solution free of 
any singularities, and thus provides a candidate for the Janus solution with 
maximal supersymmetry. To analyze the regularity properties of the 
solution, we decompose $w$ into real coordinates, 
\bea
w= x+iy \hskip 1in x \in {\bf R}, \qquad 0 \leq y \leq {\pi \over 2} \label{stripp}
\eea
The harmonic functions  are given by
\bea
h_1 & = & 2 \sin y \left ( e^x + r e^{-x} \right ) 
\no \\
h_2 & = & 2  \cos y \left  ( e^x + e^{-x} \right )  \label{janusharm}
\eea
and the numerator and denominator of the dilaton solution formula (\ref{dilsol})
take the form,
\bea
2 h_1  |\p_w h_2 |^2 - h_2 W & = & 4 N(x,y) \sin y
\no \\
2  h_2 |\p_w h_1 |^2 - h_1 W & = & 4 D(x,y) \cos y
\eea
where 
\bea
W= - 2 (1+r) \sin 2y
\eea
and the functions $N$ and $D$ are given by,
\bea
N(x,y) & = & 
    e^{3 x} + r e^{-3x} + (1+2r) e^x + (2+r) e^{-x} - (1 -r) (e^x - e^{-x} ) \cos 2y
\quad \\
D(x,y) & = & 
    e^{3x} + r^2 e^{-3x} + r(1+2r) e^{-x} + (2+r) e^x - (1-r) (e^x - r e^{-x} )\cos 2y
\no
\eea
The factors $\cos y$ and $\sin y$ cancel between numerators and denominators in the 
dilaton formula (\ref{dilsol}), and we are left with
\bea
\label{dilatonsol}
e^{4 \phi } = e^{2 \phi _+ + 2 \phi_-}
\left ( {re^x + re^{-x} \over e^x + r e^{-x} } \right ) { N(x,y) \over D(x,y)}
\eea
Note that under $\phi _+ \leftrightarrow \phi _-$, we have $r \leftrightarrow 1/r$.
The dilaton solution (\ref{dilatonsol}) is invariant under this transformation
upon simultaneously letting $x \leftrightarrow -x$ and leaving $y$ unchanged.
Thus, we may restrict to $|r|\geq 1$ without loss of generality.

\subsubsection{Regularity of the dilaton}

Clearly, the first factor on the right hand side of (\ref{dilatonsol})
will be singularity free for all $x$ if and only if $r > 0$, a relation we
shall henceforth  assume. Combining this with $|r|\geq 1$ from the 
preceding subsection, regularity thus restricts us to $r \geq 1$.
Next, the numerator  $N(x,y)$ and denominator $D(x,y)$ will be free of zeros, 
for any real value  of $x,y$,  provided
\bea
\label{ineq1}
(e^x + e^{-x} )^3 + (r-1) \left ( e^{-3x} + 2e^x + e^{-x} \right ) 
& \geq & (r-1) \left  |e^x - e^{-x} \right | 
\no \\
(e^x + e^{-x})^3 + (r-1) \left ( (r+1) e^{-3x} + (2r+3) e^{-x} + e^x \right ) 
& \geq & (r-1) \left  |e^x - r e^{-x} \right | \qquad
\eea
Since $e^x + e^{-x} \geq 2$ for all $x$, and $r \geq 1$, it is sufficient to require that 
\bea
8 + (r-1) (e^x+ e^{-x}) & \geq &  (r-1) \times \left |e^x - e^{-x} \right |
\no \\
8 + (r-1) (e^x + r e^{-x} ) & \geq & (r-1) \left  |e^x - r e^{-x} \right |
\eea
These inequalities hold for all $x$ and $r \geq 1$.
Thus, we conclude that $N(x,y), D(x,y) >0$

\begin{figure}[tbph]
\begin{center}
\epsfxsize=6in
\epsfysize=3in
\epsffile{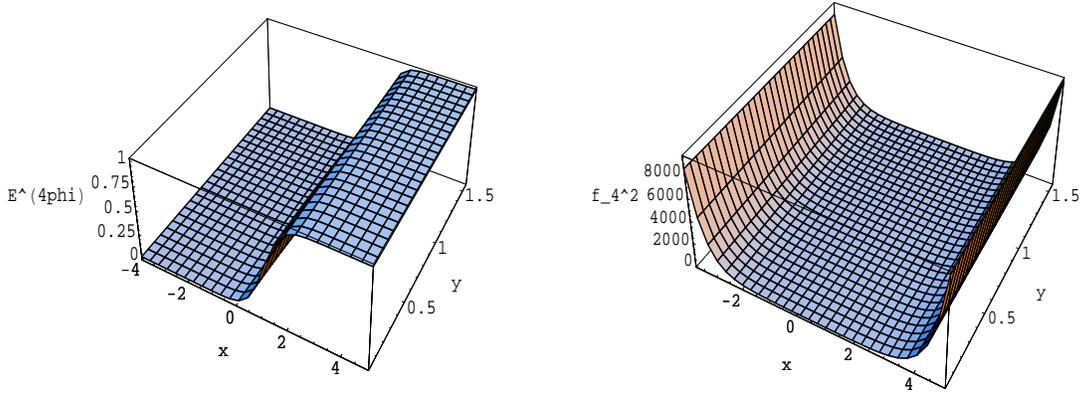}
\label{figure1}
\caption{Three-dimensional plot of the dilaton $e^{4\phi}$ (left) and the 
metric factor $f_{4}^{2}$ (right) for the supersymmetric Janus solution 
as a function of $x$ and $y$, for $r=4$.}
\end{center}
\end{figure}

\subsubsection{Regularity of the metric functions}

The $\Sigma$-metric factor $\rho$ is given by
\bea
\rho^8 = {4 (1+r)^2 N(x,y) D(x,y) \over (e^x + r e^{-x})^3 (e^x + e^{-x})^3}
\eea
The $S^2$ metric factors are given by
\bea
\label{Januss2metrics}
\rho^2 f_1 f_2 &=& 4 \nu (1+r) \sin 2y
\\
\rho^2 (f_1^2 - f_2^2) &=& - 8 (1+r) { (e^x + e^{- x}) D(x,y) \cos^2 y 
- (e^x + r e^{-x}) N (x,y) \sin^ 2 y 
\over \sqrt{r N(x,y) D(x,y) (e^x + e^{-x})(e^x + r e^{-x})} } \quad 
\no \eea
These expressions are never singular, as a result of the positivity of $N(x,y)$ and $D(x,y)$. 
The $S^2$-metric factors shrink to zero size only on the boundaries of $\Sigma$ 
defined by the lines $y =0$ and $y={\pi \over 2}$. The second equation in (\ref{Januss2metrics})
then shows that $f_1$ and $f_2$ cannot 
simultaneously vanish on $\p \Sigma$, since either $y=0$ or $y = \pi/2$ and only a single
term survives in the numerator.

\begin{figure}[tbph]
\begin{center}
\epsfxsize=6in
\epsfysize=3in
\epsffile{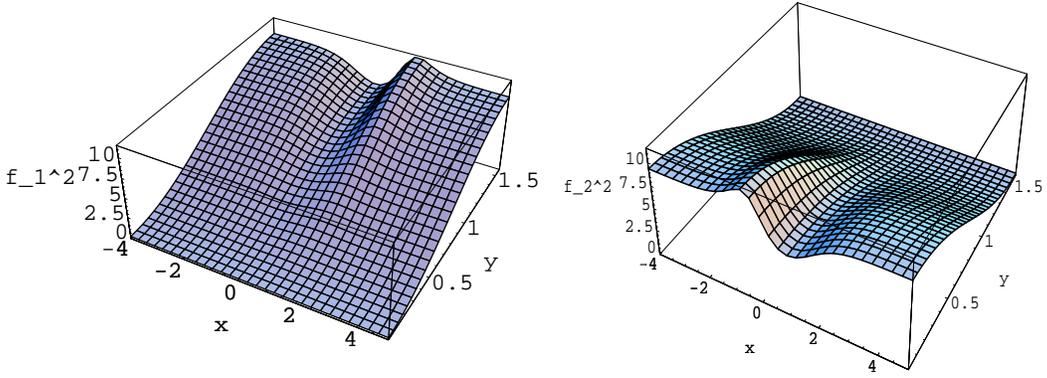}
\label{figure2}
\caption{Three-dimensional plot of the metric factor $f_{1}^{2}$  and $f_{2}^{2 }$ 
for the supersymmetric Janus solution as function of $x$ and $y$, for $r=4$.}
\end{center}
\end{figure}

\begin{figure}[tbph]
\begin{center}
\epsfxsize=4in
\epsfysize=3in
\epsffile{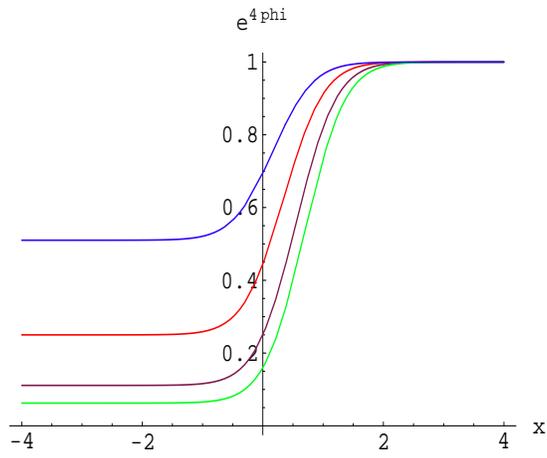}
\label{figure3}
\caption{Plot of  the dilaton $e^{4\phi}$ at $y=0$ as a function of $x$, 
for  $r=1.5$ (blue), $r=2$ (red),  $r=3$ (magenta) and $r=4$ (green).}
\end{center}
\end{figure}

The $AdS_4$ metric factor is given by
\bea
\rho^2 f_4^2 &=& 4 \sqrt{ {N(x,y) D(x,y) \over (e^x + r e^{-x})(e^x + e^{-x})}}
\eea
which is non-singular and nowhere vanishing.
Taking the limit $r \rightarrow 1$, we recover the $AdS_5 \times S^5$
metric factors of (\ref{AdSsol1}).

\subsection{The Half-BPS Janus holographic dual interface theory} 

The holographic interpretation of the original Janus solution is given by an 
interface  conformal field theory. The dual four-dimensional field theory  
lives on two four-dimensional half spaces glued together at a 
three-dimensional interface.  
Although the Janus  solution (\ref{janusharm})  is more complicated than the  
original \cite{Bak:2003jk} and the $\cN =1$ supersymmetric  Janus solution 
\cite{D'Hoker:2006uu}, it shares many features with these solutions, 
as we shall show next. 
The asymptotic behavior of the metric functions can easily be obtained 
using the parametrization of the strip (\ref{stripp}). In the limit $x\to \pm \infty$  
one gets
\begin{eqnarray}
\rho^{2} &=&  \sqrt{2}(1+r)^{1\over 2}+ o(e^{-2|x|}) 
\no \\
f_{1}^{2}&=&  4 \sqrt{2}(1+r)^{1\over 2} \sin^{2}y + o(e^{-2|x|}) 
\no \\
f_{2}^{2}&=&  4 \sqrt{2}(1+r)^{1\over 2} \cos^{2}y + o(e^{-2|x|}) 
\no \\
f_{4}^{2}&=&  {2 \sqrt{2}\over (1+r)^{1\over 2}} e^{2|x|}+ o(1) 
\label{adsasym}
\end{eqnarray}
The boundary of the bulk geometry can be obtained by extracting the part 
of the space where the metric becomes infinite. It follows from (\ref{adsasym}) 
that  the metric for $AdS_{4 }$ blows up when $x\to \pm \infty$. 
There is however an additional component to the boundary. 
Employing the Poincar\'e patch metric for $AdS_{4}$.
\begin{equation}
ds_{AdS_{4}}^{2}= {1\over z^{2}} \Big(-dt^{2}+ dx_{1}^{2}+dx_{2}^{2}+ dz^{2}\Big)
\end{equation}
it is obvious that there is another boundary component at $z\to 0$. 
The ten-dimensional asymptotic metric, in the limit $x\to \pm \infty, z\to 0$, 
is  given by\footnote{Note that the metric on the strip is given by 
$ds^2_{\Sigma}=4 \rho^2 (dx^2+dy^2)$ due to our slightly unconventional  
normalization of the two-dimensional frames $e^z, e^{\bar z}$.}
\bea
ds^2 & \sim  &  {1\over z^{2} \mu^{2}} \bigg ( z^{2 }d\mu^{2} 
+{dx_{1}^{2}+dx_{2}^{2}- dt^{2} + dz^{2}\over 2(1+r)} 
\no \\ && \qquad
+ z^{2} \mu^{2} ( dy^{2}+ \sin^{2}y ds^{2}_{S_1^2} + \cos^{2}y ds^{2}_{S_2^2})\bigg)  
+o(\mu^{2}) 
\label{asymmet}
\eea
where a new local coordinate $\mu = e^{\mp  x}$ was introduced.  The limit 
$x\to \pm \infty$ corresponds to $\mu \to 0$.  It follows from (\ref{asymmet}) 
that the boundary of the bulk geometry has three components: $x\to \pm \infty$ 
corresponds to two four-dimensional half spaces, which are glued together 
at a three-dimensional interface at $z\to 0$. The structure of the boundary 
is therefore the same as in the original Janus solution and defines an 
interface field theory.
The asymptotic behavior of the dilaton is
\bea
e^{2\phi} & = & e^{2 \phi_+} +o(e^{-2 x}), \; {\rm as} \;  x\to +\infty
\no \\   
e^{2\phi} & = & e^{2 \phi_-}  +o(e^{-2 x}), \; {\rm as}\; x\to -\infty,
\eea
Hence the super Yang-Mills  theory in the two half spaces has  two different 
values of the coupling constant $g_{YM}$,  as in the original Janus solution. 
The NSNS and RR 2-form gauge potentials behave as follows near the boundary,  
\bea
\Re (B_{(2)}) & \sim &  e^{-3|x| } \sin y \;  \hat e^{45}
\no \\
\Im (B_{(2)}) & \sim & e^{-3|x| } \cos y \;  \hat e^{67} 
\label{astbeh}
\eea
Their dependence on the $S^5$ 
corresponds to lowest Kaluza-Klein modes on the $S^5$ of the
anti-symmetric rank 2 tensor field and is  associated with a scalar 
field of dimension $\Delta=3$
\cite{Kim:1985ez,Gunaydin:1984fk} in $\cN=4$ super-Yang-Mills.

\smallskip

The behavior (\ref{astbeh}) leads to a insertion of the dual operator 
which is localized at the interface.  This agrees with the interpretation 
of the solution as a Janus interface CFT with an interface term given 
by (\ref{interfl}). The detailed  analysis is the same as the one given in 
\cite{D'Hoker:2006uu} and will be repeated in the following  for completeness.

\smallskip

In the following we will focus on one of the four-dimensional half spaces 
and use the local coordinate $\mu=e^{\pm x}$ defined above 
(not to be confused 
with the harmonic function $\mu$ introduced and used in subsection 9.1). 
The boundary is reached when $\mu  z \to 0$.   The complete boundary  
corresponds to two
4-dimensional half spaces joined by a ${\bf R}^3$ interface located at $z=0$. 

\smallskip

The AdS/CFT correspondence relates 10-dimensional Type IIB supergravity 
fields to gauge invariant operators on the $\cN =4$ super Yang-Mills side. 
In the following we  briefly review some aspects of this map. 
The Poincar\'e metric of Euclidean $AdS_5$ is given by
\be
ds^2= {1\over z^2} \left (dz^2 + \sum_i dx_i^2 \right )
\ee
Near the boundary of $AdS_5$, where $z\to 0$, 
a scalar field $\Phi_m$ of mass $m$  behaves as,
\be
\Phi _m (z,x) \, \sim \, \phi_{non-norm}(x) z^{4-\Delta} +  \phi_{norm}(x) z^{\Delta}  
\label{stateoper}
\ee
where $m^2=\Delta(\Delta-4)$. The non-normalizable mode corresponds to
insertion in the Lagrangian of an operator ${\cal O}_\Delta$ with scaling 
dimension $\Delta$.
The boundary source can be determined from (\ref{stateoper}) by
\be
\phi_{non-norm}(x) = \lim_{z\to 0}   z^{\Delta-4} \Phi(z,x) \label{nonnorma}
\ee
If $\phi_{non-norm}$ vanishes, a non-zero $\phi_{norm}$ corresponds to a
non-vanishing expectation value 
\bea 
\langle{\cal O}_\Delta\rangle= \phi_{norm}
\eea 
of the operators ${\cal O}_\Delta$ on the Yang-Mills side.
The asymptotic behavior near the boundary of the 2-form fields 
as $\mu \to 0$ is given by
\bea
b_1(\mu)&= &  {\rm const} \;   \mu^3 + o(\mu^5)
\no \\
b_2(\mu) & = & {\rm const} \;  \mu^3+o(\mu^5)
\eea
The state operator correspondence  (\ref{stateoper}) seems to suggest 
that there is no source for the $\Delta=3$ operator dual to the 3-form fields,  
since  the non-normalizable  mode is not turned on.
However, this conclusion is premature. For the Janus metric the
appropriate rescaling of the field needed to extract the non-normalizable mode is
given by
\bea
c_{non-norm}&=& \lim_{\epsilon\to 0} \epsilon^{\Delta-4} c(\mu) \no \\
&=& \lim_{\epsilon \to  0} {1\over \epsilon } \mu^3   \;  {\rm const}\no \\
&=& \lim_{\mu z\to 0}   {\  \mu^2 \over z} \; {\rm const} 
\label{cnonnorm}
\eea
where $\epsilon= \mu z$ was used.
For a point on the boundary which is away from the three-dimensional interface
one has $z\neq 0$ and it follows from (\ref{cnonnorm}) that the  source
for the dual operator vanishes away from the interface.
However for the interface one has $z=0$ and $c_{non-norm}$ in
(\ref{cnonnorm}) diverges. This behavior indicates  the presence of a
delta function source for the dual $\Delta=3$ operator on the interface, since 
the integral over a small disk around the interface $ \int d\mu\; dz z\;
c_{non-norm}$  is finite.   The localized operator on the interface is the 
interface counterterm discussed in section \ref{two}, which is necessary 
to restore $\cN =4$ interface supersymmetry. 

\bigskip\bigskip

\noindent{\large \bf Acknowledgment}

\bigskip

\noindent This work 
was supported in 
part by National Science Foundation (NSF) grant PHY-04-56200.

\newpage

\appendix

\section{Clifford algebra basis adapted to the Ansatz}
\setcounter{equation}{0}
\label{appA}

We choose a basis for the Clifford algebra which is well-adapted to the 
$AdS_4 \times S_1^2 \times S_2^2 \times \Sigma$ Ansatz, with the frame
labeled as in (\ref{frame1}),
\bea
\G^m & = & \g^m \otimes I_2 \otimes I_2 \otimes I_2 \hskip 1.1in m =0,1,2,3
\no \\
\G^{i_1} & = & \g_{(1)} \otimes \g^{i_1} \otimes I_2 \otimes I_2 \hskip 1in i_1=4,5
\no \\
\G^{i_2} & = & \g_{(1)} \otimes \sigma ^3 \otimes \g^{i_2} \otimes I_2 \hskip 1in i_2 = 6,7
\no \\
\G^a \, & = & \g_{(1)}  \otimes \sigma ^3 \otimes \sigma ^3 \otimes \gamma ^a \hskip 1in a=8,9
\eea
where  a convenient basis for the lower dimensional Clifford algebras is as follows, 
\bea
i \g^0 = \sigma ^2 \otimes I_2 & \hskip 1in & \g^4 = \g^6 = \g ^8 = \sigma ^1
\no \\
\g^1 = \sigma ^1 \otimes I_2 & \hskip 1in & \g^5 = \g^7 = \g ^9 = \sigma ^2
\no \\
\g^2 = \sigma ^3 \otimes \sigma ^2 & \hskip 1in &
\no \\
\g^3 = \sigma ^3 \otimes \sigma ^1& \hskip 1in &
\eea
We shall also need the chirality matrices on the various components of 
$AdS_4 \times S_1^2 \times S_2^2 \times \Sigma$, and they are chosen as follows,
\bea
\gamma _{(1)} & = & 
    -i \G^{0123} = \sigma ^3 \otimes \sigma ^3 \otimes I_2 \otimes I_2 \otimes I_2
\no \\
\gamma _{(2)} & = & -i \G^{45} ~~ = I_2 \otimes I_2 \otimes \sigma ^3 \otimes I_2 \otimes I_2
\no \\
\gamma _{(3)} & = & -i \G^{67} ~~ = I_2 \otimes I_2 \otimes I_2 \otimes \sigma ^3 \otimes I_2
\no \\
\gamma _{(4)} & = & -i \G^{89} ~~ = I_2 \otimes I_2 \otimes I_2 \otimes I_2 \otimes \sigma ^3 
\eea
The 10-dimensional chirality matrix in this basis is given by
\bea
\G^{11} = \G^{0123456789}   = \g_{(1)} \g_{(2)} \g_{(3)} \g_{(4)}
\eea
The complex conjugation matrices in each component are defined by
\bea
\left ( \g^m \right ) ^* = + B_{(1)} \g ^m B_{(1)} ^{-1}
& \hskip .5in &   (B_{(1)})^* B_{(1)} = + I_2 \hskip .6in B_{(1)} = i \g_{(1)} \g^2 
\no \\
\left ( \g^{i_1} \right ) ^* = - B_{(2)} \g ^{i_1} B_{(2)} ^{-1}
& &   (B_{(2)})^* B_{(2)} = - I_2 \hskip .6in B_{(2)} =  \g^5  
\no \\
\left ( \g^{i_2} \right ) ^* = - B_{(3)} \g ^{i_2} B_{(3)} ^{-1}
& &   (B_{(3)})^* B_{(3)} = - I_2 \hskip .6in B_{(3)} =  \g^7 
\no \\
\left ( \g^a \right ) ^* = - B_{(4)} \g ^a B_{(4)} ^{-1}
& &   (B_{(4)})^* B_{(4)} = - I_2 \hskip .6in B_{(4)} =  \g^9 
\eea
where in the last column we have also listed the form of these matrices
in our particular basis. The 10-dimensional complex conjugation matrix $\cB$
is defined by $(\G^M)^* = \cB \G^M \cB^{-1} $ and $\cB \cB^* = I$, and in this
basis is given by
\bea
\cB =  - \G^{2579}  
    =i  \g_{(1)} \g^2 \otimes \gamma ^5 \otimes \gamma ^6 \otimes \g^9 
    = i  B_{(1)} \otimes B_{(2)} \otimes \left ( \gamma _{(3)} B_{(3)} \right ) \otimes B_{(4)}
\eea

\section{The geometry of Killing spinors}
\setcounter{equation}{0}
\label{appB}

In this Appendix, we review the relation between the Killing spinor
equation and the parallel transport equation in the presence of a flat connection 
with torsion on $S^2$ and $AdS_4$.

\subsection{The sphere $S^2$}

On $S^2$, the Killing spinor equation is given by, 
\bea
\label{KS1}
\left ( \nabla _i  +\eta  \half \sigma _i \sigma _3 \right )\ep & = & 0 \hskip 1in  i=1,2
\eea
where $\sigma ^1, \sigma ^2, \sigma ^3$ are the standard Pauli matrices, and
$\nabla_i$ is the spin connection on $S^2$. (An equivalent equation is obtained
by letting $\sigma _i \sigma _3 \to i \sigma _i$ and $\ep ' = e^{- i \pi \sigma _3/4} \ep$.)
Integrability of this system of equations on the round sphere requires $\eta = \pm 1$.

\smallskip

The relevant flat connection with torsion is given by the Maurer-Cartan forms 
$\omega ^{(t)}$ on $SO(3)$,  in the spinor representation of $SO(3)$, 
\bea
&&
\omega ^{(t)} = U^\dagger dU 
= {1 \over 4} \omega ^{(t)} _{IJ} \, \sigma ^{[I} \sigma ^{J]} 
\hskip 1in U \in SU(2)
\label{MC1}
\\ &&
d\omega ^{(t)} + \omega ^{(t)} \wedge \omega ^{(t)} =0
\label{MC2}
\eea
where $\{ \sigma _I , \sigma _J \} = 2 \delta _{IJ}$, with $I,J=1,2,3$.
The Maurer-Cartan equation (\ref{MC2}) expresses the flatness of the connection
$\omega ^{(t)}$, which in turn reflects the fact that $SO(3)$, like any Lie group, 
is paralellizable. Next, we view $S^2$ as the coset space $S^2 = SO(3)/SO(2)$,
and decompose the directions of the cotangent space of $SO(3)$ accordingly, 
\bea
\label{MC3}
\omega ^{(t)} = { i \over 2} \omega \sigma _3
+ {1 \over 2} e_i \sigma _i \sigma _3
\hskip 1in 
\left \{ \matrix{ 
\omega  \equiv \omega ^{(t)} _{12} & \cr 
e_i \equiv \omega ^{(t)} _{i3} & \quad  i=1,2 \cr} \right .
\eea
where $e_i$ is the canonical frame, and $\omega$ the canonical $SO(2)$
connection on $S^2$. The Maurer-Cartan equations imply the absence of torsion, 
and the constancy of curvature.
The parallel transport equation for a 2-component spinor $\ep_+$ is simply
\bea
\label{MC4}
\left ( d + \omega ^{(t)} \right ) \ep_+ =0
\eea
On the one hand,  using the identification $\omega ^{(t)}= U^\dagger dU$, 
it may be solved trivially by
\bea
\ep _+ = U^\dagger \ep _0
\eea
where $\ep_0$ is a constant spinor. On the other hand, using the canonical 
decomposition (\ref{MC3}), and the expression for the covariant derivative in 
terms of forms, $e_i \nabla _i = d + i  \omega \, \sigma _3/2$, it is clear that 
the equation coincides with the Killing spinor equation with $\eta =+1$.
The solution $\ep_-$ to the Killing equation for $\eta =-1$ may be obtained
from the solution $\ep_+$, by $\ep _- = \sigma _3 \, \ep_+$.

\subsection{Minkowski $AdS_4$}

The above construction may be generalized to all spheres and their 
hyperbolic counterparts. Here, we present the case of Minkowski signature 
$AdS_4 = SO(2,3)/SO(1,3)$.
The Clifford algebra of $SO(2,3)$ is built from the Clifford generators
$\gamma ^\mu$,  of the Lorentz group $SO(1,3)$,
\bea
\{ \gamma ^\mu , \gamma ^\nu\} = 2 \eta ^{\mu \nu }
\hskip 1in 
\eta = {\rm diag} [-+++]
\eea
for $\mu, \nu =0,1,2,3$, supplemented with the chirality matrix, 
$\gamma ^{\sharp } \equiv \gamma ^{0123} $,
\bea
\{ \gamma ^ {\bar \mu} , \gamma ^{\bar \nu} \} = 2 \bar \eta ^{\bar \mu \bar \nu}
 \hskip 1in 
\bar \eta = {\rm diag} [--+++]
\eea
for $\bar \mu, \bar \nu = \sharp , 0,1,2,3$.
The corresponding Maurer-Cartan form on $SO(2,3)$ is given by
\bea
\omega ^{(t)} = V ^{-1} dV = {1 \over 4} \omega ^{(t)} _{\bar \mu \bar \nu}
\gamma ^{\bar \mu \bar \nu} \hskip 1in V \in Sp(2,2) \sim SO(2,3)
\eea
It obviously satisfies the Maurer-Cartan equations, $d \omega ^{(t)} 
+ \omega ^{(t)} \wedge \omega ^{(t)}=0$.
We decompose $\omega ^{(t)}$ onto the $SO(1,3)$ and $AdS_4$
directions of cotangent space,
\bea
\omega ^{(t)} = { 1 \over 4} \omega _{\mu \nu} \gamma ^{\mu \nu}
+ {i \over 2} e_\mu \gamma ^\mu \gamma _{(4)} 
\hskip 1in 
\left \{ \matrix{ 
\omega _{\mu \nu}   \equiv \omega ^{(t)} _{\mu \nu } & \mu,\nu =0,1,2,3 \cr 
e_\mu \equiv \omega ^{(t)} _{\mu \sharp} & \mu=0,1,2,3 \cr} \right .
\eea
The Maurer-Cartan equations $ d \omega ^{(t)} + \omega ^{(t)} \wedge \omega ^{(t)}=0$ 
for $\omega ^{(t)}$ imply that the absence of torsion and that the constancy of curvature. 
The Killing spinor equation coincides with the equation for parallel transport,
\bea
\left ( d + V^{-1} dV \right ) \ep =
\left ( d + { 1 \over 4} \omega _{\mu \nu} \gamma ^{\mu \nu}
+ {i \over 2} \eta e_\mu \gamma ^\mu \gamma _{(4)}  \right ) \ep =0
\eea
For $\eta=+1$, the general solution is given by $\ep _+ = V^{-1} \ep _0$
and $\ep_0$ is constant, while for $\eta =-1$, the solution is $\ep_- = \gamma ^\sharp \ep _+$.
Note that $V$ is a symplectic matrix,
so that $V^t J V = J$, and this allows us to define an invariant inner 
product on the spinors.

\newpage

\section{The derivation of the BPS equations}
\setcounter{equation}{0}
\label{appD}

We begin by collecting some identities that will be useful during the 
reduction of the BPS equations over the Ansatz of subsection 4.1. The Clifford 
algebra matrices needed are,
\bea
\G ^a & = & \gamma_{(1)} \gamma_{(2)} \gamma_{(3)} \g^a
\no \\
\G^{45a} &=& i \gamma_{(1)} \gamma_{(3)} \g^a 
\no\\ 
\G^{67a} &=& i \gamma_{(1)} \gamma_{(2)} \g^a
\no \\
\G^{0123a} & = & i \gamma_{(2)} \gamma_{(3)} \g^a
\no \\
\ep^{ab} \G^{4567b} & = & - \gamma_{(1)}  \ep^{ab} \g^b
\eea
We shall also need the following decompositions of $\ep$ and $\cB ^{-1} \ep ^*$,
\bea
\ep &=& \sum_{\eta_1, \eta_2, \eta_3} \chi^{\eta_1, \eta_2, \eta_3} \otimes
\zeta_{\eta_1, \eta_2, \eta_3} 
\no \\
\cB^{-1} \varepsilon^* &=& \sum_{\eta_1, \eta_2, \eta_3} \chi^{\eta_1, \eta_2 ,\eta_3} \otimes 
* \zeta _{\eta _1, \eta _2, \eta _3} 
\eea
where we use the abbreviation,
\bea 
\label{star}
* \zeta _{\eta _1, \eta _2, \eta _3} 
& = &  - i \sigma ^2 \eta _1 \eta _2 \eta _3 \zeta _{\eta _1, \eta _2 , -\eta _3}^*
\no \\
{}* \zeta & = & \tau ^{332} \otimes \sigma ^2 \zeta ^*
\eea
in $\tau$-matrix notation.

\subsection{The dilatino equation} 

The dilatino equation is, 
\bea
0=  i P_A  \G^A \cB^{-1} \ep ^* -{i\over 24} \G \cdot  G \ep 
\eea
Reduced to the Ansatz of subsection 4.1, we have the following simplifications,
\bea
P_A \G^A &=&p_a  \G^a 
\no\\
\G \cdot  G &=& 3! (g_a \G^{45a} + i \, h_a \G^{67a})
\eea
The dilatino equation now becomes,
\bea
0 & = & i p_a \G^a \sum_{\eta_1 ,\eta_2 ,\eta_3} \chi^{\eta_1 ,\eta_2 ,\eta_3} 
(-i \sigma ^2)  \eta _1 \eta _2 \eta _3 \zeta_{\eta_1, \eta_2,- \eta_3} ^*
\no \\ && 
- {i\over 4} (g_a \G^{45a} + i \, h_a \G^{67a}) \sum_{\eta_1, \eta_2, \eta_3} 
\chi^{\eta_1, \eta_2, \eta_3} \otimes \zeta_{\eta_1, \eta_2, \eta_3}
\eea
Using the above form of the $\G$-matrices, the action of
the chirality matrices on $\chi$, and flipping the signs in the summation over 
$\eta$ so as to have a common factor of $\chi$, we obtain, 
\bea
0= \sum_{\eta_1, \eta_2, \eta_3} \chi^{\eta_1, \eta_2, \eta_3} 
\otimes \bigg( - p_a \sigma^a \sigma ^2  \eta _1 \eta _2 \eta _3 
\zeta_{-\eta_1, -\eta_2, \eta_3} ^*  
+ {1\over 4} g_a \sigma^{a} \zeta_{-\eta_1, \eta_2, -\eta_3} 
+ {i\over 4} h_a \sigma^{a} \zeta_{-\eta_1, -\eta_2, \eta_3}
\bigg)
\eea
Since the $\chi^{\eta_1, \eta_2, \eta_3}$ are linearly independent we require the vanishing of
\bea
0 = - p_a \sigma^a \sigma^2 \eta_1 \eta_2 \eta_3 \zeta_{- \eta_1, - \eta_2, \eta_3}^*  
+ {1\over 4} g_a \sigma^{a} \zeta_{-\eta_1, \eta_2, -\eta_3} 
+ {i\over 4} h_a \sigma^{a} \zeta_{-\eta_1, -\eta_2, \eta_3}
\eea
This can be recast economically using the $\tau$-matrix notation,
\bea
0 = p_a \tau^{(223)}  \sigma^a \sigma^2 \zeta^*
+ {1\over 4} g_a \tau^{(101)}  \sigma^{a} \zeta
+ {i\over 4} h_a \tau^{(110)}  \sigma^{a} \zeta
\eea
Upon multiplication on the left by $\tau^{(223)}$, we recover (\ref{dilatino1}).
\smallskip

\subsection{The gravitino equation}

The gravitino equation is
\bea
0 & = & d \ep + \omega \ep + \phi \ep + g \cB^{-1} \ep^* 
\no \\
\omega &=& {1 \over 4} \omega_{AB} \Gamma^{AB} \no\\
\phi &=& -{i \over 2}Q + {i \over 480} (\Gamma \cdot F_{(5)}) e_A \G^A \no\\
g &=& - {1 \over 96} e_A \bigg( \G^A (\G \cdot G) + 2 (\G \cdot G) \G^A \bigg)
\eea

\subsubsection{The calculation of $\omega$}

The spin connection components are $\omega ^a {}_b$, whose explicit form 
we shall not need, and 
\bea
\label{spincon}
\omega^m {}_n = \hat \omega^m {}_n
&\qquad&
\omega^m {}_a = e^m {\p_a f_4 \over f_4}
\no\\
\omega^{i_1} {}_{i_2} = \hat \omega^{i_1} {}_{i_2}
&\qquad&
\omega^{i_1} {}_a = e^{i_1} {\p_a f_1 \over f_1}
\no\\
\omega^{i_2} {}_{j_2} = \hat \omega^{i_2} {}_{j_2}
&\qquad&
\omega^{i_2} {}_a = e^{i_2} {\p_a f_2 \over f_2}
\eea
The hats refers to the canonical connections on $AdS_4$, $S^2_1$, $S^2_2$ respectively.  
Projecting the spin-connection along the various directions we have
\bea
(m) &\qquad& \nabla_m^\prime \ep + \half {D_a f_4 \over f_4} \G_{m} \G^{a} \ep 
\no\\
(i_1) &\qquad& \nabla_{i_1}^\prime \ep + \half {D_a f_1 \over f_1} \G_{i_1} \G^{a} \ep
\no\\
(i_2) &\qquad& \nabla_{i_2}^\prime \ep + \half {D_a f_2 \over f_2} \G_{i_2} \G^{a} \ep
\no\\
(a) &\qquad& \nabla_a \ep
\eea
where the prime on the covariant derivative indicates that only the connection 
along $AdS_4$, $S^2_1$ or $S^2_2$ respectively is included.  
Using the Killing spinor equations (\ref{KS}) we can eliminate the primed covariant derivatives, 
which yields
\bea
(m) &\qquad& {1 \over 2 f_4} \G_{m}  
\sum_{\eta_1 ,\eta_2 ,\eta_3} \eta_1 \chi^{\eta_1, \eta_2, \eta_3} \otimes 
\zeta_{\eta_1, \eta_2, \eta_3} + \half {D_a f_4 \over f_4} \G_{m} \G^{a} \ep 
\no\\
(i_1) &\qquad& {i \over 2f_1} \G_{i_1} \g_{(1)} 
\sum_{\eta_1, \eta_2, \eta_3} \eta_2 \chi^{\eta_1, \eta_2, \eta_3} \otimes 
\zeta_{\eta_1, \eta_2, \eta_3} + \half {D_a f_1 \over f_1} \G_{i_1} \G^{a} \eps 
\no\\
(i_2) &\qquad& {i \over 2f_2} \G_{i_2} \g_{(1)} \g_{(2)} 
\sum_{\eta_1, \eta_2, \eta_3} \eta_3 \chi^{\eta_1, \eta_2, \eta_3} \otimes 
\zeta_{\eta_1, \eta_2, \eta_3} + \half {D_a f_2 \over f_2} \G_{i_2} \G^{a} \ep
\eea
Using the equation $\G^a = \gamma_{(1)} \gamma_{(2)} \gamma_{(3)} \sigma^a$, we have
\bea
(m) &\qquad& \G_m \sum_{\eta_1, \eta_2, \eta_3} \chi^{\eta_1, \eta_2, \eta_3} \otimes
\bigg( {1 \over 2 f_4} \eta_1 \zeta_{\eta_1, \eta_2, \eta_3} 
+ \half {D_a f_4 \over f_4} \sigma^{a} \zeta_{- \eta_1, - \eta_2, - \eta_3} \bigg) 
\no\\
(i_1) &\qquad& \G_{i_1} \sum_{\eta_1 ,\eta_2 ,\eta_3} \chi^{\eta_1, \eta_2, \eta_3} \otimes
\bigg( {i \over 2 f_1} \eta_2 \zeta_{- \eta_1, \eta_2, \eta_3} 
+ \half {D_a f_1 \over f_1} \sigma^{a} \zeta_{-\eta_1, -\eta_2, -\eta_3} \bigg) 
\no\\
(i_2) &\qquad& \G_{i_2} \sum_{\eta_1, \eta_2, \eta_3} \chi^{\eta_1 ,\eta_2, \eta_3} 
\otimes
\bigg( {i \over 2 f_2} \eta_3 \zeta_{-\eta_1, -\eta_2, \eta_3} + \half {D_a f_2 \over f_2} 
\sigma^{a} \zeta_{-\eta_1, -\eta_2, -\eta_3} \bigg)
\eea
where we have pulled a factor of $\G_M$ out front.  It will turn out that 
all terms in the gravitino equation contain $\G_M \chi^{\eta_1, \eta_2 ,\eta_3}$, 
and we will require the coefficients to vanish independently, just as we did for 
the dilatino equation.  The coefficient of 
$\G_M \chi^{\eta_1 ,\eta_2 ,\eta_3}$ can be expressed in the $\tau$-matrix 
notation as
\bea
\label{redomega}
(m) &\qquad& {1 \over 2 f_4} \tau^{(300)}  \zeta 
+ \half {D_a f_4 \over f_4} \tau^{(111)}  \sigma^{a} \zeta 
\no\\
(i_1) &\qquad& {i \over 2 f_1} \tau^{(130)}  \zeta 
+ \half {D_a f_1 \over f_1} \tau^{(111)}  \sigma^{a} \zeta 
\no\\
(i_2) &\qquad& {i \over 2 f_2} \tau^{(113)}  \zeta 
+ \half {D_a f_2 \over f_2} \tau^{(111)}  \sigma^{a} \zeta
\eea

\subsubsection{The calculation of $\phi$}

The $Q$ part is trivial.  The $F_{(5)}$ part is
\bea
{i \over 480} (\G \cdot F_{(5)}) \G^A e_A \ep
= - {i \over 2} f_a \G^{0123a} \G^A e_A \ep
\eea
Projecting along the various directions, we have
\bea
(m) &\qquad& \G_m \sum_{\eta_1, \eta_2, \eta_3} \chi^{\eta_1, \eta_2, \eta_3} 
\otimes {1 \over 2}  f_a \sigma^{a} \zeta_{\eta_1, - \eta_2, - \eta_3} 
\no\\
(i_1) &\qquad& - \G_{i_1} \sum_{\eta_1, \eta_2, \eta_3} \chi^{\eta_1, \eta_2 ,\eta_3} 
\otimes {1 \over 2}  f_a \sigma^{a} \zeta_{\eta_1, - \eta_2, - \eta_3} 
\no\\
(i_2) &\qquad& - \G_{i_2} \sum_{\eta_1, \eta_2, \eta_3} \chi^{\eta_1, \eta_2, \eta_3} 
\otimes {1 \over 2}  f_a \sigma^{a} \zeta_{\eta_1, - \eta_2, - \eta_3} 
\no\\
(a) &\qquad& \sum_{\eta_1, \eta_2, \eta_3} \chi^{\eta_1, \eta_2, \eta_3} \otimes 
\bigg( - {iQ \over 2}  \zeta_{\eta_1, \eta_2, \eta_3} 
+ {1 \over 2} f_b \sigma^{b} \sigma_a \zeta_{- \eta_1, \eta_2, \eta_3} \bigg)
\eea
Using the $\tau$-matrix notation,  we can write the coefficient of 
$\G_M \chi^{\eta_1, \eta_2 ,\eta_3}$ in the form
\bea
\label{redphi}
(m) &\qquad& {1 \over 2}  f_a \tau^{(011)} \otimes \sigma^{a} \zeta \\
(i_1) &\qquad& - {1 \over 2}  f_a \tau^{(011)} \otimes \sigma^{a} \zeta \no\\
(i_2) &\qquad& - {1 \over 2}  f_a \tau^{(011)} \otimes \sigma^{a} \zeta \no\\
(a) &\qquad& - {i q_a \over 2}  \zeta 
+ {1 \over 2} f_b \tau^{(100)} \otimes \sigma^{b} \sigma_a  \zeta
\no
\eea
where $q_a$ is defined by $Q = q_a e^a$.

\subsubsection{The calculation of $g$}

The relevant expression is as follows, 
\bea
g \cB^{-1} \epsilon^* = - {3! \over 96} e_A \bigg(g_a (\G^A \G^{45a} + 2 \G^{45a} \G^A) 
+ i h_a (\G^A \G^{67a} + 2 \G^{67a} \G^A) \bigg) \cB^{-1} \epsilon^*
\eea
A few useful equations are as follows,
\bea
\G^a \G^{45b} + 2 \G^{45b} \G^a &=& \G^{45} (3 \delta^{ab} - \G^{ab}) 
= i \gamma_{(2)} (3 \delta^{ab} -  \sigma^{ab}) 
\no\\
\G^a \G^{67b} + 2 \G^{67b} \G^a &=& \G^{67} (3 \delta^{ab} - \G^{ab}) 
= i \gamma_{(3)} (3 \delta^{ab} -  \sigma^{ab})
\eea
Projecting along the various directions we obtain
\bea
(m) &\quad& - \G_m \sum_{\eta_1, \eta_2, \eta_3} \chi^{\eta_1 ,\eta_2 ,\eta_3} 
\otimes {1 \over 16} (- i g_a \sigma^{a} * \zeta_{- \eta_1, \eta_2 ,- \eta_3} 
+ h_a \sigma^{a} * \zeta_{- \eta_1, - \eta_2, \eta_3}) 
\\
(i_1) &\quad& - \G_{i_1} \sum_{\eta_1, \eta_2, \eta_3} \chi^{\eta_1, \eta_2 ,\eta_3} 
\otimes {1 \over 16} (3 i g_a \sigma^{a} * \zeta_{- \eta_1, \eta_2, - \eta_3} 
+ h_a \sigma^{a} * \zeta_{- \eta_1, - \eta_2, \eta_3})
 \no\\
(i_2) &\quad& - \G_{i_2} \sum_{\eta_1, \eta_2, \eta_3} \chi^{\eta_1, \eta_2, \eta_3}
 \otimes {1 \over 16} (- i g_a \sigma^{a} * \zeta_{- \eta_1, \eta_2, - \eta_3} 
 - 3 h_a \sigma^{a} * \zeta_{- \eta_1, - \eta_2, \eta_3}) 
 \no\\
(a) &\quad& - \sum_{\eta_1, \eta_2, \eta_3} \chi^{\eta_1, \eta_2, \eta_3} 
\otimes {1 \over 16} \bigg(3 i g_a * \zeta_{\eta_1, - \eta_2, \eta_3} 
- i g_b \sigma^{ab} * \zeta_{\eta_1, - \eta_2, \eta_3} 
\no \\ 
&& \hskip 1.5in 
- 3 h_a * \zeta_{\eta_1, \eta_2, - \eta_3} 
+ h_a \sigma^{ab} * \zeta_{\eta_1, \eta_2, - \eta_3} \bigg) \no
\eea
Using the $\tau$-matrix notation, we can write the coefficient of 
$\G_M \chi^{\eta_1, \eta_2, \eta_3}$ in the form
\bea
\label{redg}
(m) && 
- {1 \over 16} (- i g_a \tau^{(101)}  \sigma^{a} * \zeta 
+ h_a \tau^{(110)}  \sigma^{a} * \zeta) 
\\
(i_1) && 
- {1 \over 16} (3 i g_a \tau^{(101)} \sigma^{a} * \zeta 
+ h_a \tau^{(110)}  \sigma^{a} * \zeta) 
\no\\
(i_2) && 
- {1 \over 16} (- i g_a \tau^{(101)}  \sigma^{a} * \zeta 
- 3 h_a \tau^{(110)}  \sigma^{a} * \zeta) 
\no\\
(a) && 
- {1 \over 16} \bigg(3 i g_a \tau^{(010)}  * \zeta 
- i g_b \tau^{(010)}  \sigma^{ab} * \zeta 
- 3 h_a \tau^{(001)}  * \zeta 
+ h_a \tau^{(001)}  \sigma^{ab} * \zeta \bigg) \no
\eea

\subsubsection{Assembling the complete gravitino BPS equation}

Now we combine the three equations $(\ref{redomega})$, $(\ref{redphi})$, 
and $(\ref{redg})$ to obtain the reduced gravitino equations.  
We again argue that the $\Gamma_M \chi^{\eta_1 \eta_2 \eta_3}$ 
are linearly independent which leads to the equations
\bea
(m) &\qquad& 
0 = {1 \over 2 f_4} \tau^{(300)}  \zeta 
+ \half {D_a f_4 \over f_4} \tau^{(111)}  \sigma^{a} \zeta
+ {1 \over 2}  f_a \tau^{(011)}  \sigma^{a} \zeta 
\no\\&& \hskip 0.4in
+ {1 \over 16} i g_a \tau^{(101)}  \sigma^{a} * \zeta 
- {1 \over 16} h_a \tau^{(110)}  \sigma^{a} * \zeta
\no\\
(i_1) &\qquad& 
0 = {i \over 2 f_1} \tau^{(130)}  \zeta 
+ \half {D_a f_1 \over f_1} \tau^{(111)}  \sigma^{a} \zeta
- {1 \over 2}  f_a \tau^{(011)}  \sigma^{a} \zeta
\no\\&& \hskip 0.4in
- {3 \over 16} i g_a \tau^{(101)} \sigma^{a} * \zeta 
- {1 \over 16} h_a \tau^{(110)}  \sigma^{a} * \zeta
\no\\
(i_2) &\qquad& 
0 = {i \over 2 f_2} \tau^{(113)}  \zeta 
+ \half {D_a f_2 \over f_2} \tau^{(111)}  \sigma^{a} \zeta
- {1 \over 2}  f_a \tau^{(011)}  \sigma^{a} \zeta
\no\\&& \hskip 0.4in
+ {1 \over 16} i g_a \tau^{(101)} \sigma^{a} * \zeta 
+ {3 \over 16} h_a \tau^{(110)}  \sigma^{a} * \zeta
\no\\
(a) &\qquad& 
0 = \bigg( D_a  + {i \over 2} \hat \omega _a   \sigma^{3} \bigg) \zeta
- {i \over 2} q_a  \zeta + {1 \over 2} f_b \tau^{(100)}  \sigma^{b} \sigma_a  \zeta
\no\\&& \hskip 0.4in
- {3 \over 16} i g_a \tau^{(010)} * \zeta + {1 \over 16} i g_b \tau^{(010)}  \sigma^{ab} * \zeta 
\no\\&& \hskip 0.4in
+ {3 \over 16} h_a \tau^{(001)}  * \zeta - {1 \over 16} h_b \tau^{(001)}  \sigma^{ab} * \zeta
\eea
where $\hat \omega _a = (\hat \omega _{89})_a$ is the spin connection along $\Sigma$.
In the first three equations, we have dropped an overall factor of $\G_M$.  
In the last equation, we have used the connection formula $(\ref{spincon})$ 
for the covariant derivative and the fact $\Gamma^{89} = i \, \sigma^3$.  
Eliminating the star using the definition $(\ref{star})$, 
$* \zeta = \tau^{(332)} \otimes \sigma^2 \zeta^*$.  The system of gravitino BPS 
equations is then
\bea
(m) &\qquad& 
0 = {1 \over 2 f_4} \tau^{(300)}  \zeta 
+ \half {\p_a f_4 \over f_4} \tau^{(111)}  \sigma^{a} \zeta
+ {1 \over 2}  f_a \tau^{(011)}  \sigma^{a} \zeta 
\no\\&& \hskip 0.4in
+ {1 \over 16} i g_a \tau^{(233)}  \sigma^{a} \sigma^2 \zeta^* 
+ {1 \over 16} h_a \tau^{(222)}  \sigma^{a} \sigma^2 \zeta^*
\no\\
(i_1) &\qquad& 
0 = {i \over 2 f_1} \tau^{(130)} \zeta 
+ \half {\p_a f_1 \over f_1} \tau^{(111)}  \sigma^{a} \zeta
- {1 \over 2}  f_a \tau^{(011)}  \sigma^{a} \zeta
\no\\&& \hskip 0.4in
- {3 \over 16} i g_a \tau^{(233)}  \sigma^{a} \sigma^2 \zeta^* 
+ {1 \over 16} h_a \tau^{(222)}  \sigma^{a} \sigma^2 \zeta^*
\no\\
(i_2) &\qquad& 
0 = {i \over 2 f_2} \tau^{(113)}  \zeta 
+ \half {\p_a f_2 \over f_2} \tau^{(111)}  \sigma^{a} \zeta
- {1 \over 2}  f_a \tau^{(011)} \sigma^{a} \zeta
\no\\&& \hskip 0.4in
+ {1 \over 16} i g_a \tau^{(233)}  \sigma^{a}  \sigma^2 \zeta^*
 - {3 \over 16} h_a \tau^{(222)}  \sigma^{a} \sigma^2 \zeta^*
\no\\
(a) &\qquad& 
0 = \bigg( \p_a I_8 + {i \over 2} \hat \omega_a  \sigma^{3} \bigg) \zeta
- {i \over 2} q_a  \zeta + {1 \over 2} f_b \tau^{(100)}  \sigma^{b} \sigma_a  \zeta
\no\\&& \hskip 0.4in
- {3 \over 16} g_a \tau^{(322)}  \sigma^2 \zeta^* 
+ {1 \over 16} g_b \tau^{(322)}  \sigma^{ab} \sigma^2 \zeta^* 
\no\\&& \hskip 0.4in
+ {3 \over 16} i h_a \tau^{(333)}  \sigma^2 \zeta^* 
- {1 \over 16} i h_b \tau^{(333)}  \sigma^{ab} \sigma^2 \zeta^*
\eea
Upon multiplying the $(m)$, $(i_1)$ and $(i_2)$ equations by $\tau ^{(111)}$,
we recover (\ref{gravitino1}).

\newpage 

\section{Solving the bilinear constraints}
\setcounter{equation}{0}
\label{appE}

In view of the chirality constraint,  certain bilinears, with insertions of an arbitrary 
Hermitian $\tau$-matrix $M$, and arbitrary $\sigma$-matrix $\sigma ^p$, for $p=0,1,2,3$,
vanish automatically,
\bea
\label{cb}
\zeta ^\dagger  M  \sigma ^p \zeta = 0,  \qquad 
    & \qquad \hbox{if} \qquad &
    \left \{ M \sigma ^p , \tau^{(111)} \sigma ^3 \right \} =0
\eea

\subsection{Construction of vanishing bilinears}

We form bilinears with the help of $\tau$-matrices $T$ such that
\bea
\left ( T \tau ^{(333)} \right ) ^t = - T \tau ^{(333)}
\hskip 1in 
\left ( T \tau ^{(322)} \right ) ^t = - T \tau ^{(322)}
\eea
Using the antisymmetry of these matrices, it is manifest that the 
following combinations, involving the fluxes, will  vanish,
\bea
0 & = & \zeta ^t T g_a \tau ^{(322)} \sigma ^2 \sigma ^a \zeta =
\zeta ^t T h_a \tau ^{(333)} \sigma ^2 \sigma ^a \zeta
\no \\
0 & = & \zeta ^\dagger T g_a \tau ^{(322)} \sigma ^2 \sigma ^a \zeta^* =
\zeta ^\dagger T h_a \tau ^{(333)} \sigma ^2 \sigma ^a \zeta^* 
\no \\
0 & = & \zeta ^\dagger T g_a \tau ^{(322)} (3 \delta ^{ab} + \sigma ^{ab}) \zeta^* =
\zeta ^\dagger T h_a \tau ^{(333)} (3 \delta ^{ab} + \sigma ^{ab}) \zeta^* 
\eea
When $T$ commutes with $\tau^{(111)}$, these identities already follow from 
the chirality constraint (\ref{cb}). Thus, new identities will
be obtained only for $T$ anti-commuting with $\tau^{(111)}$, namely
\bea
T \in \cT = \left  \{ \tau^{(310)}, ~  \tau^{(301)}, ~  \tau^{(201)}, ~ \tau^{(210)} \right \}
\eea
Using these combinations in the dilatino BPS equation, as well as in $\sigma ^3$ 
times the dilatino BPS equation, we obtain two constraints, 
\bea
\p_a B \, \zeta ^\dagger T \sigma ^a \zeta 
=\p_a B \, \zeta ^\dagger T \sigma ^3 \sigma ^a \zeta = 0
\hskip 1in T \in \cT
\eea
Using the fact that $\sigma ^3 \sigma ^a = i \epsilon ^{ab} \sigma ^b$, 
with $\epsilon ^{12}=1$, we see that $\zeta ^\dagger T \sigma ^a \zeta$ 
dotted into $\p_a B$ as well as into $\epsilon ^{ab} \p_b B$ vanishes. 
Whenever $\p_a B \not= 0$, this implies (\ref{Tcon}).

\subsubsection{The gravitino equations algebraic in $\zeta$}

We now construct another set of bilinear constraints.  
We multiply to the left the $(m)$, $(i_1)$ and $(i_2)$ algebraic gravitino equations 
given in (\ref{gravitino1}) by 
$\zeta ^\dagger T \sigma ^p$, for $p=0,3$, and use the fact that the $g_a, h_a$
terms cancel and obtain, after some minimal simplifications, 
\bea
(m) &\qquad& 
0 = - i \zeta ^\dagger T \sigma ^p \tau^{(211)}  \zeta 
+  \p_a f_4 \zeta ^\dagger T \sigma ^p \sigma^{a} \zeta
+ f_4  f_a \zeta ^\dagger T \sigma ^p \tau^{(100)}  \sigma^{a} \zeta 
\no\\
(i_1) &\qquad& 
0 =  \zeta ^\dagger T \sigma ^p \tau^{(021)}  \zeta 
+  \p_a f_1 \zeta ^\dagger T \sigma ^p  \sigma^{a} \zeta
- f_1  f_a \zeta ^\dagger T \sigma ^p \tau^{(100)}  \sigma^{a} \zeta
\no\\ 
(i_2) &\qquad& 
0 =  \zeta ^\dagger T \sigma ^p \tau^{(002)}  \zeta 
+  \p_a f_2 \zeta ^\dagger T \sigma ^p  \sigma^{a} \zeta
- f_2 f_a \zeta ^\dagger T \sigma ^p \tau^{(100)}  \sigma^{a} \zeta
\eea
The terms multiplying $\p_a f_1, \p_a f_2, \p_a f_4$ cancel via (\ref{Tcon}). 
Since for all $\tau \in \cT$,
$T \tau ^{(100)}$ anti-commutes with $\tau^{(111)}$, it follows
from (\ref{cb}) that the last term of each line vanishes, leaving
 three sets of new equations, for $p=0,3$,
\bea
 \zeta ^\dagger T \sigma ^p \tau^{(211)}  \zeta 
=
 \zeta ^\dagger T \sigma ^p \tau^{(021)}  \zeta
=
\zeta ^\dagger T \sigma ^p \tau^{(002)}  \zeta
=0
\eea
or more explicitly, the constraints (\ref{Ucon}). 

\subsection{Solution to the $T$- and $U$-constraints}

We begin by solving the $U$-constraints. Since the BPS equations are 
linear in $\zeta$ and $\zeta ^*$, the space of all $\zeta$ forms a 
vector space, in which linear combinations with real coefficients of solutions
produce again solutions. Since  $U \sigma ^p$, for $p=0,3$ is Hermitian, 
it is immediate, from (\ref{Ucon}), that  any two solutions $\zeta $ and $\zeta '$ must satisfy
\bea
( \zeta ') ^\dagger U \sigma ^p   \zeta  =0 \hskip 1in p=0,3
\eea
Since the constraints hold for the insertions of both $\sigma ^0$ and $\sigma ^3$, 
they hold separately on the chiral components $\zeta _\pm$ of $\zeta$, which obey
$\sigma ^3 \zeta _\pm = \pm \zeta _\pm$. The set of 
$U$ matrices is invariant under multiplication by $\tau ^{(011)}$. 
Thus, the $U$-constraints may be recast in the following form,
\bea
\label{Uconred}
(\zeta_\pm ') ^\dagger U_0 \left ( 1 \mp \tau ^{(111)} \right ) 
\left ( 1 + s \tau ^{(011)} \right ) \zeta_\pm &  = & 0
\eea
for $s = \pm 1$ independently of $\pm$, and $U_0 \in \{ \tau ^{(001)}, ~
\tau ^{(220)}, ~ \tau ^{(212)} \}$. 
For any fixed $\zeta '$, these are linear projector equations for $\zeta_\pm $.
The projector reduces the 2-dimensional  space of $\zeta_+$ to a 1-dimensional one.

\subsubsection{Solving the two-dimensional reduced problem}

The reduced problem (\ref{Uconred}) is effectively 2-dimensional, and thus takes the form 
\bea
(\xi ')^\dagger M \xi =0, \hskip 1in \xi = \left ( \matrix{\xi _1 \cr \xi_2\cr } \right ),
\hskip 1in \xi '= \left ( \matrix{\xi _1 '\cr \xi_2 ' \cr } \right )
\eea
where  $\xi_1, \xi_2, \xi_1', \xi_2' \in \bC$, and $M$ is a  $2\times 2$ Hermitian 
traceless matrix. Diagonalizing  $M$ by a unitary 
matrix $u$, we have $\xi = u \xi_0$, $\xi '= u \xi'_0$ and $M = \lambda u \tau^3 u^\dagger$ 
for some real constant $\lambda$.
The condition $\xi _0^\dagger \tau ^3 \xi _0=0$ has a one-dimensional vector space of 
solutions, whose basis spinor may be chosen to obey 
$\tau ^1 \xi_0 = + \xi_0$. Other choices for the basis spinor are related to the above by 
a unitary transformation (and a scaling). For example, one could have chosen 
instead the basis spinor to be $\tilde \xi _0= \tau ^2 \xi_0$, so that 
$\tau ^3 \tilde \xi_0 = - \tilde  \xi_0$.

\medskip

The results may be summarized as follows. The general solution to  
$\xi ^\dagger M \xi =0$ is completely characterized by a linear projector 
condition $P \xi = + \xi$  (or equivalently  $P \xi = - \xi$), where $P$ is 
Hermitian, and satisfies   $P^2=I$, and $\{ M, P\}=0$.

\subsubsection{Solution of the $U$ constraints}

In view of the preceeding analysis, it suffices to  find a Hermitian involution 
that anti-commutes with all  the $U$ matrices. There are 2 such matrices, 
$\tau ^{(032)}$ and $ \tau ^{(123)}$, 
which, under multiplication by $\tau^{(111)}$,  are equivalent to one another.
Thus, the $U$-constraints are solved by requiring that $\zeta$ satisfy
\bea
\label{nus}
\tau ^{(032)} \zeta _\pm = \nu _\pm \zeta _\pm
\eea
Here, $\nu _+$ can be either $+1$ or $-1$ but not both. The same holds independently
for $\nu _-$.

\subsubsection{Solution of the $T$ constraints}

Assuming that $\zeta$ satisfies the chirality condition $\sigma ^3 \tau ^{(111)}= - \zeta$,
as well as the conditions (\ref{nus}), we now enforce the constraint 
$\zeta ^\dagger T \sigma ^a \zeta =0$, for $a=1,2$, and $T\in \cT$. Choosing instead
the basis $a=+,-$, and using hermiticity of $T$, we get equivalently,
\bea
\zeta _\pm^\dagger \tau ^{(310)} \left ( 1 \mp \tau ^{(111)} \right ) 
\left ( 1 + s \tau ^{(011)} \right ) \zeta _\mp = 0
\eea
for $s=\pm 1$ independently of $\pm$. This problem is again 2-dimensional and we may
take the representation $\tau ^{(310)} \to \tau^3$ and $\tau ^{(032)} \to \nu _+ \tau ^1$.
As a result, we have $\xi_+$ proportional to $\xi_0$, forcing also
$\xi _-$ to be proportional to $\xi _0$. Thus, $\xi_-$ must be proportional to $\xi_+$,
which gives the general solution for the $T$-constraint, (\ref{cJsol}), with
$\nu_+ = \nu _-$.

\newpage 

\section{Metric factors expressed solely in terms of $h_1$ and $h_2$}
\setcounter{equation}{0}
\label{appF}

In analyzing the regularity conditions of the local solutions, it will be 
useful to have alternative expressions for the metric factors.  
To help simplify the presentation, we use the following definitions,
\bea
W &\equiv& \p_w h_1 \p_{\bar w} h_2 + \p_{\bar w} h_1 \p_w h_2
\no\\
N_1 &\equiv& 2 h_1 h_2 |\p_w h_1|^2 - h_1^2 W
\no\\
N_2 &\equiv& 2 h_1 h_2 |\p_w h_2|^2 - h_2^2 W
\eea
The metric factors are derived from (\ref{metricbilin2}), first 
in terms of $\alpha$ and $\beta$, and then, using (\ref{albetsol1}),  in terms
of $\lambda$, $\kappa$, $\phi$ and $\rho$.  
Finally, one eliminates $\phi$ using (\ref{dilsol}), $\rho$ using (\ref{rhosol1}), 
and $\kappa$ and $\lambda$ using (\ref{h1h2lambda}).  In the process, one 
must choose signs for various square roots.  
For example, to obtain $\rho^2$ from the expression for $\rho^8$ in (\ref{rhosol1}), 
the first sign choice is dictated by the sign of $W$, while the second sign choice 
is determined by the sign of the product $h_1 h_2$.  Without loss of generality, 
we may choose $N_2 > 0$ and $h_1 h_2 > 0$, by reversing the signs
respectively of $h_1$ and $h_2$, if necessary. 
A sign choice for $W$ must still be made, but this
sign choice yields different formula for the $AdS_4$ metric factor $f_4$.  
For the case $W > 0$, the metric factors are given by
\bea
\rho^2 &=& e^{- \phi} {\sqrt{N_2 W} \over h_1 h_2}
\no\\
f_1^2 &=& 2 e^{- \phi} h_2^2 \sqrt {W \over N_2}
\no\\
f_2^2 &=& 2 e^{+ \phi} h_1^2 \sqrt {W \over N_1}
\no\\
f_4^2 &=& 2 e^{3 \phi} h_1^2 h_2^2 
{| \p_w h_1 \p_{\bar w} h_2 - \p_{\bar w} h_1 \p_w h_2|^2 \over N_2 \sqrt{W N_2} }
\eea
For the case $W < 0$, the metric factors are given by
\bea
\rho^2 &=& e^{- \phi} {\sqrt{N_2 |W|} \over h_1 h_2}
\no\\
f_1^2 &=& 2 e^{+ \phi} h_1^2 \sqrt {|W| \over N_1}
\no\\
f_2^2 &=& 2 e^{- \phi} h_2^2 \sqrt {|W| \over N_2}
\no\\
f_4^2 &=& 2 e^{- \phi} \sqrt{ N_2 \over |W|}
\eea

\end{document}